\documentclass[pra,aps,floatfix,twocolumn,amsmath,amssymb]{revtex4-2}
\usepackage{graphicx}
\usepackage{color}
\usepackage{bm}
\usepackage{hyperref}
\hypersetup{colorlinks=true,citecolor=blue,linkcolor=blue,urlcolor=blue}
\allowdisplaybreaks

\begin{document}

\title{Dynamics of correlation spreading
in low-dimensional transverse-field Ising models
}

\author{Ryui Kaneko}
\thanks{Current address:
Waseda Research Institute for Science and Engineering,
Waseda University, Shinjuku, Tokyo 169-8555, Japan,
and
Department of Engineering and Applied Sciences,
Sophia University, Chiyoda, Tokyo 102-8554, Japan}
\email{ryuikaneko@aoni.waseda.jp}
\affiliation{Department of Physics, Kindai University, Higashi-Osaka, Osaka 577-8502, Japan}

\author{Ippei Danshita}
\email{danshita@phys.kindai.ac.jp}
\affiliation{Department of Physics, Kindai University, Higashi-Osaka, Osaka 577-8502, Japan}

\date{\today}
 
\begin{abstract}
We investigate the dynamical spreading of spatial
correlations after a quantum quench starting from a magnetically
disordered state in the transverse-field Ising model at one (1D) and two
spatial dimensions (2D).
We analyze specifically the longitudinal and
transverse spin-spin correlation functions at equal time with use of
several methods.
From the comparison of the results in 1D obtained
by the linear spin-wave approximation (LSWA) and those obtained by the rigorous analytical
approach, we show that the LSWA can asymptotically reproduce the exact group velocity in the limit
of strong transverse fields while it fails to capture the detailed time dependence of the correlation
functions.
By applying the LSWA to the 2D case, in which the rigorous analytical
approach is unavailable, we estimate the propagation velocity to be
$Ja/(2\hbar)$
at the strong-field limit, where $J$ is the Ising interaction and $a$ is the
lattice spacing. We also utilize the tensor-network method based on the
projected-entangled pair states for 2D and quantitatively compute the
time evolution of the correlation functions for a relatively short time.
Our findings provide useful benchmarks for quantum simulation
experiments of correlation spreading and theoretical refinement of the
Lieb-Robinson bound in the future.
\end{abstract}

\maketitle

\section{Introduction}
\label{sec:intro}

Neutral atoms trapped in optical-tweezer arrays are promising platforms
for analog quantum simulations~\cite{browaeys2020,morgado2021,wu2021}.
The controllability of individual atoms with laser pulses and
of interatomic interactions via Rydberg excitations
enables one
to realize fast and high-fidelity quantum operations.
Recent rapid technological developments allow for
manipulating many Rydberg atoms in large
arrays~\cite{scholl2021,bluvstein2021,ebadi2021}
and investigating the ground state of quantum lattice systems
experimentally~\cite{bernien2017,keesling2019,deleseleuc2019,verresen2021,semeghini2021,samajdar2020,samajdar2021,wang2021,liao2021}.
Such experiments have also stimulated theoretical research on
fundamental quantum many-body systems.
For instance,
the ground-state phase diagrams of the transverse-field Ising model
and those of its strong Ising interaction limit, the PXP model,
have been intensively examined using the quantum Monte Carlo
method~\cite{lienhard2018,kaneko2021,yue2021,merali2021_arxiv}.

Rydberg-atom arrays have also given the opportunity to
study the nonequilibrium dynamics
of isolated quantum many-body systems,
which are hard to simulate numerically
with classical computers.
In particular,
the correlation-spreading dynamics of quantum Ising
models~\cite{guardado-sanchez2018,lienhard2018}
is one of the intriguing topics
that is likely to be further addressed.
At present, experiments with more than $200$ Rydberg atoms 
are feasible~\cite{scholl2021,bluvstein2021,ebadi2021},
allowing one to study unprecedentedly large lattice systems
in one (1D) and two spatial dimensions (2D).

These recent experiments on long-time dynamics
in quantum many-body systems
have motivated us to
quantitatively calculate the velocity of the correlation propagation,
which will serve as useful references for future experiments.
In general, there are two kinds of propagation velocities
for correlation spreading dynamics:
one is the phase velocity and the other is the group velocity.
The former can be captured by the first peak of the wave packet,
whereas the latter can be extracted by the envelope of the wave packet.
The group velocity is bounded from above
in nonrelativistic quantum systems,
and this upper limit is known as the Lieb-Robinson
bound~\cite{lieb1972,hastings2010_arxiv}.

While significant progress has been made
concerning rigorous inequalities
related to the Lieb-Robinson bound,
such inequalities do not necessarily offer practical reference values
for experiments.
Usually, the Lieb-Robinson bound is intended to
provide general conditions for arbitrary correlations.
Consequently,
the bound can be too loose and sometimes meaningless
when examining the propagation velocity
of particular correlation functions that are
measurable in experiments.
With this in mind,
the Lieb-Robinson bound has been improved
very recently~\cite{wang2020};
however, their method still gives a looser bound
than the exact solution if it is available.

In some cases,
direct numerical simulations on classical computers
would give much more detailed information
about correlation spreading
than rigorous inequalities for the Lieb-Robinson bound.
Such numerical data would also strengthen the validity
of experimental findings through cross-checking
experimental and theoretical results.
Indeed,
many numerical efforts have been made
to calculate the quench or sweep dynamics in 1D and 2D.
These attempts include
the time-dependent variational Monte Carlo method
with the Slater-Jastrow wave function~\cite{blass2016}
and with more sophisticated neural-network wave
functions~\cite{schmitt2018,schmitt2020,
gutierrez2022,schmitt2022,schmitt2022b,lin2022,donatella2022_arxiv},
the form factor expansions~\cite{granet2020},
the numerical linked-cluster
expansion~\cite{white2017_arxiv,richter2020,gan2020},
the tensor-network method based on matrix product states
(MPS)~\cite{schollwock2011,haegeman2016,hashizume2020,hashizume2022},
and that based on projected
entangled pair states (PEPS)~\cite{kshetrimayum2017,czarnik2019,
hubig2019,dziarmaga2021,dziarmaga2022b,schmitt2022,lin2022b}.

In this paper,
we study quench dynamics in
the transverse-field Ising model
on a chain in 1D and that on a square lattice in 2D
by using several methods,
including the tensor-network method based on PEPS and
the linear spin-wave approximation (LSWA).
We take the initial state to be the magnetically disordered product state,
which is the ground state in the strong-field limit, and calculate time
evolution of spin-spin correlations at equal time after a sudden quench
of the transverse field. We focus on the quench within a parameter
region where the ground state is magnetically disordered.
We extract the group velocity of the correlation propagation from the
spin-spin correlations for several values of the transverse field. In
the 1D case, we show that the group velocity extracted from the LSWA
results asymptotically approaches that
extracted from the rigorous analytical
results with increasing the transverse field,
while the agreement in the time dependence of the correlation
functions is limited to a 
short time
before the first peak appears.
Our results indicate that
the LSWA can quantitatively predict the propagation velocity as long as the
final transverse field is sufficiently strong.
In the 2D case, using the LSWA, we estimate the group velocity to be
$Ja/(2\hbar)$, where $J$ is the Ising interaction and $a$ is the
lattice spacing. We use the PEPS method
in a complementary way to perform
more
quantitative calculations on the time evolution of the correlation
functions for a relatively short time.

This paper is organized as follows:
In Sec.~\ref{sec:method},
we introduce the model
and all the analytical and numerical methods 
used in this study.
In Secs.~\ref{sec:results_1d} and \ref{sec:results_2d},
we present the time-dependent spin-spin correlation functions
and extract the corresponding group velocity
in 1D and 2D, respectively.
We discuss the relation between our propagation velocity
and the Lieb-Robinson bound proposed recently,
and draw our conclusions in Sec.~\ref{sec:summary}.
For simplicity, we set $\hbar=1$ throughout this paper.
 \section{Model and methods}
\label{sec:method}

We consider the transverse-field Ising model
with the periodic boundary condition defined as
\begin{align}
 \hat{H}
 =
 - J \sum_{\langle i,j\rangle} \hat{S}_i^z \hat{S}_j^z
 - \Gamma \sum_{i} \hat{S}_i^x,
\label{eq:tfising_model}
\end{align}
where $\hat{S}_i^z$ and $\hat{S}_i^x$
correspond to the $z$ and $x$ components of the $S=1/2$ Pauli spin,
$J$ represents the strength of the spin exchange interaction,
and $\Gamma$ represents the strength of the transverse field.
The symbol $\langle i,j\rangle$ means that
the sum is taken over nearest-neighbor sites.
We focus on the ferromagnetic spin exchange interaction ($J>0$)
on a chain in 1D and that on a square lattice in 2D.
Both ferromagnetic and antiferromagnetic models
are equivalent under appropriate unitary transformations
for bipartite lattices.
The ground state is ordered (disordered)
for $\Gamma<\Gamma_{\mathrm{c}}$ ($\Gamma>\Gamma_{\mathrm{c}}$),
where $\Gamma_{\mathrm{c}}$ is the transition point
given as
$\Gamma_{\mathrm{c}}/J=1/2$~\cite{pfeuty1970} in 1D
and
$\Gamma_{\mathrm{c}}/J\approx 1.522$~\cite{rieger1999,bloete2002,kaneko2021}
in 2D.
Hereafter we take $J$ as the unit of energy.
We also take the lattice constant to be unity
throughout this paper.

We investigate the quench dynamic starting from the disordered state
$|\psi_0\rangle = \otimes_{i}\, |\!\!\rightarrow\rangle_i$
at $\Gamma\rightarrow\infty$
to the disordered parameter region
$\Gamma \in (\Gamma_{\mathrm{c}},\infty)$.
We study the equal-time longitudinal and
connected transverse correlation functions
at distance $\bm{r}$,
which are defined as
\begin{align}
 C^{zz}(\bm{r},t)
 &=
 \langle\psi(t)|
 {\hat{S}}^z_{\bm{r}}  
 {\hat{S}}^z_{\bm{0}}
 |\psi(t)\rangle,
\\
 C^{xx}_{\rm connected}(\bm{r},t)
 &=
 \langle\psi(t)|
 {\hat{S}}^x_{\bm{r}}  
 {\hat{S}}^x_{\bm{0}}
 |\psi(t)\rangle
\nonumber
\\
 &\phantom{=}
 -
 \langle\psi(t)|
 {\hat{S}}^x_{\bm{r}}  
 |\psi(t)\rangle
 \langle\psi(t)|
 {\hat{S}}^x_{\bm{0}}
 |\psi(t)\rangle
\end{align}
with $|\psi(t)\rangle = e^{-i\hat{H}t} |\psi_0\rangle$,
respectively.
Hereafter, we take the lattice spacing to be unity ($a=1$).
In 1D, we obtain them
by the exact calculations via the Jordan-Wigner transformation
and by the LSWA via the Holstein-Primakoff
transformation.
In 2D,
we use the tensor-network method,
the exact diagonalization (ED) method,
and the LSWA.
We will summarize each method below.

We extract the group velocity
from the envelope of the wave packet
in the spin-spin correlation functions.
Let us first discuss
how the correlation spreading is related to
the Lieb-Robinson bound.
In a system with short-range interaction,
a commutator of any operators
$\hat{O}_{\rm A}$ and $\hat{O}_{\rm B}$
in regions A and B
satisfies the relation
\begin{align}
 \bigl\|[\hat{O}_{\rm A}(t), \hat{O}_{\rm B}]\bigr\|
 \le \mathrm{const}
 \times
 \exp\left(-\frac{L-vt}{\chi}\right),
\end{align}
where
$\hat{O}_{\rm A}(t)
= \exp\bigl(i\hat{H}t\bigr) \hat{O}_{\rm A}
\exp\bigl(-i\hat{H}t\bigr)$,
$L$ is the distance between the regions A and B,
and $\chi$ is constant~\cite{lieb1972,hastings2010_arxiv}.
The velocity $v$ corresponds to the Lieb-Robinson bound.
This relation means that
the information from the region A is
transmitted to the region B up to a time
$t\approx L/v$.
Then,
the inequality of the Lieb-Robinson bound ensures that,
for any operators
$\hat{O}_{\rm A}$ and $\hat{O}_{\rm B}$ in regions A and B
having the distance $L$,
the expectation value for a state $|\psi\rangle$
with a finite correlation length $\chi$
satisfies~\cite{bravyi2006}
\begin{align}
 &
 |
 \langle\psi(t)|
 \hat{O}_{\rm A} \hat{O}_{\rm B}
 |\psi(t)\rangle
 -
 \langle\psi(t)|
 \hat{O}_{\rm A}
 |\psi(t)\rangle
 \langle\psi(t)|
 \hat{O}_{\rm B}
 |\psi(t)\rangle
 |
\nonumber
\\
 \le \,
 &
 \mathrm{const} \times
 e^{-\frac{L-2vt}{\chi'}},
\end{align}
where $\chi'$ is a constant that depends on $\chi$.
This velocity $2v$ on the right-hand side
corresponds to twice 
the Lieb-Robinson bound.
When the correlation spreading is well described by the quasiparticle,
the group velocity ($v^{\mathrm{group}}$) of the fastest quasiparticle
is often regarded as the Lieb-Robinson
bound~\cite{calabrese2011,cheneau2012,jurcevic2014,gong2022}.

To estimate the group velocity of the fastest quasiparticle,
we calculate the slope obtained from the peak-time dependence
of the distance.
In general, the maximum group velocity
is larger than the velocity associated with the largest
correlation peak location,
and they do not have to be the same.
On the other hand,
the latter value is easy to extract
and is often regarded as
the maximum group velocity (particularly in experiments).
They do coincide for the quench dynamics
in the 1D transverse-field Ising model,
as we will see later.
Therefore, we regard
the velocity associated with the largest correlation peak location
as the maximum group velocity
and, hereafter, call it the Lieb-Robinson velocity.
To avoid confusion, we will use the term
``Lieb-Robinson bound'' to refer to the actual bound in the inequality
and the term
``Lieb-Robinson velocity'' to refer to the velocity extracted from
peak positions.
The Lieb-Robinson bound is larger than or equal to
the Lieb-Robinson velocity.

Under these circumstances,
the Lieb-Robinson velocity
gives twice the group velocity
($2v^{\mathrm{group}}$) of a
fastest
quasiparticle.
Intuitively,
the factor $2$ originates from pairs of
quasiparticles
moving to the left or right from a given point.
This quasiparticle picture
has been discussed intensively in the dynamics of the Bose-Hubbard
model~\cite{cheneau2012,barmettler2012,despres2019,takasu2020}.
In the present analysis,
we have presented the group velocity $v^{\mathrm{group}}$
of a certain single quasiparticle
estimated from one half of
the slope obtained from the peak-time dependence
of the distance.

\subsection{Exact calculations in 1D}

The analytical form of the time-dependent correlation functions can be
obtained rigorously for the 1D transverse-field Ising
model~\cite{lieb1961,pfeuty1970,barouch1971,
sachdev2011,calabrese2012,calabrese2012b,suzuki2013}.
We briefly review the detailed derivation of the time-dependent
correlation functions in Appendix~\ref{sec:app_exact_corr}
and present the final results below.

The longitudinal correlation function is represented as
a Pfaffian of a $2r\times 2r$ skew symmetric matrix:
\begin{align}
 C^{zz}(r,t)
 &=
 (-1)^{\frac{r(r-1)}{2}} \cdot
 \frac{1}{4}
 \mathrm{Pf}
 \begin{pmatrix}
  S & G \\
  -G^T & Q
 \end{pmatrix}.
\end{align}
Here elements of the matrices $S$, $Q$, and $G$
are defined as
\begin{align}
 s_{i,j}
 &=
\begin{cases}
 S_{i-1,j-1} & \text{if $i<j$},\\
 -S_{j-1,i-1} & \text{if $i>j$},\\
 0 & \text{if $i=j$},
\end{cases}
\\
 q_{i,j}
 &=
\begin{cases}
 Q_{i-1,j-1} & \text{if $i<j$},\\
 -Q_{j-1,i-1} & \text{if $i>j$},\\
 0 & \text{if $i=j$},
\end{cases}
\\
 g_{i,j}
 &=
 G_{i-1,j},
\end{align}
where the time-dependent correlation functions
$S_{i,j}$, $Q_{i,j}$, and $G_{i,j}$
are given as
\begin{align}
 S_{i,j}
 &=
 - 
 \frac{2}{L} \sum_{k>0}
 \bigl\{
   \cos[k(r_i-r_j)] [|\tilde{u}_k(t)|^2 + |\tilde{v}_k(t)|^2]
\nonumber
\\
 &\phantom{=}
 - i \sin[k(r_i-r_j)] [\tilde{u}_k(t) \tilde{v}^{*}_k(t) + \tilde{v}_k(t) \tilde{u}^{*}_k(t)]
 \bigr\},
\\
 Q_{i,j}
 &=
 +
 \frac{2}{L} \sum_{k>0}
 \bigl\{
   \cos[k(r_i-r_j)] [|\tilde{u}_k(t)|^2 + |\tilde{v}_k(t)|^2]
\nonumber
\\
 &\phantom{=}
 + i \sin[k(r_i-r_j)] [\tilde{u}_k(t) \tilde{v}^{*}_k(t) + \tilde{v}_k(t) \tilde{u}^{*}_k(t)]
 \bigr\},
\\
 G_{i,j}
 &=
 -
 \frac{2}{L} \sum_{k>0}
 \bigl\{
   \cos[-k(r_i-r_j)] [|\tilde{u}_k(t)|^2 - |\tilde{v}_k(t)|^2]
\nonumber
\\
 &\phantom{=}
 - i \sin[-k(r_i-r_j)] [\tilde{u}_k(t) \tilde{v}^{*}_k(t) - \tilde{v}_k(t) \tilde{u}^{*}_k(t)]
 \bigr\}.
\end{align}
The symbol $\sum_{k>0}$ means the sum taken over all
$k = 2\pi n/L$
with
$n = 1/2, 3/2, \dots, (L-3)/2, (L-1)/2$ for even $L$.
For the quench starting from the disordered state ($\Gamma\rightarrow\infty$),
the parameters $\tilde{u}_k(t)$ and $\tilde{v}_k(t)$ for $0<k<\pi$ are described as
\begin{align}
 \tilde{u}_k(t)
 &=
 i \frac{\tilde{b}'_k}{\omega'_k} \sin \left( 2\omega'_k \times \frac{tJ}{4} \right),
\\
 \tilde{v}_k(t)
 &=
 - i \cos \left( 2\omega'_k \times \frac{tJ}{4} \right)
 - \frac{\tilde{a}'_k}{\omega'_k} \sin \left( 2\omega'_k \times \frac{tJ}{4} \right)
\end{align}
with
\begin{align}
 \tilde{a}'_k
 &= \frac{2\Gamma}{J} + \cos k,
\\
 \tilde{b}'_k
 &= \sin k,
\\
 \omega'_k
 &= \sqrt{\frac{4\Gamma^2}{J^2} + \frac{4\Gamma}{J} \cos k + 1},
\end{align}
respectively.
Parameters with prime symbols indicate
physical quantities after the quench.
On the other hand,
the transverse correlation function is given as
\begin{align}
 C^{xx}_{\rm connected}(r,t)
 &=
 -\frac{1}{4}
 ( Q_{0,r} S_{0,r} + G_{r,0} G_{0,r}).
\end{align}
We numerically evaluate each correlation function
for sufficiently large systems.
We use the library for Pfaffian computations~\cite{wimmer2012}
in the case of the longitudinal correlation function.

\subsection{Spin-wave approximation}

We investigate a small quench
starting from the completely disordered point ($\Gamma\rightarrow\infty$)
to the parameter within a disordered phase
($\Gamma_{\mathrm{c}}^{\mathrm{classical}}\ll \Gamma < \infty$,
where
$\Gamma_{\mathrm{c}}^{\mathrm{classical}}=JD$
with $D$ being the spatial dimension).
We focus on small quantum fluctuations around
the disordered state
and map quantum Ising spins to bosons
using the linearized Holstein-Primakoff
transformation~\cite{henry2012,cevolani2016,buyskikh2016,menu2018,menu2023}.
The equal-time correlation functions for quantum spins
can be obtained by calculating those for bosons.
They serve as a good approximation
as long as the transverse magnetization is large enough
($\langle S_i^x \rangle \approx 1/2$).
We give the detailed derivation in Appendix~\ref{sec:app_sw_corr}
and show the obtained spin-spin correlation functions below.

The longitudinal correlation function
at distance $\bm{r}$
($1\le r_{\nu}\le L/2$ with $\nu=1,2,\dots,D$)
is given as
\begin{align}
 C^{zz}(\bm{r},t)
 &=
 \frac{S}{2L^D} \sum_{\bm{k}} e^{i\bm{k}\cdot\bm{r}}
 \frac{B'_{\bm{k}}}{A'_{\bm{k}} + B'_{\bm{k}}}
 (\cos 2\Omega'_{\bm{k}} t - 1),
\end{align}
where $S(=1/2)$ is the size of spin
and other parameters are defined as
\begin{align}
 \Omega'_{\bm{k}} &= {\rm sgn}(A'_{\bm{k}}) \sqrt{{A_{\bm{k}}'}^2 - {B_{\bm{k}}'}^2},
\\
 A'_{\bm{k}} &= - \frac{z}{2} J S \gamma_{\bm{k}} + \Gamma,
\\
 B'_{\bm{k}} &= - \frac{z}{2} J S \gamma_{\bm{k}},
\\
 \gamma_{\bm{k}} &= \frac{1}{D} \sum_{\nu=1}^{D} \cos {k}_{\nu}    
\end{align}
with
$z=2D$ being the coordination number.
Parameters with prime symbols correspond to
physical quantities after the quench.
On the other hand,
the transverse correlation function
at distance $\bm{r}$
($1\le r_{\nu}\le L/2$ with $\nu=1,2,\dots,D$)
is given as
\begin{align}
 &~\phantom{=}~
 C^{xx}_{\rm connected}(\bm{r},t)
\nonumber
\\
 &=
 \left|
 \frac{1}{L^D} \sum_{{\bm{k}}} e^{i{\bm{k}}\cdot{\bm{r}}}
 \frac{B'_{\bm{k}}}{2 \Omega'_{\bm{k}}}
 \left[\frac{A'_{\bm{k}}}{\Omega'_{\bm{k}}} (\cos 2\Omega'_{\bm{k}} t - 1) + i \sin 2\Omega'_{\bm{k}} t \right]
 \right|^2
\nonumber
\\
 &~\phantom{=}~
 +
 \left|
 \frac{1}{L^D} \sum_{{\bm{k}}} e^{i{\bm{k}}\cdot{\bm{r}}}
 \frac{{B_{\bm{k}}'}^2}{2{\Omega_{\bm{k}}'}^2}
 (\cos 2\Omega'_{\bm{k}} t - 1)
 \right|^2.
\end{align}
We numerically calculate each correlation function
for sufficiently large systems.

\subsection{2D tensor-network method}
\label{sec:2d_tensor_network}

\begin{figure}[t]
\centering
\includegraphics[width=1.0\columnwidth]{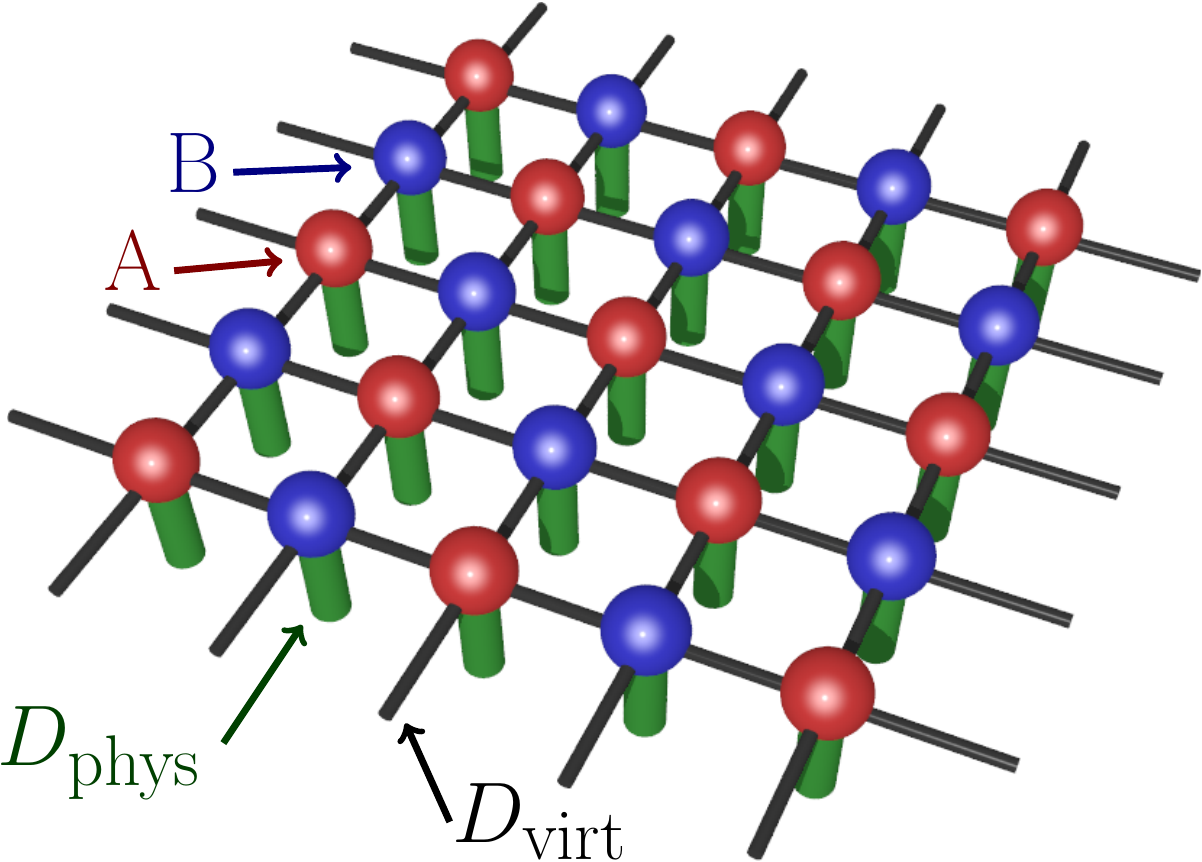}
\caption{Schematic picture of the iPEPS
having a two-site unit-cell structure.
The two sublattice sites are represented by A and B.
Each ball corresponds to a rank-five tensor,
which is located at a lattice site
and has four thin sticks and a thick stick.
The thin and thick sticks represent
the virtual and physical degrees of freedom,
and the bond dimensions of the former and the latter
are defined as $D_{\rm virt}$ and $D_{\rm phys}$, respectively.
}
\label{fig:peps}
\end{figure}

We use
the infinite projected entangled pair state
(iPEPS)~\cite{martin-delgado2001,verstraete2004a_arxiv,
verstraete2004b,verstraete2008,jordan2008,
phien2015,orus2014,orus2019}
or
the infinite tensor product state~\cite{hieida1999,okunishi2000,
nishino2001,maeshima2001,
nishio2004_arxiv}
to investigate short-time dynamics
in the infinite system.
We choose translationally invariant
iPEPS consisting of a two-site unit-cell
structure as shown in Fig.~\ref{fig:peps}.
The dimension of the local Hilbert space is $D_{\rm phys}=2$
for spin $S=1/2$.
The initial state
$|\psi_0\rangle = \otimes_{i} |\rightarrow\rangle_i$
is the ground state in the limit of $\Gamma/J\rightarrow\infty$,
which can be prepared by the virtual bond dimension $D_{\rm virt}=1$.

We apply the simple update
algorithm~\cite{jiang2008,jordan2008}
to simulate the real-time dynamics of
the transverse-field Ising model.
In this algorithm,
we approximate the real-time evolution operator
in a very short-time step $dt$ using the Suzuki-Trotter
decomposition~\cite{trotter1959,suzuki1966,suzuki1976}
and obtain the two-site gate
$
 e^{-idt\hat{H}}
 \approx
 \prod_{\langle i,j \rangle} e^{-idt\hat{H}_{ij}}
$
with
$
 \hat{H}_{ij}
 =
 - J \hat{S}_i^z \hat{S}_j^z
 - \Gamma (\hat{S}_i^x + \hat{S}_j^x) / z
$
($z=4$)
satisfying
$\hat{H} = \sum_{\langle i,j \rangle} \hat{H}_{ij}$.
The gate acts on two neighboring tensors
and increases the virtual bond dimensions.
We truncate the bond dimensions of the local tensors
using the singular value decomposition
so that the bond dimensions of iPEPS remain $D_{\rm virt}$.
Note that
the decomposition temporarily breaks the one-site
translation symmetry into a two-site one
and calls for at least a two-site unit-cell structure
even when the system is translation
invariant~\cite{jiang2008,jordan2008}.
In the actual calculations,
the second-order Suzuki-Trotter decomposition is used,
and the time step is typically chosen as $dt J = 0.001$
for a quench to a strong field $\Gamma$.
Simulations using doubled or halved $dt$ show
no significant change in the short-time dynamics
as for the present model.

\begin{figure}[t]
\centering
\includegraphics[width=1.0\columnwidth]{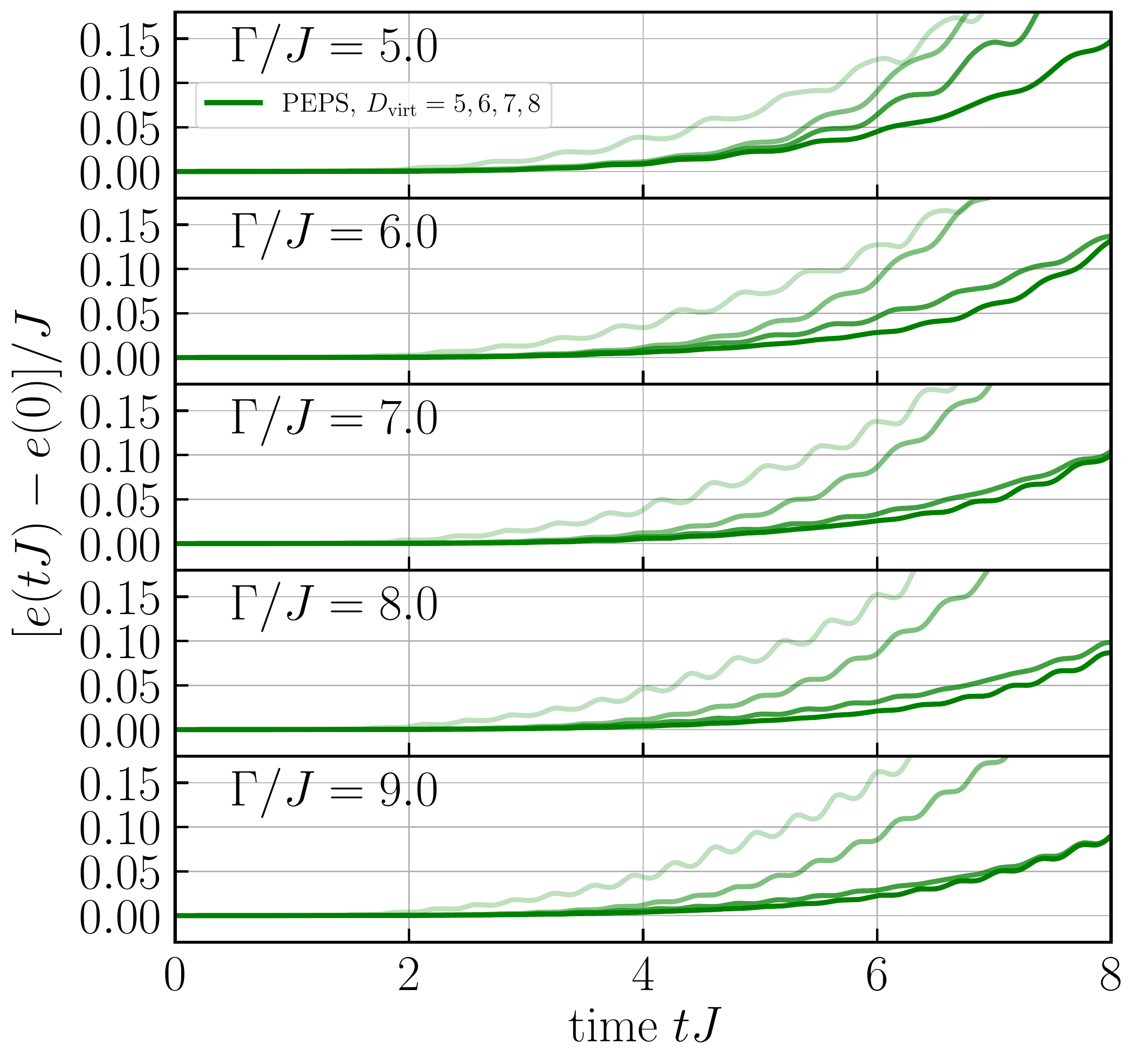}
\caption{Bond-dimension dependence of energy density
obtained by iPEPS simulations.
We consider the quench to $\Gamma/J=5$, $6$, $7$, $8$, and $9$
and subtract the energy density at $t=0$.
The lines correspond to the time-dependent energy density
for the bond dimensions $D_{\rm virt}=5$, $6$, $7$, and $8$
(from lighter to darker).
The energy is nearly conserved for $D_{\rm virt}\ge 6$
in a time frame $tJ\in [0,4]$.
}
\label{fig:2d_peps_ene}
\end{figure}

We improve the accuracy of time-evolved wave functions
by increasing the dimension of the virtual bond $D_{\rm virt}$
and confirm the convergence of physical quantities.
Previous studies~\cite{czarnik2019,dziarmaga2021}
suggest that results for bond dimensions $D_{\rm virt}\gtrsim 6$ 
already show good convergence within
a short-time frame $tJ\lesssim 4$
even in one of the most difficult cases, i.e.,
the quench to the critical point.
(Concerning the unit of time,
the energy scale is four times larger
in previous studies~\cite{czarnik2019,dziarmaga2021}
because they used the Pauli spin $\sigma^z=2S^z$.)

In the present iPEPS simulations,
we adopt the tensor-network library
TeNeS~\cite{tenes,ptns,motoyama2022}
and increase the virtual bond dimensions
up to $D_{\rm virt}=8$ for safety.
In general, as for numerical simulations of a quench dynamics,
the obtained correlations would be reliable
in a short time that the energy is conserved.
We investigate the time dependence of the energy density
in the unit of Ising interaction $J$
for different fields $\Gamma/J$
with increasing the bond dimensions $D_{\rm virt}$
(see Fig.~\ref{fig:2d_peps_ene}).
The energy density is nearly conserved 
for a short time ($tJ\lesssim 4$) when $D_{\rm virt}\ge 6$
regardless of the choice of the transverse field.

The corner transfer matrix renormalization group
method~\cite{nishino1996,nishino1997,nishino1999,
okunishi2000,orus2009,corboz2010,corboz2011,
corboz2014,phien2015,orus2014,orus2019}
is used to calculate physical quantities
in the thermodynamic limit.
We take the bond dimensions of the environment
tensors as $\chi_{\rm virt}=2(D_{\rm virt})^2$ so that physical quantities
are well converged.

\subsection{Exact diagonalization method}

The ED method is often used to get insight into
the dynamics of small quantum many-body
systems~\cite{kollath2007,laeuchli2008,goto2019,kunimi2021,yoshii2022}.
We use the \textsc{QuSpin} library~\cite{weinberg2017,weinberg2019}
for ED calculations.
We consider the system sizes up to $28$ sites
under the periodic boundary condition.
In the present setup,
both the Hamiltonian and the initial state are translationally invariant,
and the total momentum of the initial and time-evolved states
remains zero.
We restrict ourselves to the zero-momentum sector~\cite{sandvik2010}
and follow the dynamics of the state.
Instead of generating matrix elements on the fly
to reduce the memory cost,
we keep all the elements of sparse matrices
in the compressed sparse row format
to accelerate calculations.
To compute the matrix exponential applied to a vector,
we use the Taylor series expansion with error
analysis proposed by Al-Mohy and Higham~\cite{higham2010,almohy2011}.

We confirm that the ED results (up to $28$ sites)
reproduce the exact analytical results in 1D (not shown).
We mainly show the ED results in 2D (for $5\times 5$ sites)
hereafter.

 \section{Results in 1D}
\label{sec:results_1d}

We first present the time dependence of spin-spin correlations in 1D
using the exact analytical approach and the LSWA.
We extract the group velocity after a sudden quench to a strong field
from these data.

\subsection{Exact results}

\begin{figure}[t]
\centering
\includegraphics[width=1.0\columnwidth]{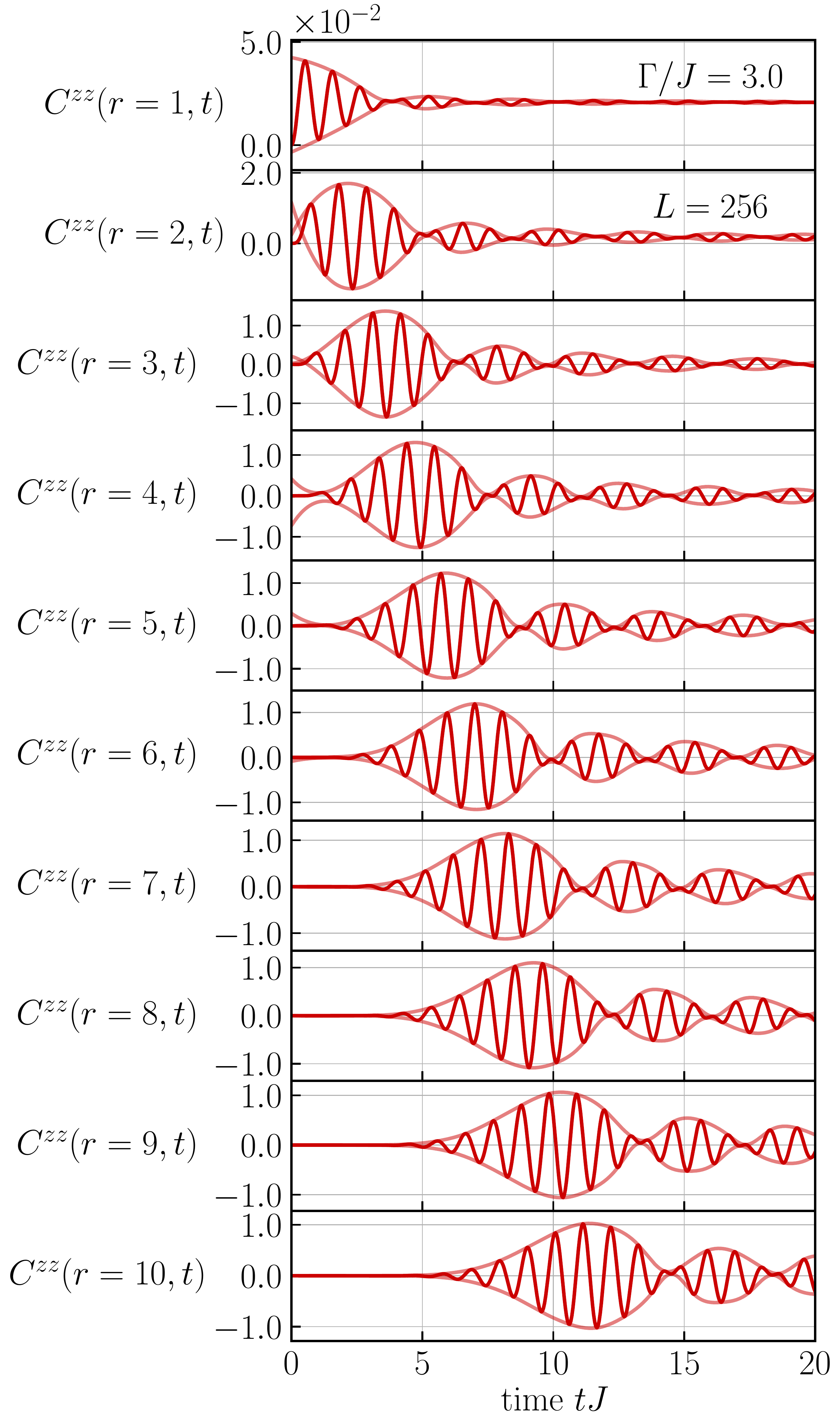}
\caption{Exact equal-time longitudinal correlation functions in 1D.
We consider the quench to $\Gamma/J=3$ for a system size $L=256$
and show the short-time dynamics for distances $r=1$,
$2$, $\dots$, and $10$.
The envelope of each correlation function is a guide to the eye.
The upper (lower) part of the envelope at each distance
is obtained by
first searching all the local maxima (minima) in the correlation
and then interpolating them
using a one-dimensional cubic B-spline curve~\cite{deboor1978}.
}
\label{fig:1d_exact_zz}
\end{figure}

\begin{figure}[t]
\centering
\includegraphics[width=1.0\columnwidth]{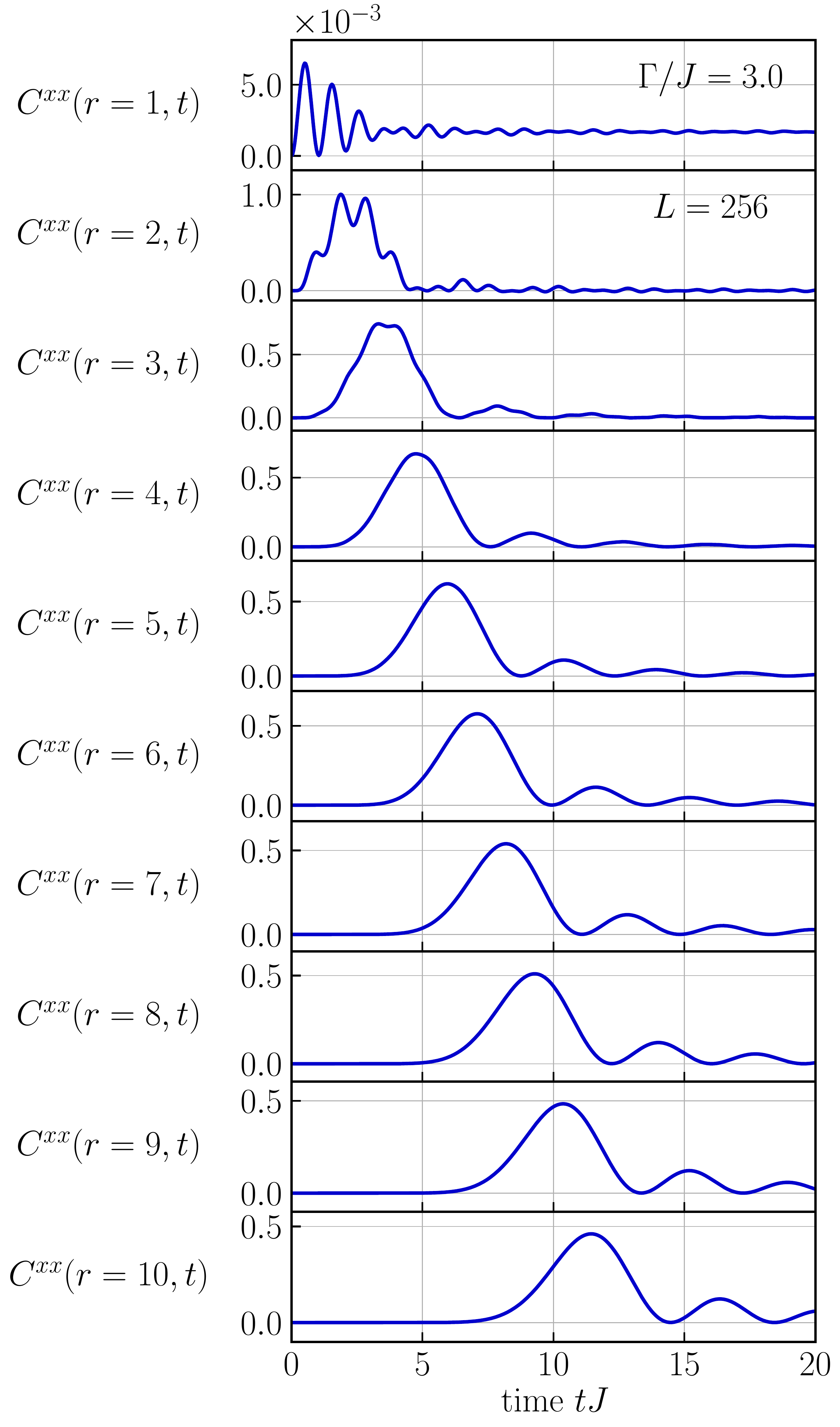}
\caption{Exact equal-time transverse correlation functions in 1D.
The parameters are the same as those in Fig.~\ref{fig:1d_exact_zz}.
}
\label{fig:1d_exact_xx}
\end{figure}

\begin{figure}[t]
\centering
\includegraphics[width=1.0\columnwidth]{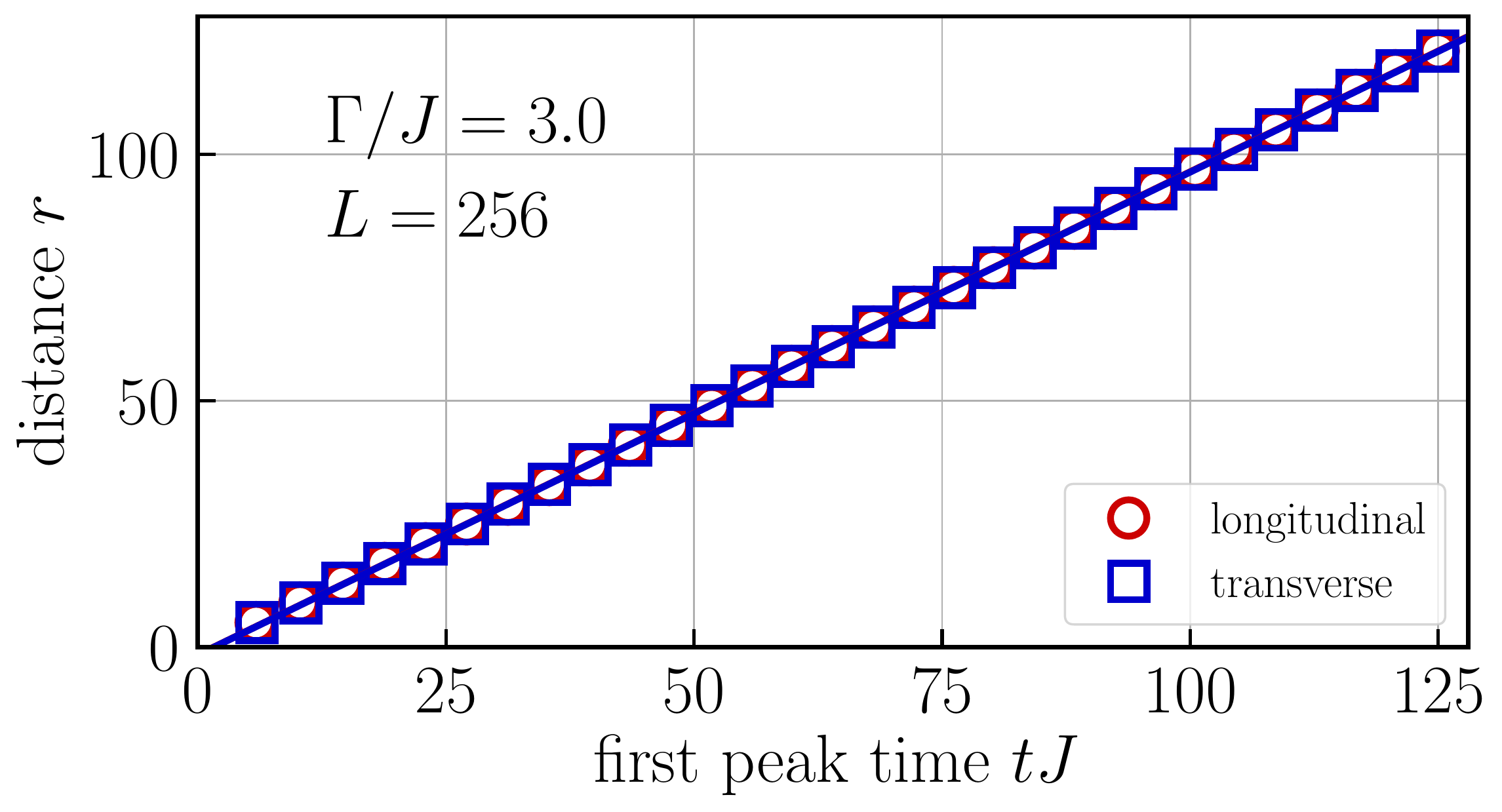}
\caption{First-peak time dependence of distance
for exact correlations in 1D.
We show data points when the distance is a multiple of four.
The group velocity is estimated from
one half of the slope.
}
\label{fig:1d_exact_rt}
\end{figure}

\begin{figure}[t]
\centering
\includegraphics[width=1.0\columnwidth]{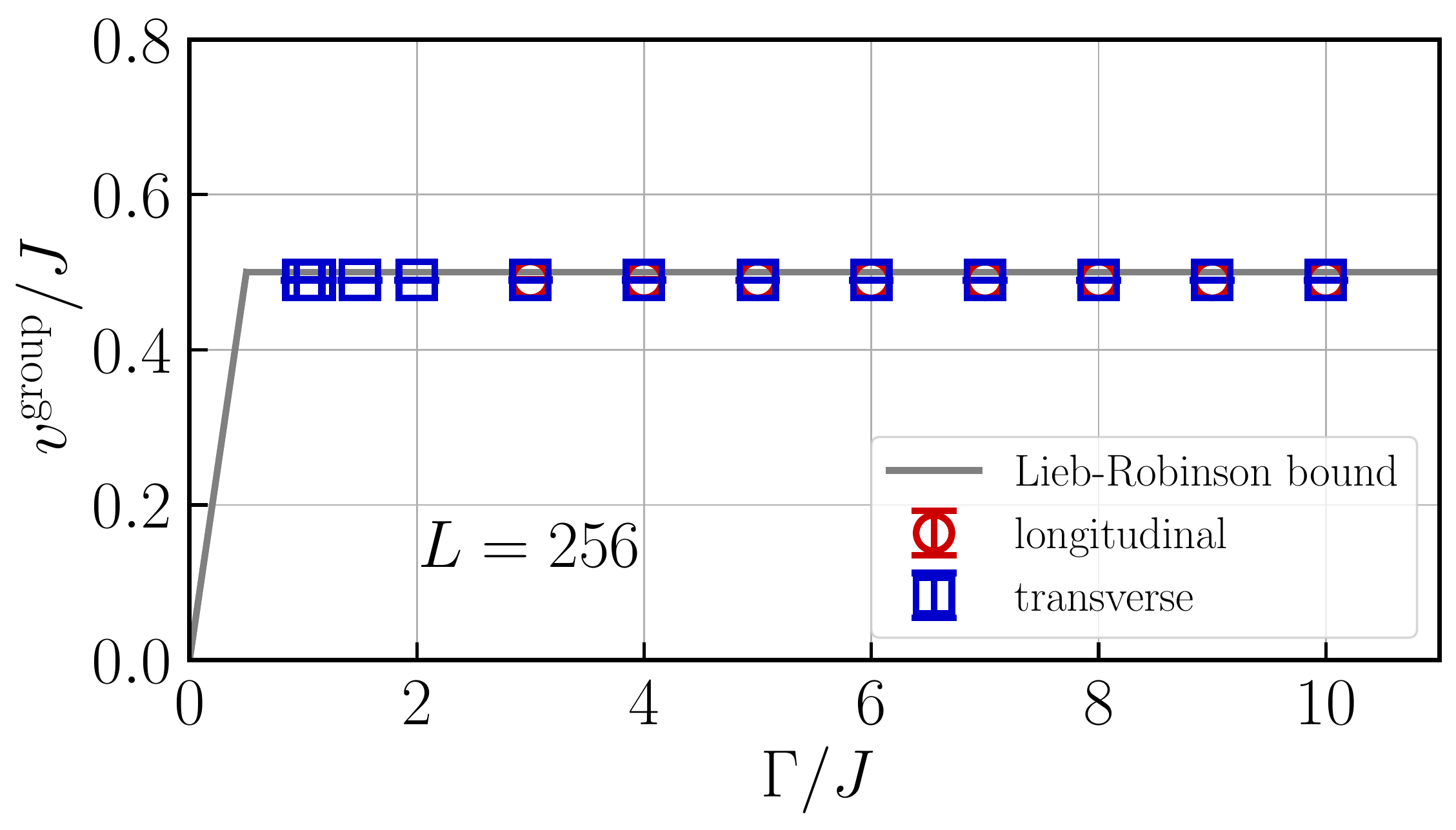}
\caption{Field dependence of the group velocity in 1D.
The exact Lieb-Robinson velocity
$v^{\mathrm{LR}}/J=1/2$
is shown as a reference.
For $\Gamma/J<3$,
we only show the group velocity estimated from
the transverse correlations
because the envelopes of longitudinal correlations
become unclear for a weaker field.
}
\label{fig:1d_exact_v}
\end{figure}

\begin{figure}[t]
\centering
\includegraphics[width=1.0\columnwidth]{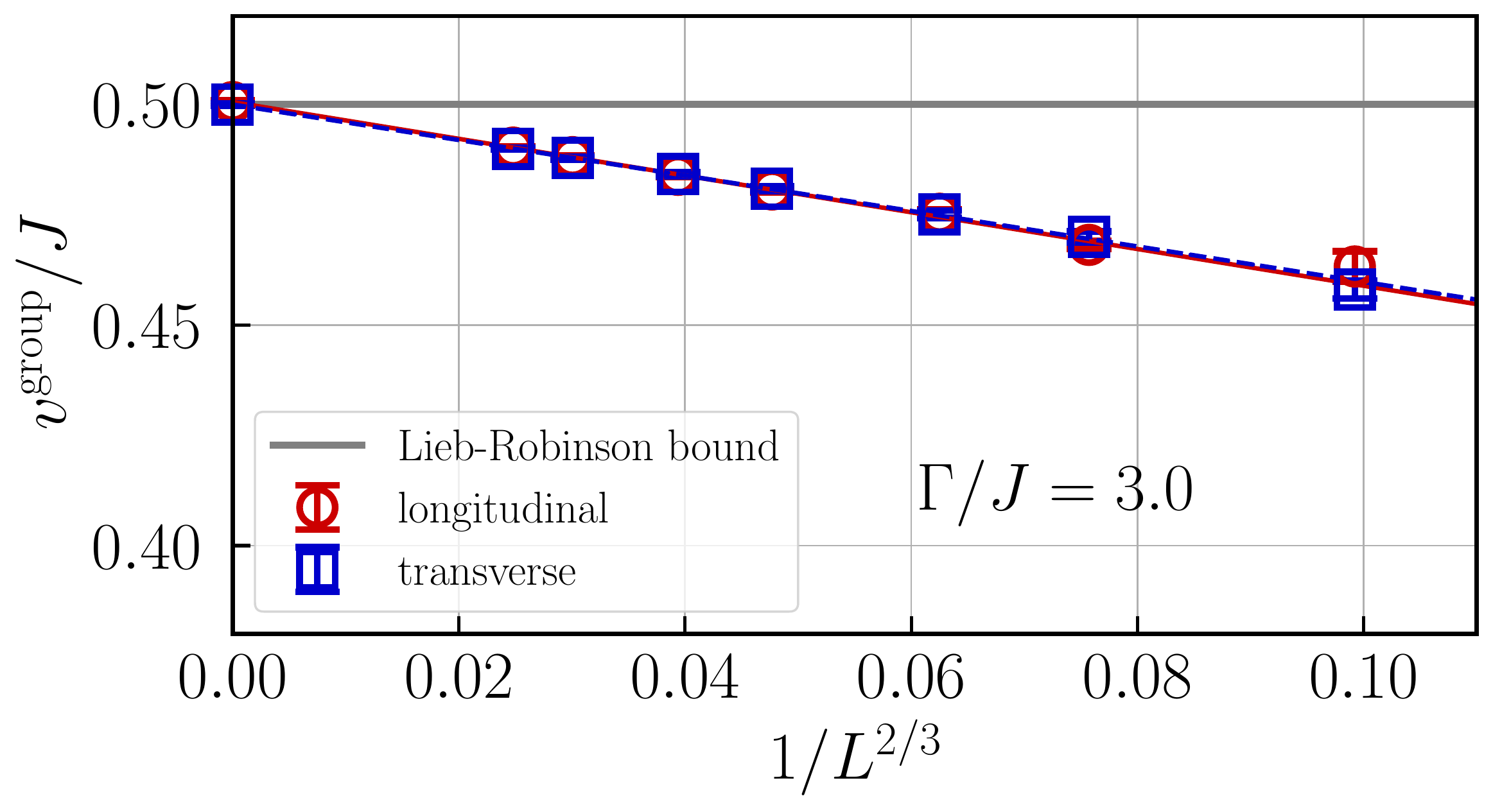}
\caption{Size scaling of the group velocity in 1D.
The group velocity is well fitted by $1/L^{2/3}$
with $L$ being the length of a chain
and is extrapolated to the value of
the exact Lieb-Robinson velocity
$v^{\mathrm{LR}}/J=1/2$.
}
\label{fig:1d_exact_fss}
\end{figure}

We show the exact equal-time longitudinal
correlation functions in Fig.~\ref{fig:1d_exact_zz}.
At an early time ($tJ\ll r$),
the intensity of correlation is nearly zero.
On the other hand, when $tJ\gtrsim r$,
the correlation starts to develop
and exhibits rapid oscillations.
For each distance,
the earliest peak in the envelope
of the wave packet has the largest intensity.
The peak time of the largest envelope peak
moves almost linearly with the distance,
suggesting the light-cone-like spreading of correlations.

We also show the exact equal-time transverse
correlation functions in Fig.~\ref{fig:1d_exact_xx}.
In contrast to the longitudinal correlations,
the rapid oscillations appear only for short distances
($r\lesssim 3$) and are negligibly small
for most of the distances.
Besides, the intensity of the transverse correlation is much smaller
than that of the longitudinal one.
On the other hand, the peak time of the transverse correlation
almost coincides with that of the largest envelope peak
in the longitudinal correlation.
The transverse correlation decays rapidly
just before and after the peak time.

To estimate the propagation velocity,
we first extract the peak time of
the envelope of correlations
as a function of distance.
We show the corresponding time and distance in
Fig.~\ref{fig:1d_exact_rt}.
The data for longitudinal and transverse correlations
overlap very well.
The distance is nearly proportional to the peak time
for both correlations.
For each field $\Gamma/J$ and size $L$,
we estimate the group velocity $v^{\mathrm{group}}$
from one half of the slope
so that it corresponds
directly to
the speed of one of quasiparticle pairs
moving to the left or right.
Since the data points are slightly out of the straight line
at the very short and long distances,
we discard those for $r\le 5$ and $r\ge L/2-5$
when extracting the velocity.

To see how the group velocity behaves as a function of
the transverse field, we first examine
a sufficiently large system ($L=256$)
as shown in Fig.~\ref{fig:1d_exact_v}.
Both velocities estimated from longitudinal and transverse correlations
are nearly $0.5J$ for all fields $\Gamma/J\ge 3$.
The group velocity of the spin-spin correlations
agrees with the exact Lieb-Robinson velocity
in the 1D transverse-field Ising model
(see Appendix~\ref{sec:app1_max_vgroup}
for the derivation of the exact value).
This fact suggests that
the quasiparticles with the fastest propagation velocity
among the various correlation functions
are directly responsible for the spreading of spin-spin correlations.

Although the estimated velocity is very close to $0.5J$,
it is slightly smaller than the exact value in finite-size systems.
To check the size dependence and confirm the convergence,
we perform the finite-size scaling of the estimated velocity.

For this purpose,
let us first discuss how the finite-size effect appears.
The spin-spin correlation functions
in the 1D transverse-field Ising model
are described by
the single-particle correlation functions
of fermionic quasiparticles.
In the thermodynamic limit,
they are given by the Bessel functions~\cite{sachdev2011}.
The size dependence of the Bessel functions
has been carefully investigated in the case of long-time dynamics
of the 1D Bose-Hubbard model~\cite{barmettler2012},
as well as in that of
the 1D transverse-field Ising model~\cite{igloi2000}.
The distance $r$ dependence of the peak time $t$ is given as
\begin{align}
 t \approx \frac{1}{v_{\infty}} ( r + \epsilon r^{1/3} ),
\end{align}
where $v_{\infty}$ is the velocity at large distances,
and $\epsilon$ is a constant
related to the peak position of the Bessel
function~\cite{barmettler2012}.
The instantaneous velocity
$v(r) = [t(r+1) - t(r)]^{-1}$
at each time $t$
is independent of distances and becomes $v_{\infty}$
if $\epsilon=0$,
but it is slightly modified in the presence of finite $\epsilon$.
For $\epsilon\not=0$, the instantaneous velocity is obtained as
\begin{align}
 v(r)
 \approx v_{\infty} \left( 1 - \frac{\epsilon}{3} r^{-2/3} \right).
\end{align}
Since the farthest distance for a chain of length $L$ is
$r=L/2$ ($\propto L$),
we may safely assume that
the deviation between the finite-size and infinite-size
velocities $\Delta v(L)$ follows the relation
\begin{align}
\label{eq:1d_exact_size_dep_v}
 \Delta v(L)
 :=
 \left|
 v\left(\frac{L}{2}\right)
 - \lim_{L\rightarrow\infty} v\left(\frac{L}{2}\right)
 \right| \propto L^{-2/3}
\end{align}
for $L\gg 1$.

We then extrapolate the finite-size group velocities
to the thermodynamic limit using
Eq.~\eqref{eq:1d_exact_size_dep_v}.
We estimate the error bars using the covariance
obtained from weighted least-squares regression.
As shown in Fig.~\ref{fig:1d_exact_fss},
all the data points lie on an expected straight line
for both correlations.
The extrapolated group velocity
at $\Gamma/J=3$ is
$v^{\mathrm{group}}/J = 0.5005(4)$
[$v^{\mathrm{group}}/J = 0.4999(3)$]
for the longitudinal (transverse) correlations
and almost converges to the exact Lieb-Robinson velocity
($v^{\mathrm{LR}}/J=0.5$)
of the 1D transverse field Ising model
within the error bar of the extrapolation.
We have also confirmed that
the estimated velocity converges to the exact one
for all the other transverse fields that we have studied
($\Gamma/J \ge 1$).
Therefore, the fastest correlation spreading
can be measured by the spin-spin correlations
in the 1D transverse-field Ising model.

\subsection{Results by the LSWA}

\begin{figure}[t]
\centering
\includegraphics[width=1.0\columnwidth]{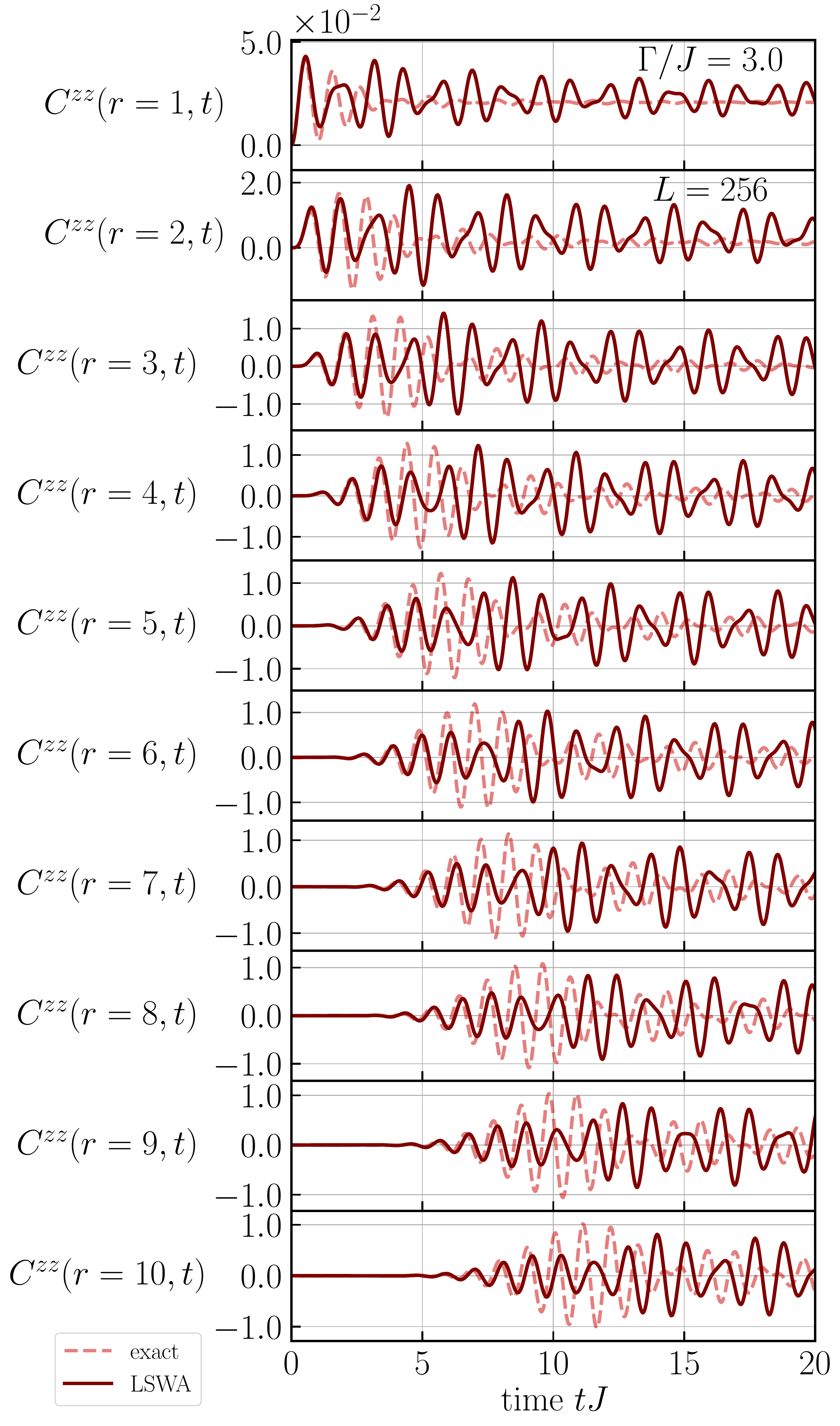}
\caption{Equal-time longitudinal correlation functions
obtained by the LSWA in 1D.
The parameters are the same as those in Fig.~\ref{fig:1d_exact_zz}.
We show the exact correlations (dashed line) for comparison.
}
\label{fig:1d_sw_zz}
\end{figure}

\begin{figure}[t]
\centering
\includegraphics[width=1.0\columnwidth]{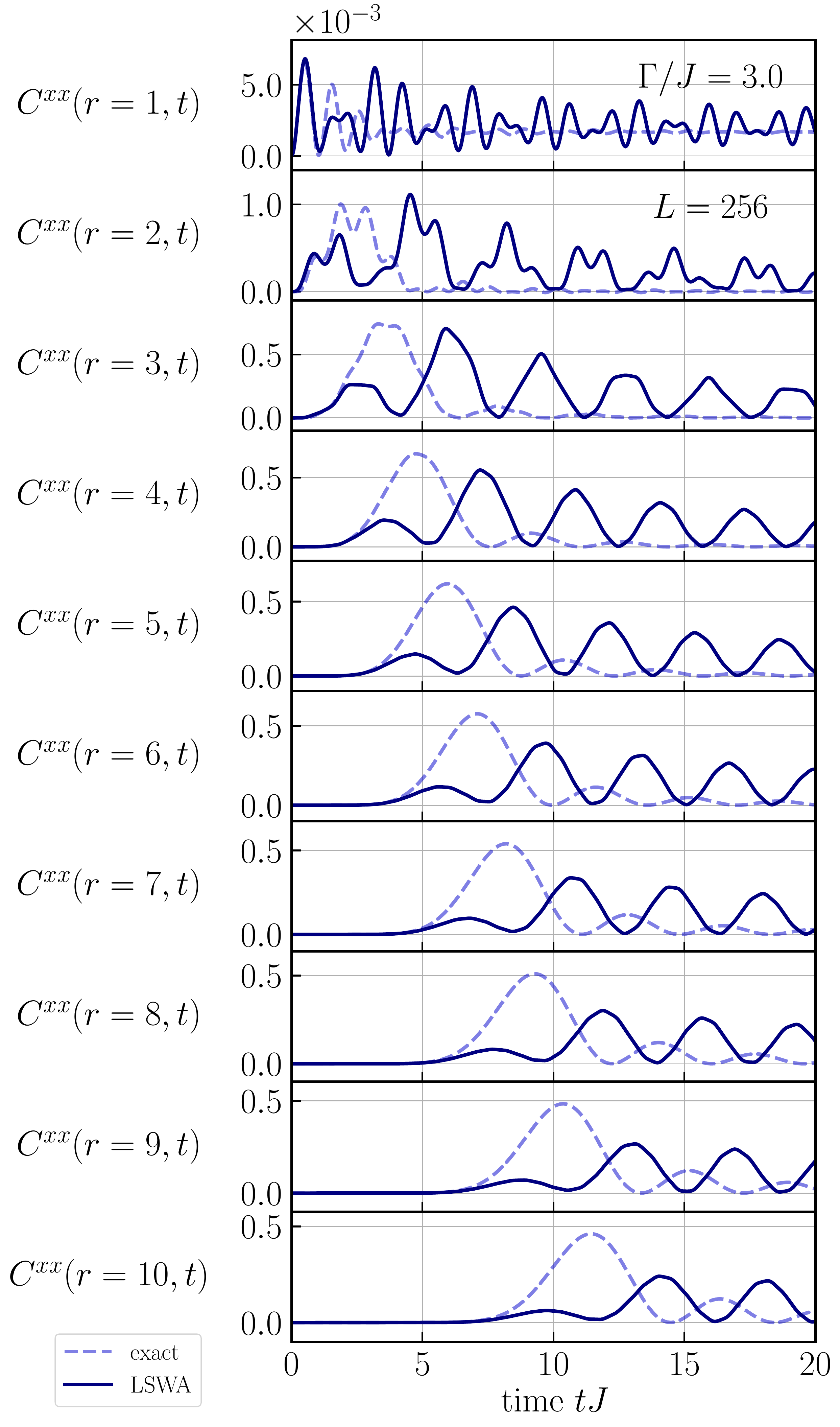}
\caption{Equal-time transverse correlation functions
obtained by the LSWA in 1D.
The parameters are the same as those in Fig.~\ref{fig:1d_exact_zz}.
We show the exact correlations (dashed line) for comparison.
}
\label{fig:1d_sw_xx}
\end{figure}

\begin{figure}[t]
\centering
\includegraphics[width=1.0\columnwidth]{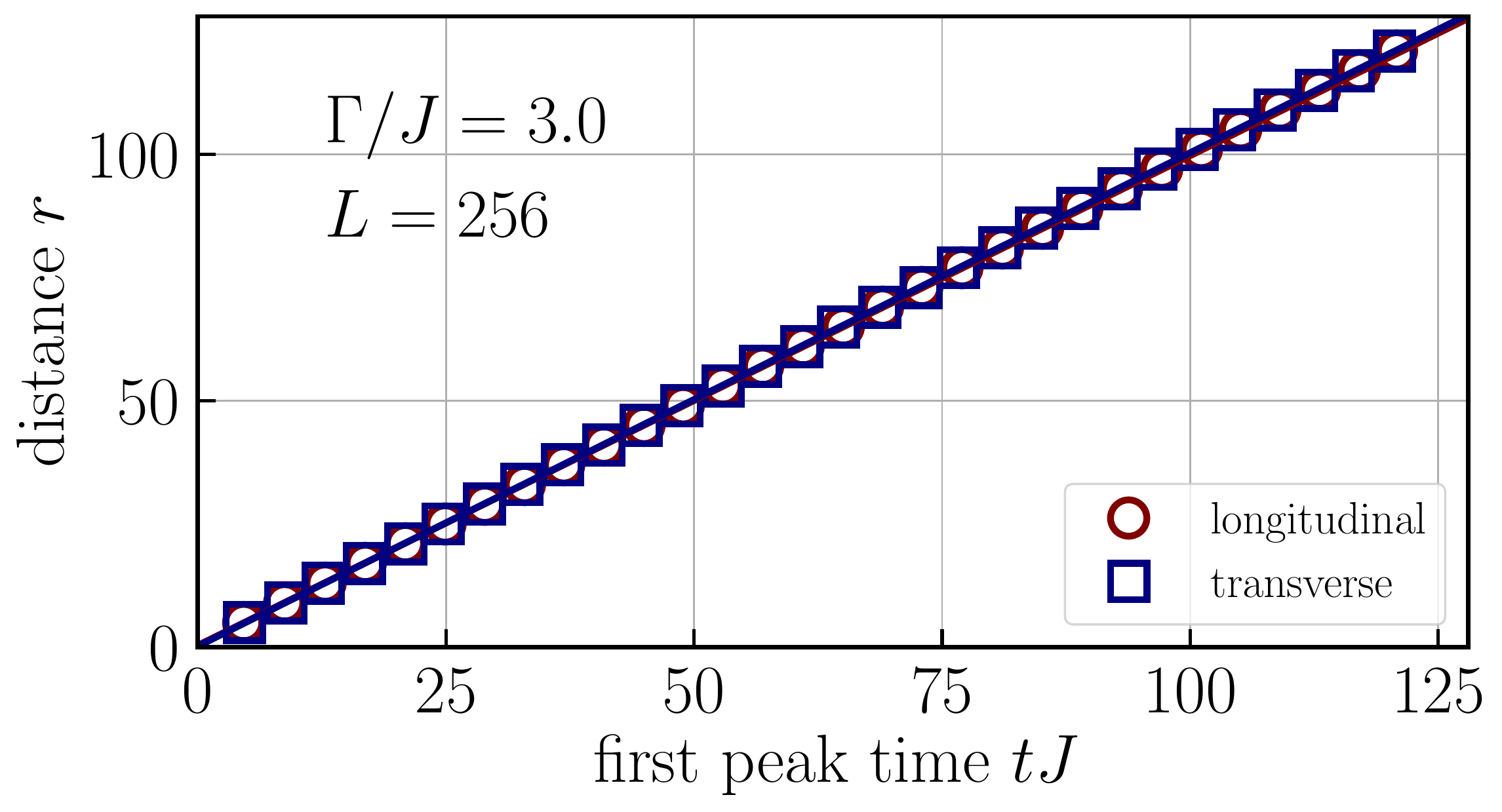}
\caption{First-peak time dependence of distance
for correlations obtained by the LSWA in 1D.
We show data points when the distance is a multiple of four.
}
\label{fig:1d_sw_rt}
\end{figure}

\begin{figure}[t]
\centering
\includegraphics[width=1.0\columnwidth]{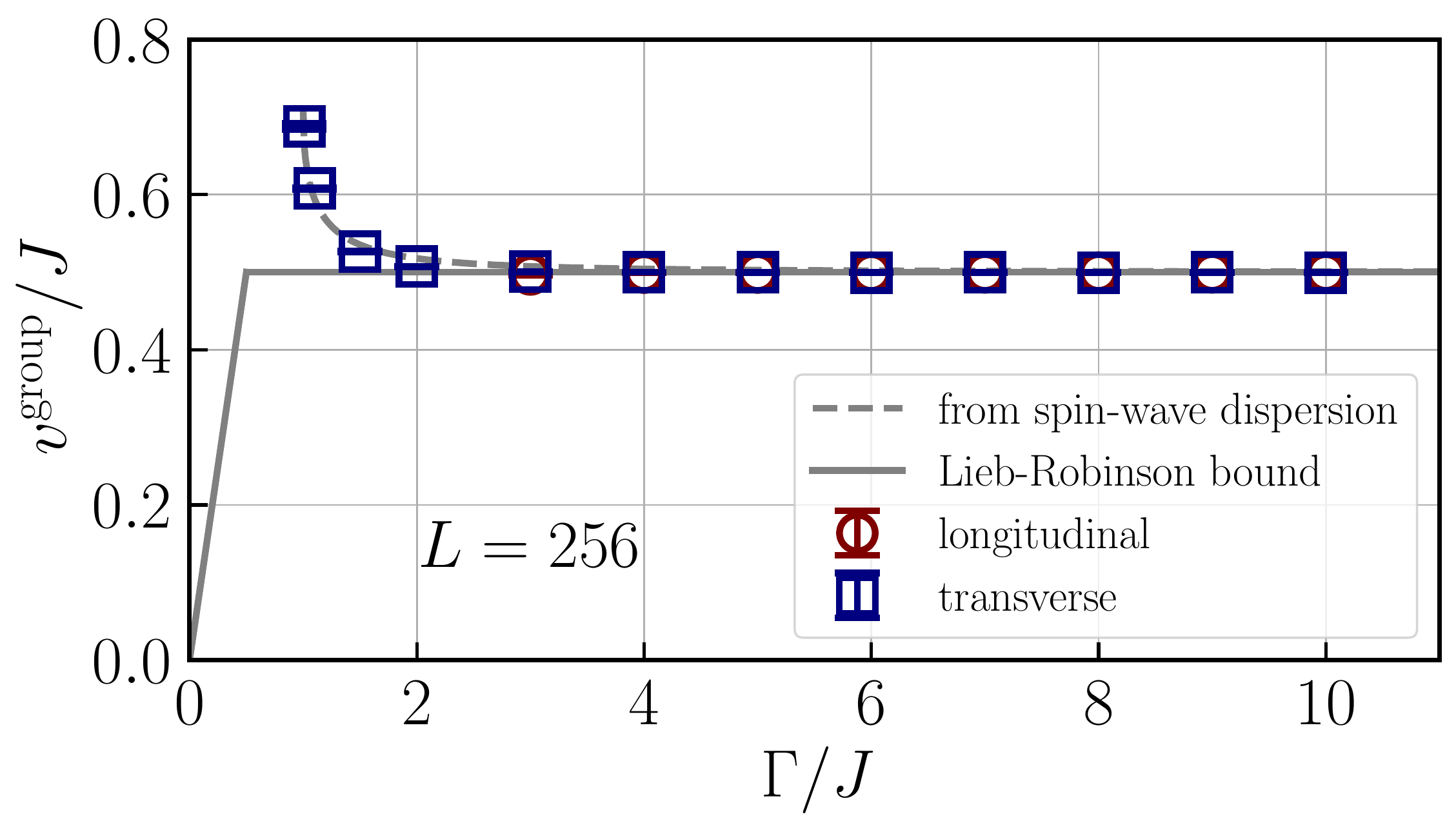}
\caption{Field dependence of the group velocity
obtained by the LSWA in 1D.
The exact Lieb-Robinson velocity
$v^{\mathrm{LR}}/J=1/2$
and
the maximum group velocity
$
v^{\mathrm{SW}}/J
= [ 1 + \sqrt{1-(J/\Gamma)^2} ]^{-1/2}
/ \sqrt{2}
$
estimated from the spin-wave dispersion
(see Appendix~\ref{sec:app_sw_corr_disp_max_group_vel})
are shown as references.
For $\Gamma/J<3$,
we only show the group velocity estimated from
the transverse correlations
because the envelopes of longitudinal correlations
become unclear for a weaker field.
}
\label{fig:1d_sw_v}
\end{figure}

\begin{figure}[t]
\centering
\includegraphics[width=1.0\columnwidth]{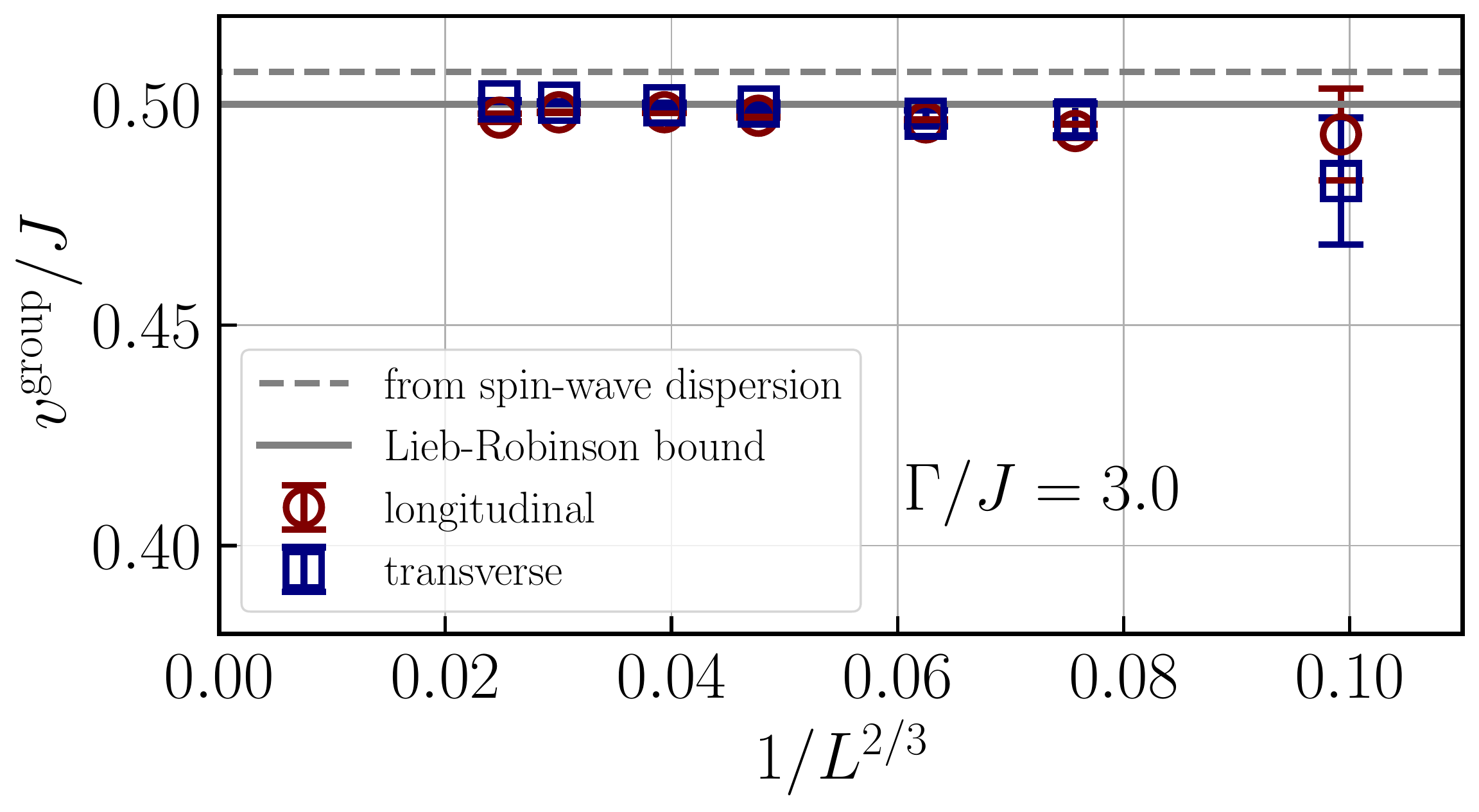}
\caption{Size dependence of the group velocity
estimated by the LSWA in 1D.
The size dependence is smaller than the exact case
(see Fig.~\ref{fig:1d_exact_fss}).
}
\label{fig:1d_sw_fss}
\end{figure}

To examine how good the LSWA is
as for the correlation spreading,
we calculate the equal-time spin-spin correlation functions
by the LSWA
and compare the results with those of the exact analysis.
In general,
the LSWA gets better
with increasing spatial
dimensions~\cite{huse1988,soukoulis1991,runge1992}
because it takes into account a correction
to the leading order of the mean-field approximation.
Here we will demonstrate that
the group velocity of the correlation propagation
obtained by the LSWA
agrees well with the exact one even in the lowest 1D.

We show the longitudinal correlation functions in
Fig.~\ref{fig:1d_sw_zz}.
As in the case of the exact analysis,
the correlations are suppressed for $tJ\lesssim r$
and begin to develop for $tJ\gtrsim r$
at a given distance $r$.
The LSWA quantitatively reproduces
the period of oscillations
and 
the intensity of exact correlations
up to about $tJ\approx r$.
In the short time ($tJ\lesssim r$),
a very small number of quasiparticle excitations
would come into play,
and the LSWA becomes more accurate
in this dilute regime.

On the other hand, the transverse correlation functions
appear to be accurate up to the point
where they begin to increase
(see Fig.~\ref{fig:1d_sw_xx}).
In contrast to the exact analytical result,
where the earliest peak has the largest intensity,
the LSWA predicts that
the second earliest peak has the largest intensity.
Nevertheless, the time of maximum intensity
does not differ significantly
between the exact and approximate results.
The first-peak time is typically about $2tJ$ early,
while
the time of maximum intensity is typically about $2tJ$ late
for all distances
in the case of the LSWA.
These effects
do not change the propagation velocity significantly.
Therefore, the group velocity estimated
by the LSWA is expected to be close
to the exact one.

As in the case of exact analysis,
we observe the suppression of rapid oscillations
in the transverse correlations
using the LSWA.
This phenomenon can be easily understood
in the magnon picture.
The original transverse correlation
corresponds to the density-density correlation
of magnons.
The density operator is less susceptible to
the effects of phases.
On the other hand,
the original longitudinal correlation
corresponds to the single-particle correlation
of magnons, which directly feels the effects of phases.
Therefore, the transverse (longitudinal) correlation
tends to exhibit less (more) oscillations.
Such effects have been intensively examined
in the correlation spreading of the Bose-Hubbard
model~\cite{cheneau2012,barmettler2012,despres2019,nagao2019,kaneko2022}.

Likewise,
the LSWA also predicts
that the intensity of the transverse correlation
is smaller than that of the longitudinal one.
They are approximately given as
$
|C^{zz}(\bm{r},t)|
= \mathcal{O}(
J/\Gamma)
$
and
$ 
|C^{xx}_{\rm connected}(\bm{r},t)|
= \mathcal{O}(
J^2/\Gamma^2)
$,
respectively
(see
Appendixes~\ref{sec:app_sw_corr_longitudinal}
and \ref{sec:app_sw_corr_transverse}).

Having
assessed
the accuracy of the LSWA,
we extract the group velocity from the 1D correlations.
We first investigate the distance dependence of peak time
for a sufficiently large system ($L=256$),
as shown in Fig.~\ref{fig:1d_sw_rt}.
Again, both correlations show almost the same result,
and the distance is nearly proportional to the peak time.
We estimate the velocity using the data for $5<r<L/2-5$.

We summarize the field dependence of the group velocity
in Fig.~\ref{fig:1d_sw_v}.
Both group velocities estimated from the longitudinal
and transverse correlations are nearly $0.5J$
irrespective of the choice of the transverse field
for $\Gamma/J\gtrsim 3$.
Note that the LSWA group velocity is expected to deviate from
the exact one at $\Gamma/J\lesssim 2$ because too many quasiparticles are
created due to such a large quench.

Finally, we have confirmed the size dependence of the estimated
group velocity.
As shown in Fig.~\ref{fig:1d_sw_fss},
the LSWA shows much smaller size dependence
than the exact analysis in Fig.~\ref{fig:1d_exact_fss}.
The velocity is nearly converged for $L\ge 48$
and is extrapolated to $0.5J$,
corresponding to the Lieb-Robinson velocity.

 \section{Results in 2D}
\label{sec:results_2d}

Next, we examine the time-dependent correlations in 2D
using the ED method,
the tensor-network method based on iPEPS,
and the LSWA.
As in the case of 1D,
we estimate the group velocity after a sudden quench to a strong field.

\subsection{Results by the LSWA}
\label{sec:results_2d_sw}

\begin{figure}[t]
\centering
\includegraphics[width=1.0\columnwidth]{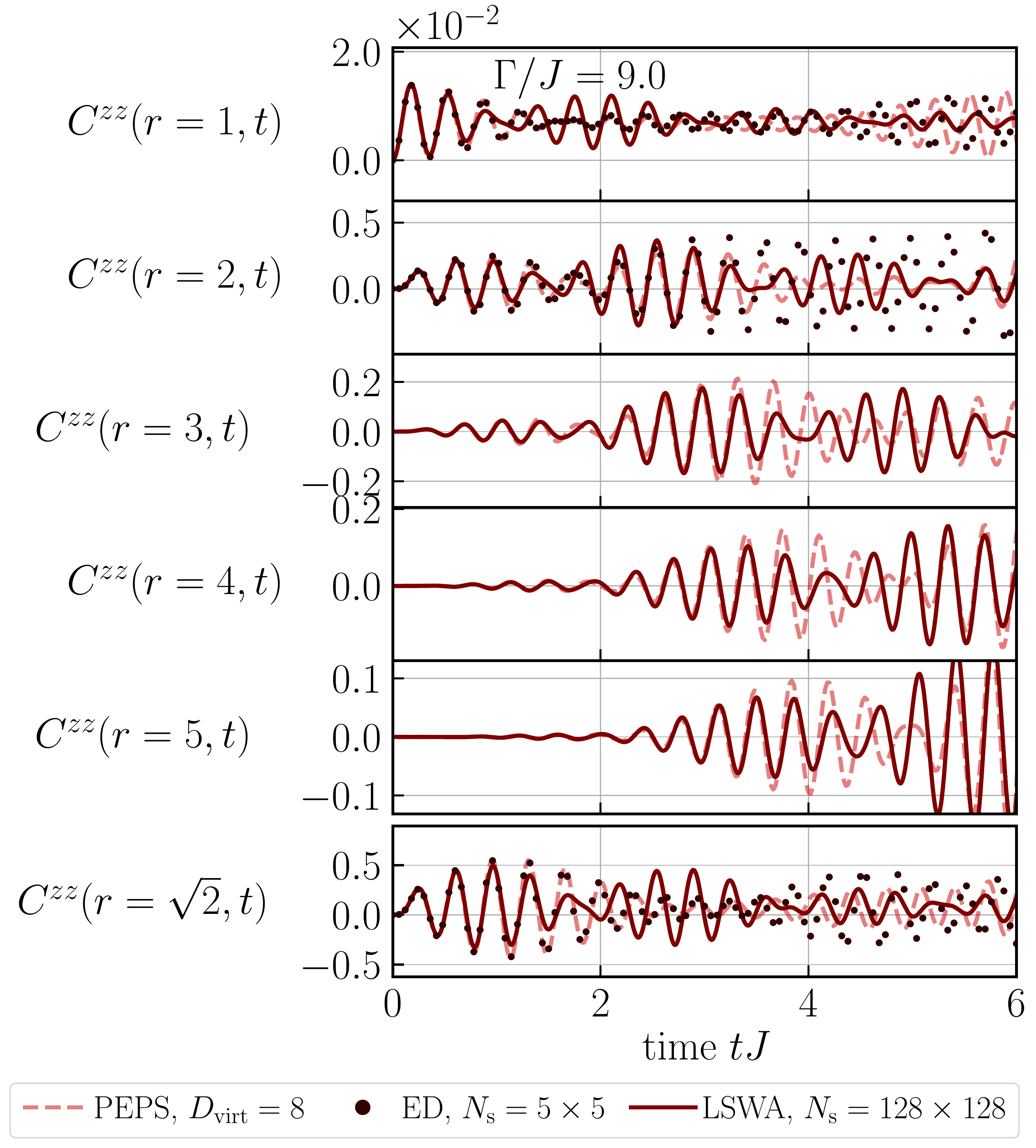}
\caption{Equal-time longitudinal correlation functions
obtained by the LSWA, the ED method, and the tensor-network
method in 2D.
We consider the quench to $\Gamma/J=9$
for a finite system of $N_{\mathrm{s}}=L^2$, $L=128$
by the LSWA (solid line),
for a finite system of $N_{\mathrm{s}}=L^2$, $L=5$
by ED simulations (small circles),
and 
for the infinite system with the bond dimensions
$D_{\rm virt}=8$
by iPEPS simulations (dashed line).
}
\label{fig:2d_sw_zz}
\end{figure}

\begin{figure}[t]
\centering
\includegraphics[width=1.0\columnwidth]{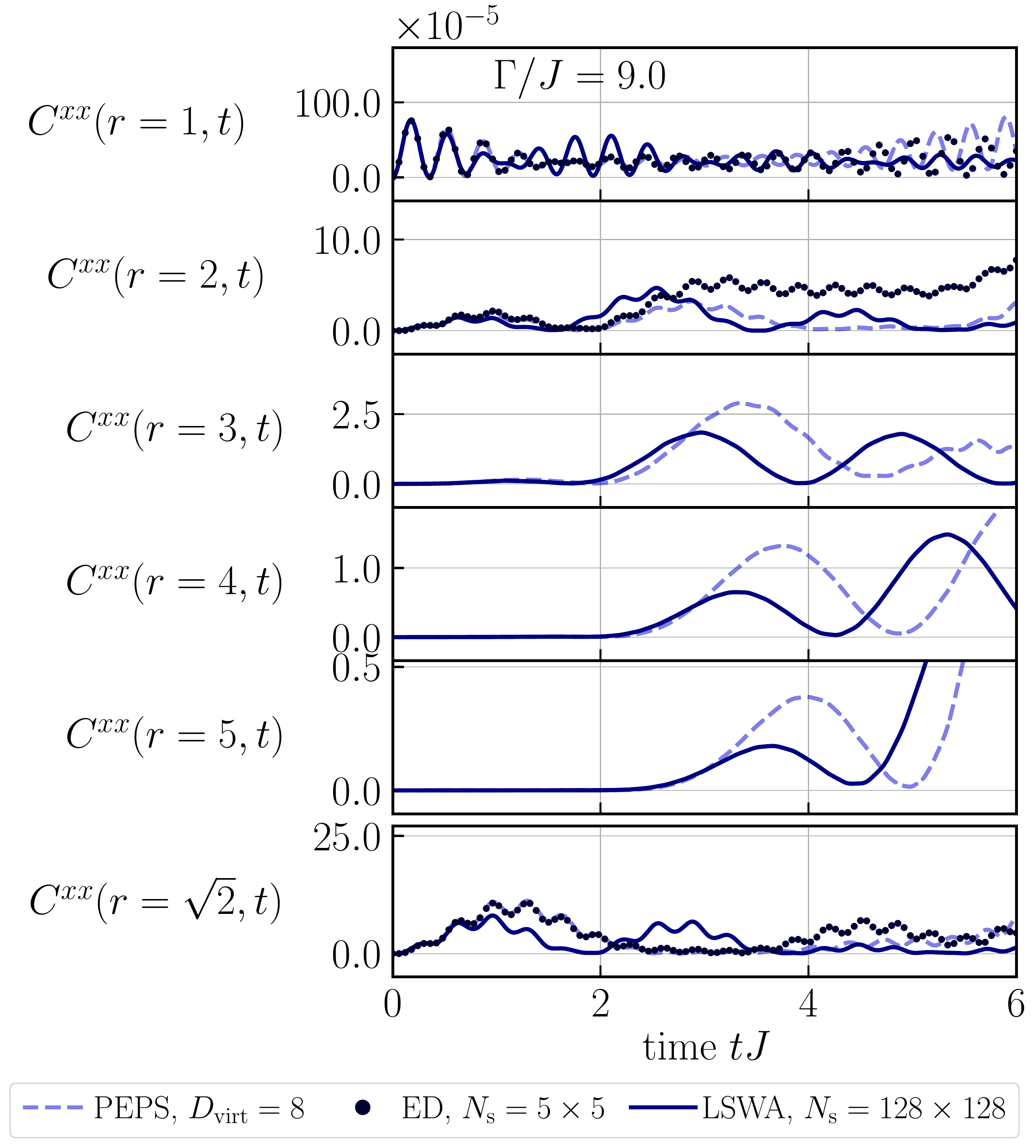}
\caption{Equal-time transverse correlation functions
obtained by the LSWA, the ED method, and the tensor-network
method in 2D.
The parameters are the same as those in Fig.~\ref{fig:2d_sw_zz}.
}
\label{fig:2d_sw_xx}
\end{figure}

\begin{figure}[t]
\centering
\includegraphics[width=1.0\columnwidth]{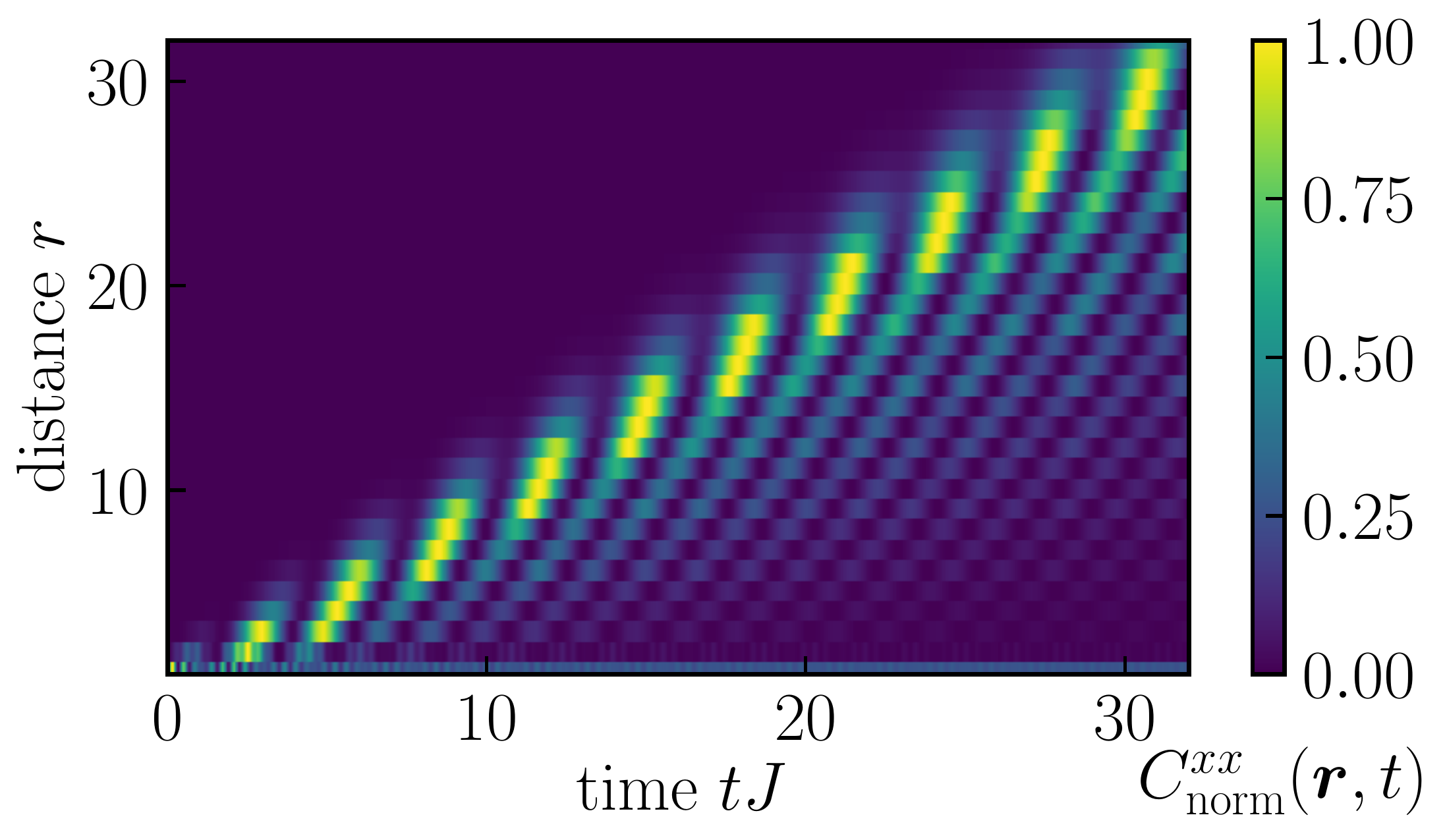}
\caption{Contour plot of normalized intensity of
equal-time transverse correlation functions
as a function of time and distance
obtained by the LSWA in 2D.
The parameters
$L=128$ and $\Gamma/J=9$
are the same as those in Fig.~\ref{fig:2d_sw_zz}.
We show the normalized correlation function
$
C^{xx}_{\rm norm}(\bm{r},t) :=
C^{xx}_{\rm connected}(\bm{r},t) /
\max_{t\in [0,L/(2J)]} C^{xx}_{\rm connected}(\bm{r},t)
$
($\in [0,1]$)
along the horizontal axis [$\bm{r}=(r,0)$]
for each distance up to $r=32$.
}
\label{fig:2d_sw_xx_2d}
\end{figure}

\begin{figure}[t]
\centering
\includegraphics[width=1.0\columnwidth]{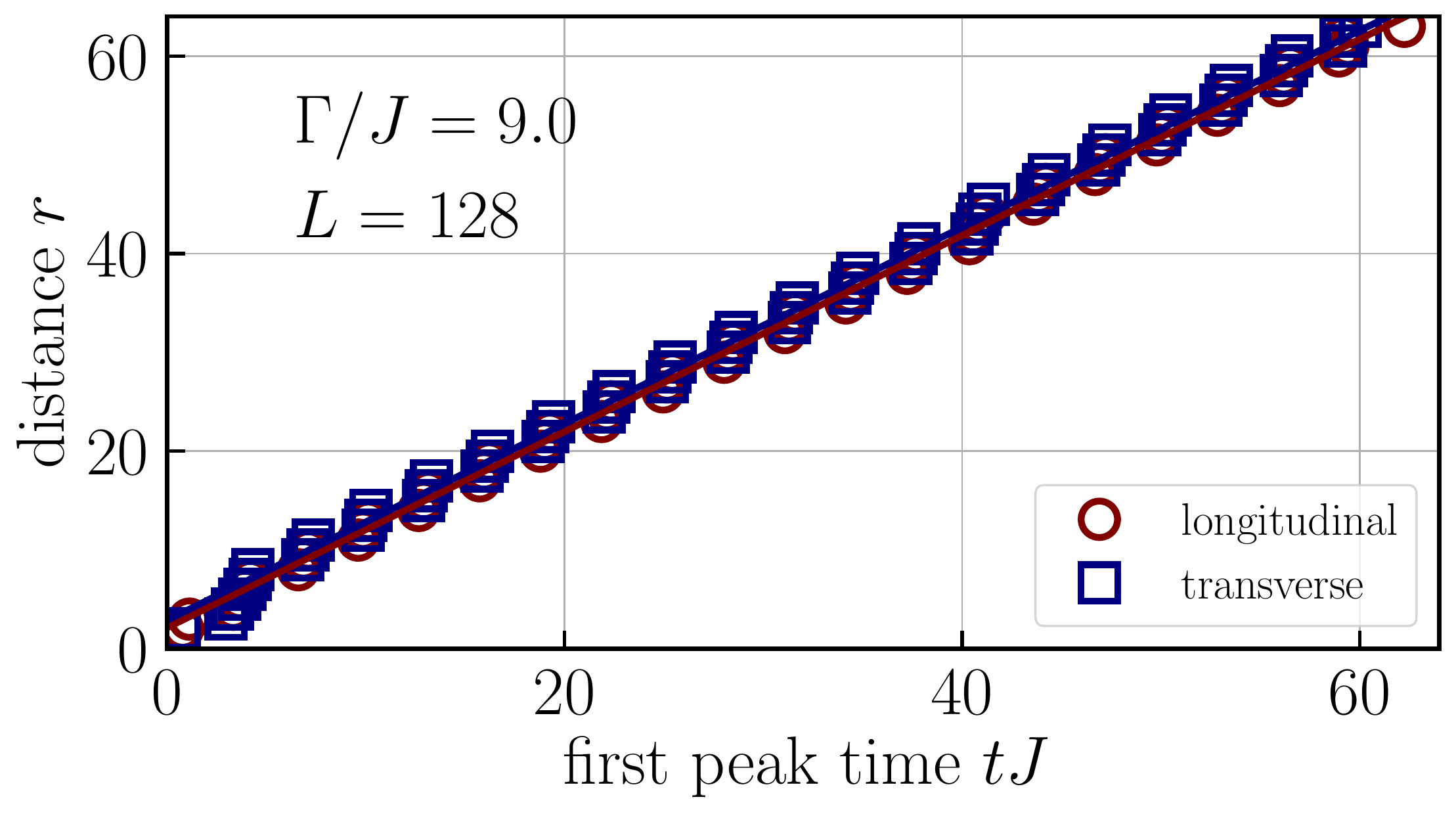}
\caption{Dominant-peak time dependence of distance
for correlations obtained by the LSWA in 2D.
}
\label{fig:2d_sw_rt}
\end{figure}

\begin{figure}[t]
\centering
\includegraphics[width=1.0\columnwidth]{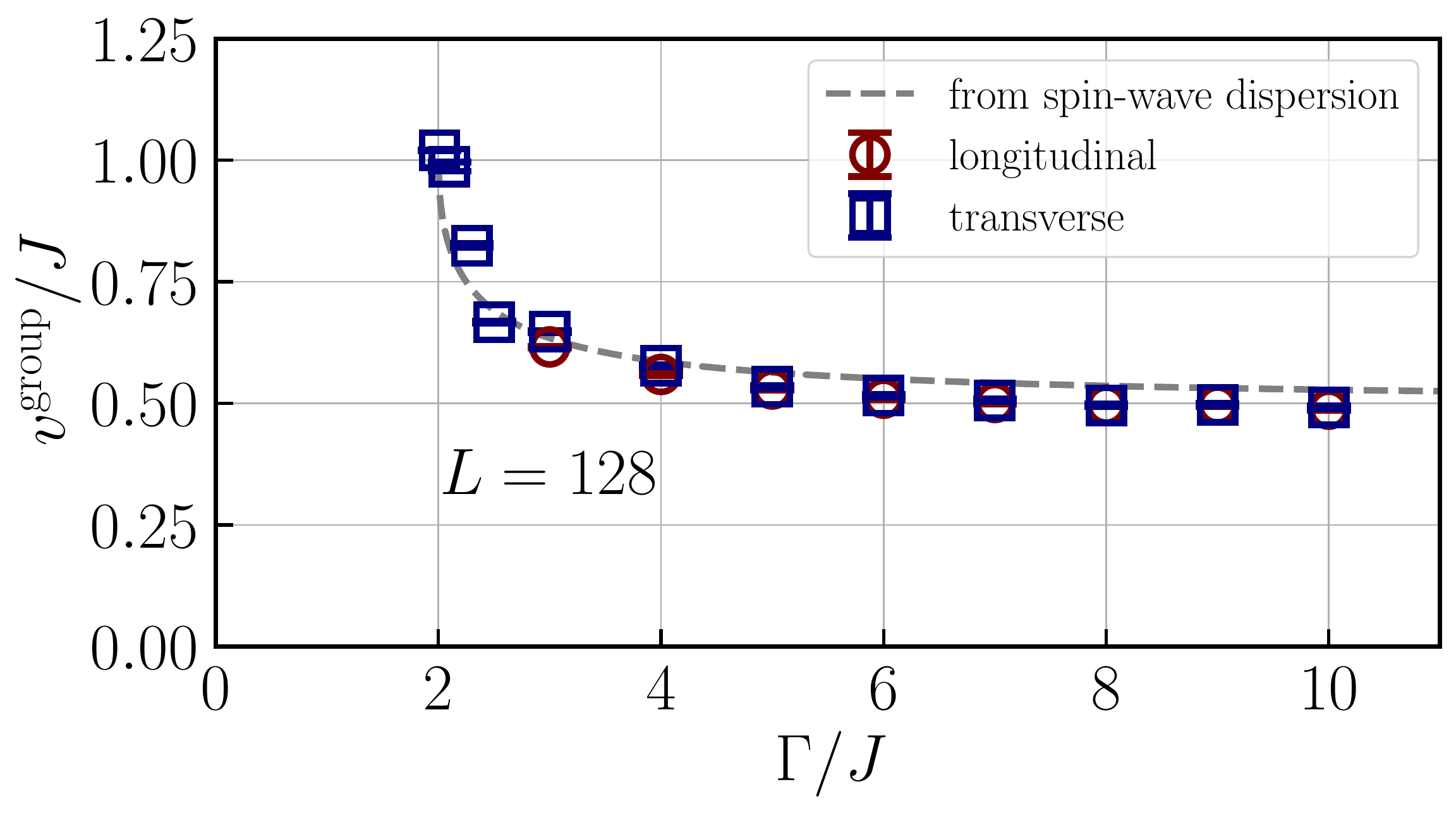}
\caption{Field dependence of the group velocity
obtained by the LSWA in 2D.
The maximum group velocity
$
v^{\mathrm{SW}}/J
= ( 1 - J/\Gamma + \sqrt{1-2J/\Gamma} )^{-1/2}
/ \sqrt{2}
$
estimated from the spin-wave dispersion
(see Appendix~\ref{sec:app_sw_corr_disp_max_group_vel})
is shown as a reference.
For $\Gamma/J<3$,
we only show the group velocity estimated from
the transverse correlations
because the envelopes of longitudinal correlations
become unclear for a weaker field.
}
\label{fig:2d_sw_v}
\end{figure}

\begin{figure}[t]
\centering
\includegraphics[width=1.0\columnwidth]{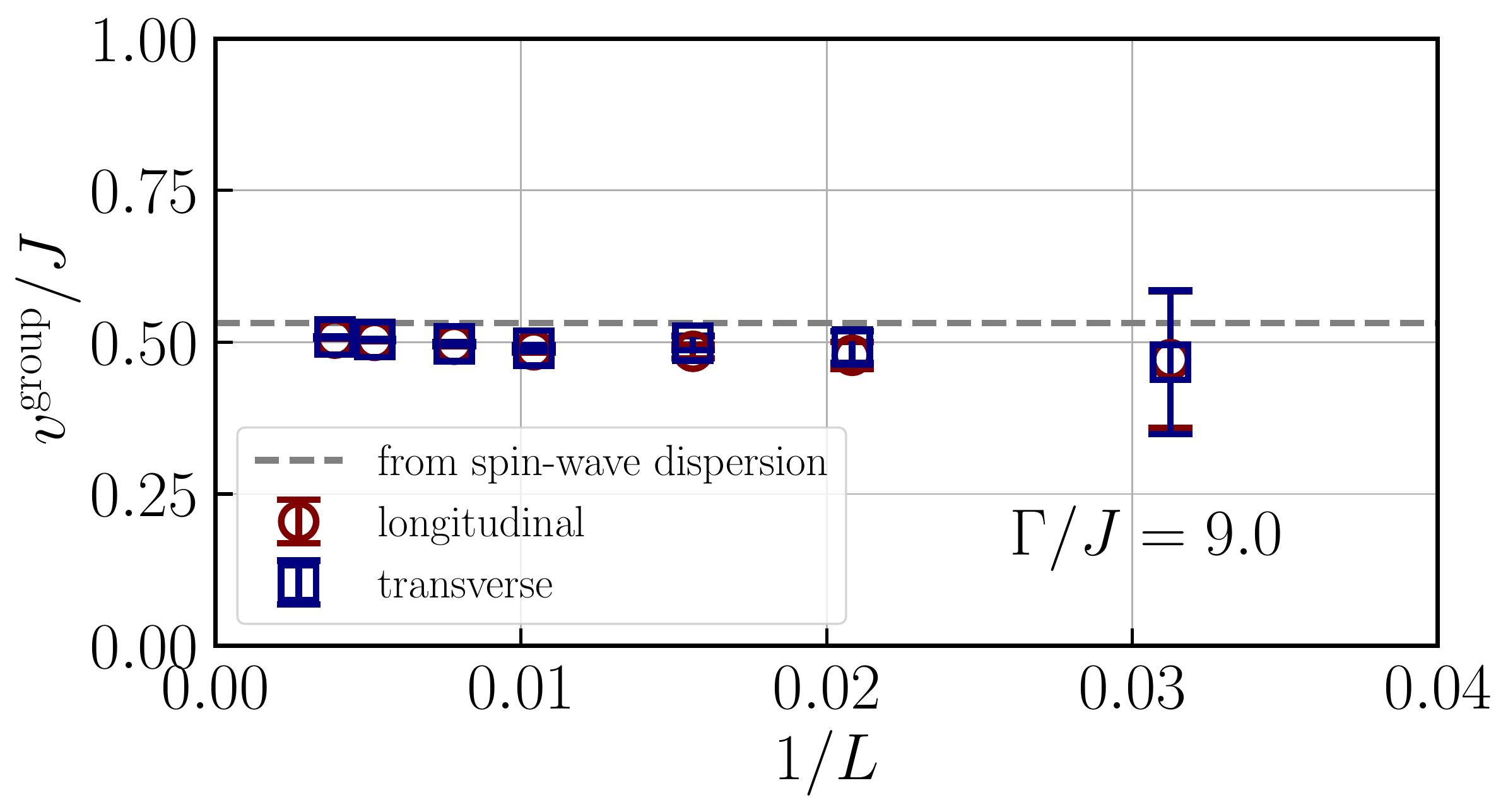}
\caption{Size dependence of the group velocity
estimated by the LSWA in 2D.
}
\label{fig:2d_sw_fss}
\end{figure}

We apply the LSWA
to calculate the spin-spin correlation functions
and to extract the group velocity.
In the case of the 1D transverse-field Ising model,
the LSWA reproduces the exact results
to the extent that the group velocity of the correlation propagation
quantitatively agrees at a sufficiently strong field.
We will demonstrate that
it reproduces the 2D correlations
obtained by the nearly exact simulations
much better than in 1D.
It also allows us to estimate the group velocity
from the correlations at farther distances
than the ED and iPEPS simulations,
as we will demonstrate below.

We compare the longitudinal correlation functions
obtained by the
ED method and the LSWA
in Fig.~\ref{fig:2d_sw_zz}.
The LSWA well reproduces
the
correlations obtained by the ED method
up to the point
where the second peak of the envelope appears
[see, e.g., $C^{zz}(r=2,t)$ in Fig.~\ref{fig:2d_sw_zz}].
The period of oscillations almost coincides
between the
ED method and the LSWA.
As expected in the LSWA
in higher spatial dimensions,
the agreement in 2D looks much better than in 1D
(compare Fig.~\ref{fig:1d_sw_zz} and Fig.~\ref{fig:2d_sw_zz}).

The longitudinal correlations exhibit rapid oscillations
as in the case of 1D.
On the other hand,
in contrast to the 1D case,
where the earliest envelope peak has the largest intensity,
it does not always
exhibit the largest intensity in 2D.
The order of the envelope peaks with the largest intensity
varies with distance in 2D,
which would make it more difficult
to extract the group velocity.
This observation may be ascribed to the complex
interference effects in 2D.

The transverse correlation function
obtained by the LSWA
also qualitatively reproduces the ED result
(see Fig.~\ref{fig:2d_sw_xx}).
In contrast to the longitudinal correlations,
the rapid oscillations are much weaker for $r\gtrsim 3$.

To clarify how the correlation develops for a longer time
and to examine the complex interference effects in 2D,
we depict the normalized intensity
of the transverse correlations
as a function of time and distance
in Fig.~\ref{fig:2d_sw_xx_2d}.
In general,
the LSWA performs better in the dilute regime,
corresponding to the region $r\gtrsim tJ$.
Within this range,
we observe a stronger intensity near the line
satisfying $r\approx tJ$.
However, areas of high intensity are not continuously connected
and are rather separated in small pieces.
Such pieces are bundled together
forming the boundary of the light cone.
When we focus on the short-time and short-distance region,
we can only look at the first small area of high intensity.
If we use such data,
we would incorrectly estimate the group velocity.
Indeed,
as we will see later
in Sec.~\ref{sec:results_2d_peps},
the velocity obtained by the iPEPS method
for a relatively short time
has a considerable degree of ambiguity.

To estimate the group velocity in 2D,
we collect the peak times and distances
in Fig.~\ref{fig:2d_sw_rt}.
Both correlations exhibit the consistent results.
Although the jagged behavior
caused by the complex interference effects
is observed in the data points,
the distance becomes nearly proportional to the peak time
for sufficiently large systems.
We extract the group velocity
from one half of the slope
so that it corresponds directly to
the velocity of one quasiparticle.

We show the field dependence of the group velocity
along the horizontal axis
for a large system ($N_{\mathrm{s}}=L^2$, $L=128$)
in Fig.~\ref{fig:2d_sw_v}.
At a very strong transverse field,
the velocity turns out to be nearly $0.5J$.
The velocity is likely to increase
with decreasing the transverse field.
This observation is qualitatively consistent with the result
obtained in perturbation theory
(see Appendix~\ref{sec:app_sw_corr_disp_max_group_vel}).
The velocity
estimated from correlations
is basically on the curve
represented by
$
v^{\mathrm{SW}}/J
= ( 1 - J/\Gamma + \sqrt{1-2J/\Gamma} )^{-1/2}
/ \sqrt{2}
$,
which is determined by the derivative
of the spin-wave dispersion
(see Appendix~\ref{sec:app_sw_corr_disp_max_group_vel}).

We finally check the size dependence
of the estimated group velocity
in Fig.~\ref{fig:2d_sw_fss}.
As in the case of 1D,
the velocity does not depend on the size
significantly for $L\gtrsim 48$
and converges to the value close to $0.5J$.
Therefore, the LSWA
predicts
that the speed of spin-spin correlation spreading
is $v^{\mathrm{group}} \approx 0.5J$
for a small quench to $\Gamma\gg J$
in the 2D transverse-field Ising model.

\subsection{Tensor-network results}
\label{sec:results_2d_peps}

\begin{figure}[t]
\centering
\includegraphics[width=1.0\columnwidth]{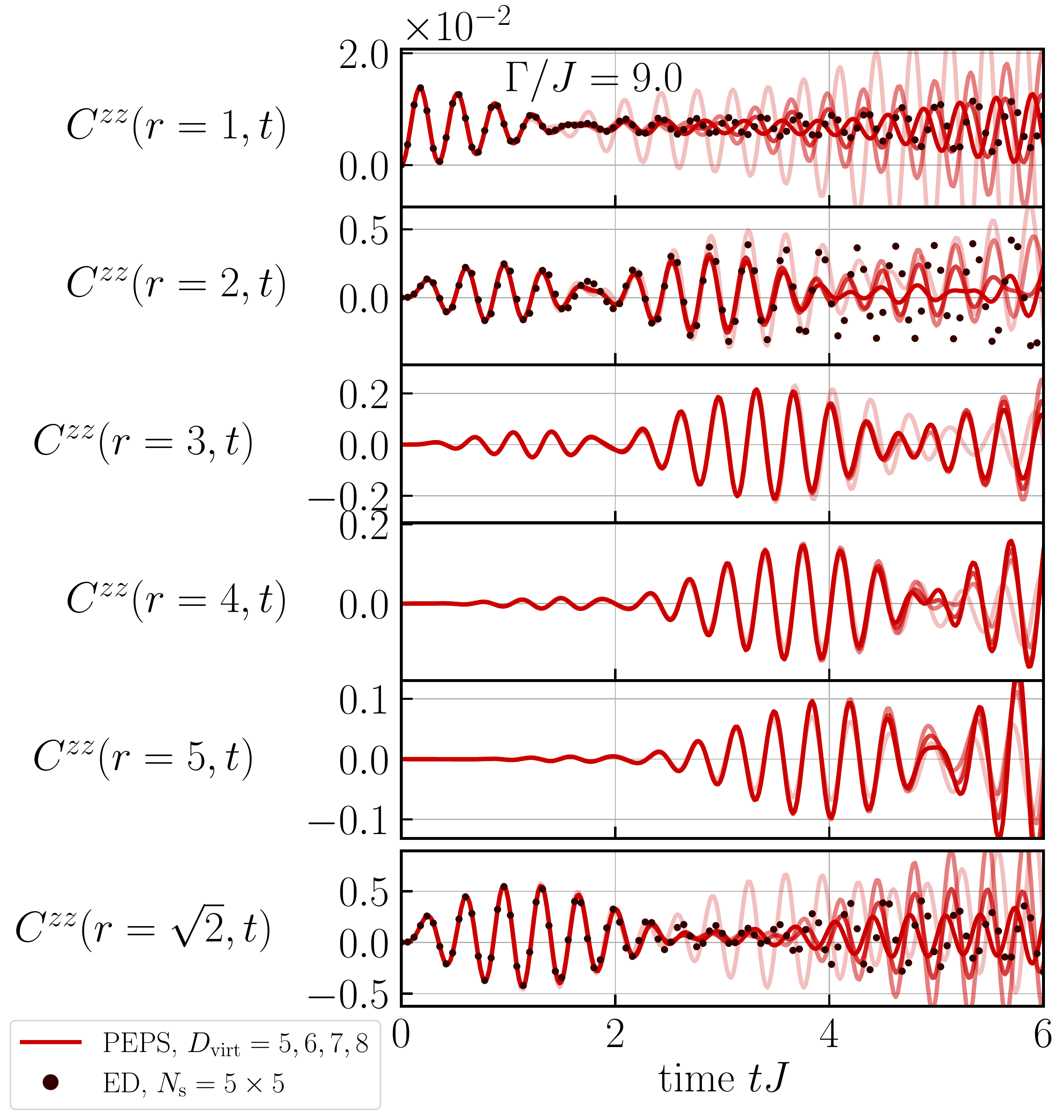}
\caption{Equal-time longitudinal correlation functions in 2D.
We consider the quench to $\Gamma/J=9$
for the infinite system with the bond dimensions
$D_{\rm virt}=5$, $6$, $7$, and $8$
(solid lines from lighter to darker)
by iPEPS simulations
and
for a finite system of $N_{\mathrm{s}}=L^2$, $L=5$
(small circles)
by ED simulations.
We show the short-time dynamics for distances $r=1$,
$2$, $\dots$, $5$, and $\sqrt{2}$.
Both data agree very well for $t/J\lesssim 4$.
}
\label{fig:2d_peps_zz}
\end{figure}

\begin{figure}[t]
\centering
\includegraphics[width=1.0\columnwidth]{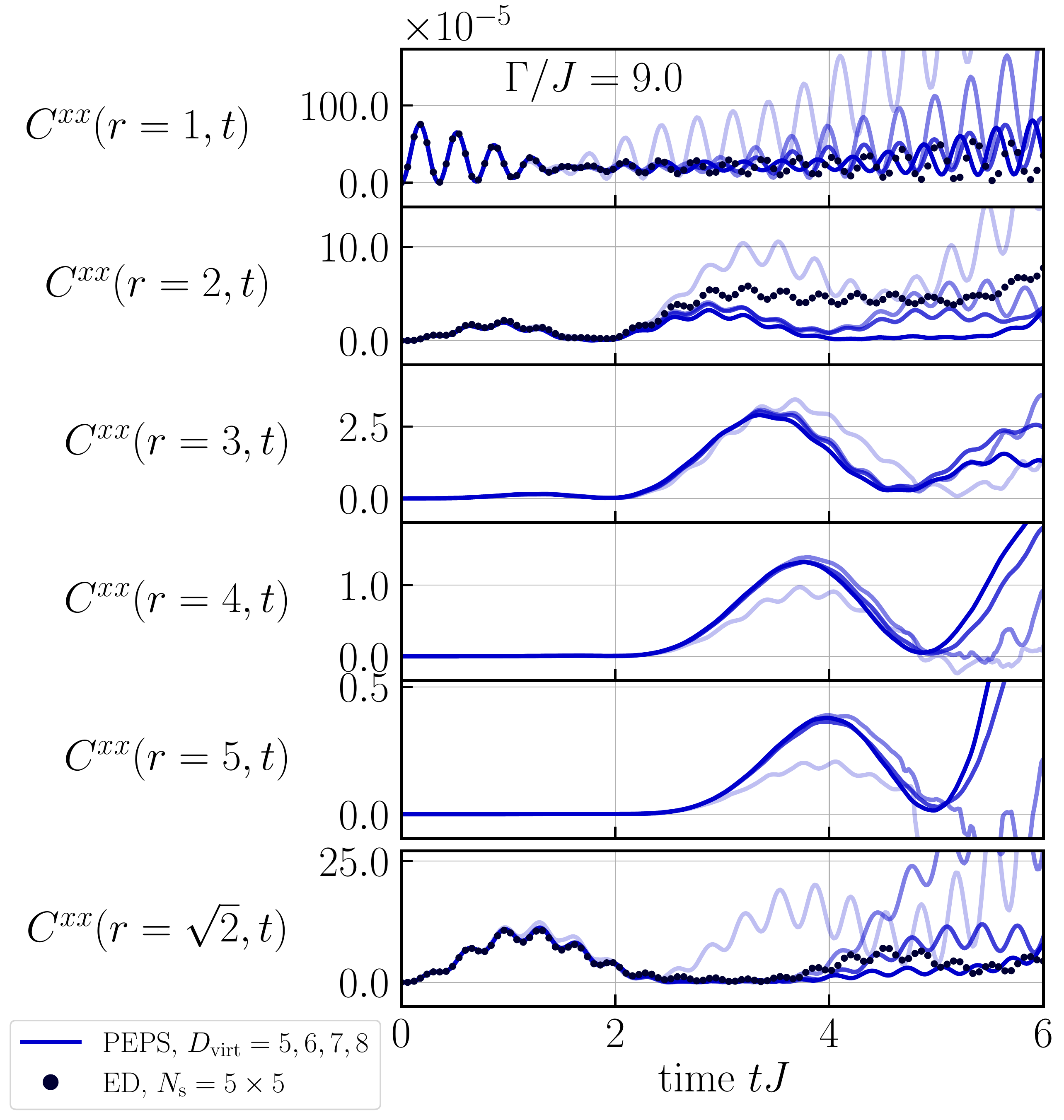}
\caption{Equal-time transverse correlation functions in 2D.
The parameters are the same as those in Fig.~\ref{fig:2d_peps_zz}.
}
\label{fig:2d_peps_xx}
\end{figure}

\begin{figure}[t]
\centering
\includegraphics[width=1.0\columnwidth]{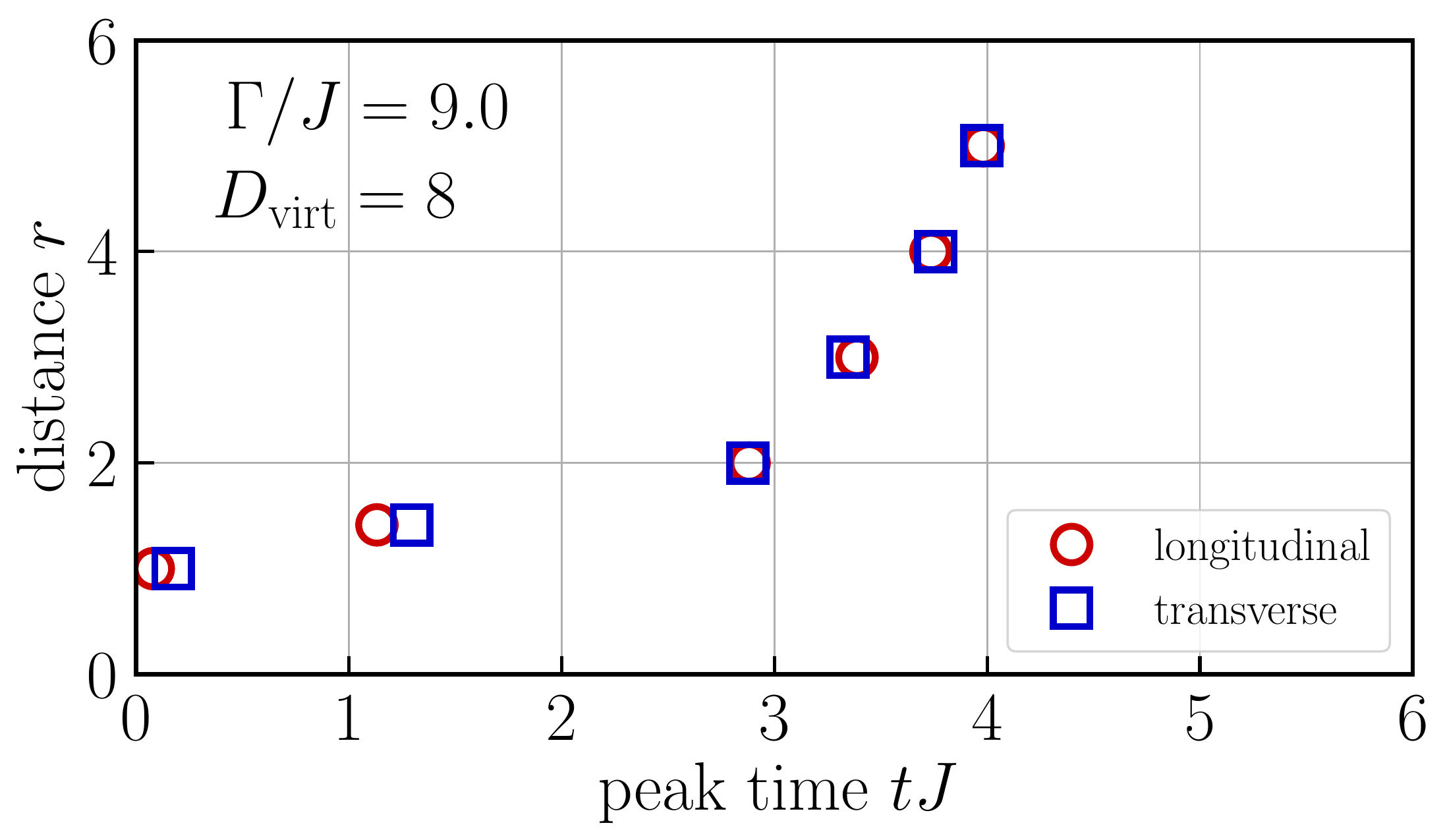}
\caption{Dominant-peak time dependence of distance
for correlations obtained by iPEPS simulations ($D_{\rm virt}=8$) in 2D.
The group velocity is estimated from
the value $r/[2t(r)]$ for distances $r=3$, $4$, and $5$.
}
\label{fig:2d_peps_rt}
\end{figure}

\begin{figure}[t]
\centering
\includegraphics[width=1.0\columnwidth]{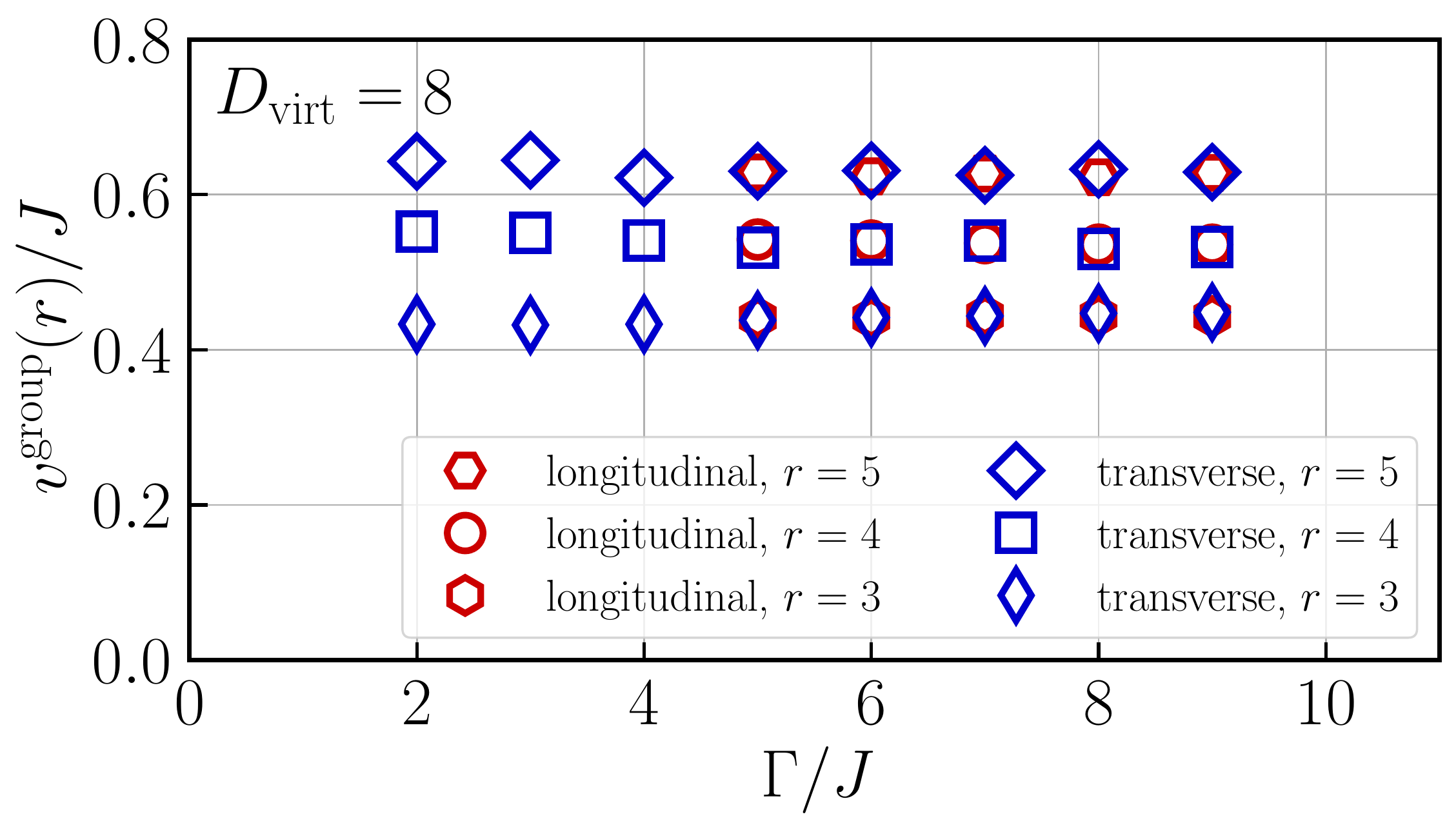}
\caption{Field dependence of the group velocity
estimated from iPEPS simulations ($D_{\rm virt}=8$) in 2D.
For $\Gamma/J<5$,
we only show the group velocity estimated from
the transverse correlations
because the envelopes of longitudinal correlations
become unclear for a weaker field.
}
\label{fig:2d_peps_v}
\end{figure}

As a complementary method to the LSWA,
we use the tensor-network method based on the iPEPS
to calculate the spin-spin correlation functions.
We will see that the tensor-network method
has an advantage in calculating
the time dependence
of correlations more accurately than the LSWA.

Before presenting the correlations obtained by the iPEPS simulations,
let us comment on the time range of the applicability
of the method.
As we have discussed in Sec.~\ref{sec:2d_tensor_network},
as for numerical simulations of a quench dynamics,
the obtained correlations would be reliable
in a short time that the energy is conserved.
In our case,
the energy density is found to be nearly conserved 
for a short time ($tJ\lesssim 4$) when $D_{\rm virt}\ge 6$
(see Fig.~\ref{fig:2d_peps_ene}).
Therefore, we will present the correlations
within this time frame hereafter.

We show the longitudinal correlation functions
obtained by the ED and iPEPS simulations
in Fig.~\ref{fig:2d_peps_zz}.
The ED method can deal with small systems in 2D
and gives the correlations up to $r\approx 2$ at the farthest.
For these distances ($r\lesssim 2$)
and short times ($tJ\lesssim 4$),
the data by the ED and iPEPS methods completely overlap.
Since the iPEPS method directly handles
the infinite system,
the ED method appears to provide the correlations
that can almost be regarded as those at the thermodynamic limit
in this regime.
The iPEPS method can predict the peak positions of
correlations at slightly farther distances
and still conserve the energy for $tJ\lesssim 4$.
The peak in the envelope of correlation
hits $tJ\approx 4$ when $r=5$,
and thus the correlations up to
$r=5$ would be reliable for the velocity estimation.

As in the case of the LSWA,
the longitudinal correlations exhibit rapid oscillations.
Moreover, in 2D, the tensor-network method also predicts that
the earliest envelope peak does not always
correspond to the peak having the largest intensity
(see Fig.~\ref{fig:2d_peps_zz}).
This observation suggests that
the complex interference effects in 2D
are not the artifact of the LSWA.

We also examine the transverse correlation functions
in Fig.~\ref{fig:2d_peps_xx}.
The ED and iPEPS methods
provide almost the same correlations
for $r\lesssim 2$ and $tJ\lesssim 4$.
Again,
the iPEPS method is applicable to farther distances 
up to $r=5$.
The rapid oscillations are quickly suppressed
for $r\gtrsim 3$, as in the case of 1D
and also as in the LSWA for 2D.
The peak positions of the transverse correlations
are nearly the same as those of the envelope peak
in the longitudinal correlations.

The qualitative behavior of correlations obtained
by the tensor-network method and the LSWA
is similar (see Figs.~\ref{fig:2d_sw_zz}
and \ref{fig:2d_sw_xx}).
The peak time of the correlations does not differ
significantly between the two methods.
Although the first-peak time is a little ahead in the LSWA,
the peak-time difference is typically $0.5tJ$ in 2D,
which is smaller than $2tJ$ in 1D.
Because the LSWA is applicable to
a much longer time,
it is more suitable for estimating
the group velocity.
On the other hand, the 
time dependencies
of correlations
agree well between the ED and tensor-network methods,
whereas they slightly differ between the ED method and the LSWA.
Therefore, the tensor-network method
is more appropriate to obtain quantitative data.

To estimate the group velocity,
we pick up the peak time for each distance
from these correlations obtained by iPEPS simulations,
as shown in Fig.~\ref{fig:2d_peps_rt}.
Since the data obtained by the bond dimensions
$D_{\rm virt}=6$, $7$, and $8$ are well converged,
we present the result for the largest bond dimension $D_{\rm virt}=8$.
Within the range of time where the iPEPS simulations
are considered to be reliable, it is hard to tell
whether the light-cone-like spreading of correlations
exists or not in 2D.
However,
as we have shown by the LSWA
in Sec.~\ref{sec:results_2d_sw},
such behavior is caused by the complex interference effects in 2D;
it is highly probable that the light cone exists.
Therefore, we may assume that
the distance eventually grows linearly with the peak time
also in the iPEPS results.
We then extract the group velocity
as $v^{\mathrm{group}} = r/[2t(r)]$ for each distance $r$.
We mainly focus on the data for farther distances
($r=3$, $4$, and $5$) because
data for short distances tend to be off the light-cone behavior
in general.

The field dependence of the group velocities
along the horizontal axis
for distances $r=3$, $4$, and $5$ are given
in Fig.~\ref{fig:2d_peps_v}.
They do not vary significantly for $\Gamma/J \in [2,9]$.
Since the velocity increases with increasing the distance,
we estimate the group velocity
as the average of the smallest and largest values
with the
ambiguity
given by one half of their difference.
It is given as
$v^{\mathrm{group}}/J \in [0.43,0.65]$
for all transverse fields that we have studied
using the iPEPS method.
As we have discussed in Sec.~\ref{sec:results_2d_sw},
the LSWA
also predicts the similar velocity
$v^{\mathrm{group}}/J \approx 0.5$.
The velocities obtained by the LSWA
and those obtained by the tensor-network method
agree within the
ambiguity.

 \section{Discussion and summary}
\label{sec:summary}

Let us compare our group velocity estimated
from the spin-spin correlations
with the recent Lieb-Robinson bound.
In 1D, our estimate of the group velocity is $v^{\mathrm{group}}=J/2$.
This is the same as the exact Lieb-Robinson velocity
$v^{\rm LR}=J/2$ in the 1D transverse-field Ising model,
indicating that the spin-spin correlations propagate
at the speed of fastest quasiparticles.
On the other hand,
the recent Lieb-Robinson bound for general lattice systems
provides the speed $v^{\rm recent} = 1.51J$~\cite{wang2020}.
As was already pointed out in Ref.~[\onlinecite{wang2020}],
it is
approximately three times as large as
the exact Lieb-Robinson velocity.

In 2D, the group velocity along the horizontal axis
is estimated to be $v^{\rm horizontal}\approx J/2$ as well.
We do not know the exact 
excitation velocity
in the 2D transverse-field Ising model so far.
However, 
for a small quench within a disorder phase,
we might expect that the fastest quasiparticles
are responsible for spin-correlation spreading also in 2D.
We come to this conclusion because
the dispersion corresponding to the fastest quasiparticles
obtained in perturbation theory~\cite{pfeuty1971}
turns out to be the same as the dispersion
estimated in the LSWA
(see Appendix~\ref{sec:app_sw_corr_disp_max_group_vel}),
and
the LSWA reproduces
the spin-spin correlations obtained by the exact analysis in 1D
and those obtained by the nearly exact simulations in 2D
fairly well
(see Secs.~\ref{sec:results_1d} and \ref{sec:results_2d}).
Therefore,
as for the transverse-field Ising model,
even in 2D,
it is natural to regard the group velocity
of the spin-spin correlations
obtained by the LSWA
as
the Lieb-Robinson velocity.
From the comparison between this value
(the horizontal $v^{\rm horizontal}\approx J/2$
or the diagonal $v^{\rm diagonal}\approx J/\sqrt{2}$
velocity)
and the best currently available estimate
($v^{\rm recent,horizontal} = JX_{y=2} \approx 2.836J$
or
$v^{\rm recent,diagonal} = 2JX_{y=1/2} \approx 3.787J$,
where $X_y$ is the solution to the equation
$x \, \mathrm{arcsinh} \, x  = \sqrt{x^2+1} + y$)~\cite{wang2020},
it is likely that there is still much room
for improving the Lieb-Robinson bound in 2D.

In conclusion,
we have studied the correlation-spreading dynamics in the
transverse-field Ising model on a chain and that on a square lattice.
We have calculated the longitudinal and transverse spin-spin correlation
functions after a sudden quench starting from the disordered state to a
strong field within a disordered phase.
We have applied several analytical and numerical methods
and crossvalidated all data.

In 1D, we have compared the time-dependent correlations using
the exact analytical formulas and the LSWA.
We have found that
the group velocity of the correlation propagation extracted from
the LSWA results asymptotically approaches that
from the exact analytical formulas as the transverse field increases.
In addition,
the transverse correlation tends to exhibit less oscillations
than the longitudinal one.
This fact makes it easier to
extract the propagation velocity
without drawing the envelope of the wave packet
of the correlation
when we measure the transverse one.
Moreover, the 1D spin-spin correlations are found to propagate
at the speed of fastest quasiparticles
corresponding to the exact Lieb-Robinson velocity.

In 2D, we have calculated the correlations using
the ED method,
the tensor-network method based on iPEPS,
and the LSWA.
As in the case of 1D,
we have confirmed that the three methods reproduce nearly the same
correlations within a short-time frame.
The tensor-network method and the LSWA
allow us to calculate the correlations for much farther distances
than the ED method can deal with.
In particular,
the LSWA is convenient for estimating the propagation velocity,
whereas the tensor-network method is advantageous in calculating
the time dependence
of correlations accurately.
We have extracted the group velocity by these two methods
and obtained the value which is nearly equal to
one half of the magnitude of the Ising interaction.
The group velocity of the spin-spin correlations in 2D
turns out to be much smaller than the best currently available estimate
for the Lieb-Robinson bound~\cite{wang2020}.

Our findings on the group velocity would be helpful for
future analog quantum simulations of Rydberg-atom arrays
and stimulate further research on the Lieb-Robinson bound.
The present tensor-network method,
which can accurately calculate the dynamics in one of the most
fundamental two-dimensional quantum many-body systems, opens
the possibility of future applications
to other systems.

\begin{acknowledgments}
The authors acknowledge fruitful discussions with
Shimpei Goto,
Daichi Kagamihara,
and
Mathias Mikkelsen.
The authors thank Chen-Yue Guo for correcting typographical errors
in equations for the exact longitudinal correlation function in 1D.
This work was financially supported by JSPS KAKENHI
(Grants Nos.\ 
JP18H05228, JP21H01014, and JP21K13855),
by MEXT Q-LEAP (Grant No.\ JPMXS0118069021),
and by JST FOREST (Grant No.\ JPMJFR202T).
The numerical computations were performed on computers at
the Yukawa Institute Computer Facility and on computers at
the Supercomputer Center, the Institute for Solid State Physics,
the University of Tokyo.
\end{acknowledgments}

\appendix

\section{Details of exact calculations in 1D}
\label{sec:app_exact_corr}

\subsection{Hamiltonian}

We review the derivation of the exact form
of two-body correlation functions
after a sudden quench~\cite{lieb1961,pfeuty1970,barouch1971,
sachdev2011,calabrese2012,calabrese2012b,suzuki2013}.
For simplicity, we consider the Hamiltonian
\begin{align}
\label{eq:exact_1d_hamiltonian_g}
 {\hat{H}} &=
 - \sum_i {\hat{\sigma}}_i^z {\hat{\sigma}}_{i+1}^z
 - \tilde{g} \sum_{i} {\hat{\sigma}}_i^x,
\end{align}
which corresponds to
the Hamiltonian in Eq.~\eqref{eq:tfising_model}
with $J=4$ and 
$\Gamma=\tilde{g}J/2$.
To get the correlation functions
for the original Hamiltonian,
we have to use $\tilde{g}=2\Gamma/J$
and replace time $4t$ with $tJ$.

After the Jordan-Wigner transformation
\begin{align}
 {\hat{\sigma}}_i^x &= 2 {\hat{c}}^{\dagger}_i {\hat{c}}_i - 1,
\\
 {\hat{\sigma}}_i^z &= \prod_{j=1}^{i-1}
 (1 - 2 {\hat{c}}^{\dagger}_j {\hat{c}}_j)
 ({\hat{c}}_i + {\hat{c}}^{\dagger}_i)
\end{align}
and the Fourier transformation
\begin{align}
 {\hat{c}}_j &= \frac{1}{\sqrt{L}} \sum_k e^{-ikr_j} {\hat{c}}_k,
\\
 k &= \frac{2\pi n}{L},
\end{align}
where
$n = -(L-1)/2$, $-(L-3)/2$, $\dots$,
$-1/2$, $1/2$, $\dots$,
$(L-3)/2$, $(L-1)/2$
for even $L$
or
$n = -(L-1)/2$, $(L-3)/2$, $\dots$,
$-2$, $-1$, $0$, $1$, $2$, $\dots$,
$(L-3)/2$, $(L-1)/2$
for odd $L$,
we obtain
\begin{align}
 {\hat{H}}
 &= \sum_k ( {\hat{c}}^{\dagger}_k ~ {\hat{c}}_{-k} )
 \begin{pmatrix}
  \tilde{a}_k & -i \tilde{b}_k \\
  i \tilde{b}_k & - \tilde{a}_k
 \end{pmatrix}
 \begin{pmatrix}
  {\hat{c}}_k \\
  {\hat{c}}^{\dagger}_{-k}
 \end{pmatrix},
\\
 \tilde{a}_k &= \tilde{g} + \cos k,
\\
 \tilde{b}_k &= \sin k.
\end{align}
Using the Bogoliubov transformation
\begin{align}
 \begin{pmatrix}
  {\hat{c}}_k \\
  {\hat{c}}^{\dagger}_{-k}
 \end{pmatrix}
 &=
 \begin{pmatrix}
  u_k & iv_k \\
  iv_k & u_k
 \end{pmatrix}
 \begin{pmatrix}
  {\hat{\gamma}}_k \\
  {\hat{\gamma}}^{\dagger}_{-k}
 \end{pmatrix},
\\
 u_k &= \cos \frac{\theta_k}{2},
\\
 v_k &= \sin \frac{\theta_k}{2},
\\
 \tan \theta_k &= \frac{\sin k}{g + \cos k}
\end{align}
satisfying 
$u_{-k} = u_k$ and $v_{-k} = - v_k$,
we get
\begin{align}
 {\hat{H}} &= 2 \sum_k \omega_k \left( {\hat{\gamma}}^{\dagger}_k
 {\hat{\gamma}}_k - \frac{1}{2} \right),
\\
 \omega_k &= \sqrt{\tilde{g}^2 + 2\tilde{g} \cos k + 1}.
\end{align}
The coefficients $u_k$ and $v_k$ can be described
by $\tilde{a}_k$, $\tilde{b}_k$, and $\omega_k$ as
\begin{align}
\label{eq:exact_1d_uk}
 u_k
 &= \frac{\tilde{a}_k - \omega_k}{\sqrt{2 \omega_k (\omega_k - \tilde{a}_k) }}
 = \frac{(\tilde{a}_k - \omega_k)\sqrt{\omega_k +
 \tilde{a}_k}}{\sqrt{2\omega_k}|\tilde{b}_k|},
\\
\label{eq:exact_1d_vk}
 v_k
 &= \frac{\tilde{b}_k}{\sqrt{2 \omega_k (\omega_k - \tilde{a}_k) }}
 = \frac{{\rm sgn}(\tilde{b}_k)\sqrt{\omega_k +
 \tilde{a}_k}}{\sqrt{2\omega_k}}.
\end{align}

In this paper,
we mainly consider the quantum quench
from $\tilde{g}=g_0$ to $\tilde{g}=g<\infty$
within the disordered phase.
We write the Hamiltonian before (after) the quench
as $\hat{H}$ (${\hat{H}'}$).
For $\tilde{g}\rightarrow\infty$, we have
$u_k\rightarrow 0$ and $v_k\rightarrow {\rm sgn}(\sin k)$.

\subsection{Longitudinal correlation functions}

We evaluate the time-dependent longitudinal correlation functions
defined as
\begin{align}
 \bar{C}^{zz}(r,t)
 &=
 \langle\psi_0|
 e^{i{\hat{H}'}t}
 {\hat{\sigma}}^z_i
 {\hat{\sigma}}^z_{i+r}
 e^{-i{\hat{H}'}t}
 |\psi_0\rangle
\\
 &=
 \langle\psi_0|
 e^{i{\hat{H}'}t}
 ({\hat{c}}^{\dagger}_i + {\hat{c}}_i)
 \left[
 \prod_{j=i}^{i+r-1} (1 - 2 {\hat{c}}^{\dagger}_j {\hat{c}}_j)
 \right]
\nonumber
\\
 &\phantom{=}
 \cdot
 ({\hat{c}}^{\dagger}_{i+r} + {\hat{c}}_{i+r})
 e^{-i{\hat{H}'}t}
 |\psi_0\rangle.
\end{align}
Using the equality
$1 - 2 {\hat{c}}^{\dagger}_j {\hat{c}}_j = ({\hat{c}}^{\dagger}_j + {\hat{c}}_j)({\hat{c}}^{\dagger}_j - {\hat{c}}_j)$
and defining the operators
\begin{align}
 {\hat{A}}_i = {\hat{c}}^{\dagger}_i + {\hat{c}}_i,
 &\quad
 {\hat{B}}_i = {\hat{c}}^{\dagger}_i - {\hat{c}}_i,
\\
 {\hat{A}}_i(t) = e^{i{\hat{H}'}t} {\hat{A}}_i e^{-i{\hat{H}'}t},
 &\quad
 {\hat{B}}_i(t) = e^{i{\hat{H}'}t} {\hat{B}}_i e^{-i{\hat{H}'}t},
\end{align}
we obtain the correlation function 
\begin{align}
 \bar{C}^{zz}(r,t)
 &=
 \langle\psi_0|
 {\hat{B}}_i(t)
 {\hat{A}}_{i+1}(t) {\hat{B}}_{i+1}(t)
 {\hat{A}}_{i+2}(t) {\hat{B}}_{i+2}(t)
 \cdots
\nonumber
\\
 &\phantom{=}\cdot
 {\hat{A}}_{i+r-1}(t) {\hat{B}}_{i+r-1}(t)
 {\hat{A}}_{i+r}(t)
 |\psi_0\rangle.
\end{align}
It can be evaluated by the Pfaffian of
a $2r\times 2r$ skew symmetric matrix $A$
using the Wick's theorem:
\begin{align}
 \bar{C}^{zz}(r,t) = (-1)^{\frac{r(r-1)}{2}} \cdot {\rm Pf} A.
\end{align}
The matrix $A$ is given as
\begin{align}
 A &=
 \begin{pmatrix}
  S & G \\
  -G^T & Q
 \end{pmatrix}
\end{align}
with matrices
{\begingroup\makeatletter\def\f@size{7}\check@mathfonts
\def\maketag@@@#1{\hbox{\m@th\normalsize\normalfont#1}}\begin{align}
 S &=
 \begin{pmatrix}
  0 & S_{0,1} & S_{0,2} & \cdots & S_{0,r-2} & S_{0,r-1} \\
  -S_{0,1} & 0 & S_{1,2} & \cdots & S_{1,r-2} & S_{1,r-1} \\
  -S_{0,2} & -S_{1,2} & 0 & \cdots & S_{2,r-2} & S_{2,r-1} \\
  \vdots & \vdots & \vdots & \ddots & \vdots & \vdots \\
  -S_{0,r-2} & -S_{1,r-2} & -S_{2,r-2} & \cdots & 0 & S_{r-2,r-1} \\
  -S_{0,r-1} & -S_{1,r-1} & -S_{2,r-1} & \cdots & -S_{r-2,r-1} & 0 \\
 \end{pmatrix},
\\
 Q &=
 \begin{pmatrix}
  0 & Q_{0,1} & Q_{0,2} & \cdots & Q_{0,r-2} & Q_{0,r-1} \\
  -Q_{0,1} & 0 & Q_{1,2} & \cdots & Q_{1,r-2} & Q_{1,r-1} \\
  -Q_{0,2} & -Q_{1,2} & 0 & \cdots & Q_{2,r-2} & Q_{2,r-1} \\
  \vdots & \vdots & \vdots & \ddots & \vdots & \vdots \\
  -Q_{0,r-2} & -Q_{1,r-2} & -Q_{2,r-2} & \cdots & 0 & Q_{r-2,r-1} \\
  -Q_{0,r-1} & -Q_{1,r-1} & -Q_{2,r-1} & \cdots & -Q_{r-2,r-1} & 0 \\
 \end{pmatrix},
\\
 G &=
 \begin{pmatrix}
  G_{0,1} & G_{0,2} & G_{0,3} & \cdots & G_{0,r-1} & G_{0,r} \\
  G_{1,1} & G_{1,2} & G_{1,3} & \cdots & G_{1,r-1} & G_{1,r} \\
  G_{2,1} & G_{2,2} & G_{2,3} & \cdots & G_{2,r-1} & G_{2,r} \\
  \vdots & \vdots & \vdots & \ddots & \vdots & \vdots \\
  G_{r-2,1} & G_{r-2,2} & G_{r-2,3} & \cdots & G_{r-2,r-1} & G_{r-2,r} \\
  G_{r-1,1} & G_{r-1,2} & G_{r-1,3} & \cdots & G_{r-1,r-1} & G_{r-1,r} \\
 \end{pmatrix}.
\end{align}
\endgroup
}
Here we define time-dependent correlation functions
\begin{align}
 S_{i,j} &= \langle {\hat{B}}_i(t) {\hat{B}}_j(t) \rangle,
\\
 Q_{i,j} &= \langle {\hat{A}}_i(t) {\hat{A}}_j(t) \rangle,
\\
 G_{i,j} &= \langle {\hat{B}}_i(t) {\hat{A}}_j(t) \rangle
 = - \langle {\hat{A}}_j(t) {\hat{B}}_i(t) \rangle
\end{align}
and use that they are translational invariant.
We will obtain the explicit form of evaluating
$S_{i,j}$, $Q_{i,j}$, and $G_{i,j}$
in Appendix~\ref{sec:exact_single_particle_corr}.

\subsection{Transverse correlation functions}

We evaluate the time-dependent transverse correlation functions
defined as
\begin{align}
 \bar{C}^{xx}(r,t)
 &=
 \langle\psi_0|
 e^{i{\hat{H}'}t}
 {\hat{\sigma}}^x_i
 {\hat{\sigma}}^x_{i+r}
 e^{-i{\hat{H}'}t}
 |\psi_0\rangle
\\
 &=
 \langle\psi_0|
 e^{i{\hat{H}'}t}
 (2 {\hat{c}}^{\dagger}_i {\hat{c}}_i - 1)
 (2 {\hat{c}}^{\dagger}_{i+r} {\hat{c}}_{i+r} - 1)
 e^{-i{\hat{H}'}t}
 |\psi_0\rangle.
\end{align}
Using the equality
$1 - 2 {\hat{c}}^{\dagger}_j {\hat{c}}_j = ({\hat{c}}^{\dagger}_j + {\hat{c}}_j)({\hat{c}}^{\dagger}_j - {\hat{c}}_j)$
and the expressions for ${\hat{A}}_i$ and ${\hat{B}}_i$,
we obtain
\begin{align}
 \bar{C}^{xx}(r,t)
 &=
 \langle\psi_0|
 {\hat{A}}_i(t)
 {\hat{B}}_i(t)
 {\hat{A}}_{i+r}(t)
 {\hat{B}}_{i+r}(t)
 |\psi_0\rangle
\\
 &=
 G_{i,i}^2
 - Q_{i,i+r} S_{i,i+r}
 + (-G_{i+r,i}) G_{i,i+r}.
\end{align}
Subtracting the correlation of
the local transverse magnetization,
which is given as
$\langle{\hat{\sigma}}^x_i(t)\rangle
= \langle\psi_0| e^{i{\hat{H}'}t} {\hat{\sigma}}^x_i e^{-i{\hat{H}'}t} |\psi_0\rangle
= - \langle\psi_0| {\hat{A}}_i(t) {\hat{B}}_i(t) |\psi_0\rangle
= - G_{i,i} = - G_{0,0}$,
we get the connected correlation function
\begin{align}
 \bar{C}^{xx}_{\rm connected}(r,t)
 &:= \bar{C}^{xx}(r,t) - \langle{\hat{\sigma}}^x_i(t)\rangle \langle{\hat{\sigma}}^x_{i+r}(t)\rangle
\\
 &= - Q_{i,i+r} S_{i,i+r}
   - G_{i+r,i} G_{i,i+r}.
\end{align}
The explicit form of evaluating
$S_{i,j}$, $Q_{i,j}$, and $G_{i,j}$
will be given
in Appendix~\ref{sec:exact_single_particle_corr}.

\subsection{Single-particle correlation functions for fermions}
\label{sec:exact_single_particle_corr}

Let us focus on $t=0$ operators.
For simplicity, we restrict ourselves to the case of even $L$.
Using the Fourier transformation
${\hat{c}}_j = \frac{1}{\sqrt{L}} \sum_k e^{-ikr_j} {\hat{c}}_k$
and the Bogoliubov transformation
${\hat{c}}_k = u_k {\hat{\gamma}}_k + i v_k {\hat{\gamma}}_{-k}$,
and then splitting the sum
$\sum_k$ into the positive and negative parts
$\sum_{k>0} + \sum_{k<0}$,
we rewrite ${\hat{A}}_i$ and ${\hat{B}}_i$ as
\begin{align}
 {\hat{A}}_i &= {\hat{a}}^{\dagger}_i + {\hat{a}}_i,
\\
 {\hat{B}}_i &= {\hat{b}}^{\dagger}_i - {\hat{b}}_i
\end{align}
with
\begin{align}
 {\hat{a}}_i &=
 \frac{1}{\sqrt{L}} \sum_{k>0}
 \left[
   e^{ikr_j}  (u_k - i v_k) {\hat{\gamma}}_k
 + e^{-ikr_j} (u_k + i v_k) {\hat{\gamma}}_{-k}
 \right],
\\
 {\hat{b}}_i &=
 \frac{1}{\sqrt{L}} \sum_{k>0}
 \left[
   e^{ikr_j}  (u_k + i v_k) {\hat{\gamma}}_k
 + e^{-ikr_j} (u_k - i v_k) {\hat{\gamma}}_{-k}
 \right].
\end{align}
The operators ${\hat{a}}_i$ and ${\hat{b}}_i$ satisfy
\begin{align}
   \langle \{ {\hat{a}}_i, {\hat{a}}_j \} \rangle
 &= \langle \{ {\hat{b}}_i, {\hat{b}}_j \} \rangle
 = \langle \{ {\hat{a}}_i, {\hat{b}}_j \} \rangle
 = 0,
\\
 \langle \{ {\hat{a}}_i, {\hat{a}}^{\dagger}_j \} \rangle &=
 \langle \{ {\hat{b}}_i, {\hat{b}}^{\dagger}_j \} \rangle  = \delta_{ij},
\\
 \langle \{ {\hat{a}}_i, {\hat{b}}^{\dagger}_j \} \rangle &=
 \langle \{ {\hat{a}}^{\dagger}_i, {\hat{b}}_j \} \rangle
 =: - G^{ab}_{i-j}
\end{align}
with
\begin{align}
 G^{ab}_{i-j}
 &=
 -
 \frac{1}{\sqrt{L}} \sum_{k>0}
 \bigl[
   e^{ik(r_i-r_j)} (u_k - i v_k)^2
\nonumber
\\
 & \phantom{=}
 + e^{-ik(r_i-r_j)} (u_k + i v_k)^2
 \bigr].
\end{align}

After the quench, the Heisenberg equation for ${\hat{c}}_k(t)$ is 
given
as
\begin{align}
\label{eq:exact_1d_heisenberg_ck}
 i \frac{d}{dt} {\hat{c}}_k(t)
 &=
 - 2 \tilde{a}'_k {\hat{c}}_k(t)
 - 2i \tilde{b}'_k {\hat{c}}^{\dagger}_{-k}(t),
\\
\label{eq:exact_1d_heisenberg_akp}
 \tilde{a}'_k &= g + \cos k,
\\
\label{eq:exact_1d_heisenberg_bkp}
 \tilde{b}'_k &= \sin k,
\end{align}
where the prime symbols indicate the parameters after the quench.
As in the static case,
we can introduce $\tilde{u}_k(t)$ and $\tilde{v}_k(t)$
for the Bogoliubov transformation at time $t$
\begin{align}
\label{eq:bogoliubov_time}
 \begin{pmatrix}
  {\hat{c}}_k(t) \\
  {\hat{c}}^{\dagger}_{-k}(t)
 \end{pmatrix}
 &=
 \begin{pmatrix}
  \tilde{u}_k(t) & -\tilde{v}^{*}_k(t) \\
  \tilde{v}_k(t) & \tilde{u}^{*}_k(t)
 \end{pmatrix}
 \begin{pmatrix}
  {\hat{\gamma}}_k \\
  {\hat{\gamma}}^{\dagger}_{-k}
 \end{pmatrix}
\end{align}
satisfying
\begin{align}
 \tilde{u}_{-k}(t) &= \tilde{u}_k(t),
\\
 \tilde{v}_{-k}(t) &= -\tilde{v}_k(t),
\\
 |\tilde{u}_k(t)|^2 + |\tilde{v}_k(t)|^2 &= 1.
\end{align}
Here ${\hat{\gamma}}_k$ corresponds to
the Bogoliubov excitations before the quench.
From Eqs.~\eqref{eq:exact_1d_heisenberg_ck}
and \eqref{eq:bogoliubov_time},
$\tilde{u}_k(t)$ and $\tilde{v}_k(t)$
should satisfy
\begin{align}
 i \frac{d}{dt}
 \begin{pmatrix}
  \tilde{u}_k(t) \\
  \tilde{v}_k(t)
 \end{pmatrix}
 =
 \begin{pmatrix}
  -2\tilde{a}'_k & -2i\tilde{b}'_k \\
  2i\tilde{b}'_k & 2\tilde{a}'_k
 \end{pmatrix}
 \begin{pmatrix}
  \tilde{u}_k(t) \\
  \tilde{v}_k(t)
 \end{pmatrix}.
\end{align}
Then,
for the sudden quench ($g_0\rightarrow g$),
$\tilde{u}_k(t)$ and $\tilde{v}_k(t)$
are explicitly given as
\begin{align}
 \begin{pmatrix}
  \tilde{u}_k(t) \\
  \tilde{v}_k(t)
 \end{pmatrix}
 &=
 \begin{pmatrix}
  u_k \cos 2\omega'_k t
  + i \frac{\tilde{a}'_k u_k + \tilde{b}'_k v_k}{\omega'_k} \sin 2\omega'_k t \\
  - i v_k \cos 2\omega'_k t
  + \frac{\tilde{b}'_k u_k - \tilde{a}'_k v_k}{\omega'_k} \sin 2\omega'_k t
 \end{pmatrix},
\end{align}
where each variable is represented as
\begin{align}
 \tilde{a}_k &= g_0 + \cos k,
\\
 \tilde{b}_k &= \sin k,
\\
 \omega_k &= \sqrt{g_0^2 + 2g_0 \cos k + 1},
\\
 \omega'_k &= \sqrt{g^2 + 2g \cos k + 1},
\end{align}
and Eqs.~\eqref{eq:exact_1d_heisenberg_akp}
and \eqref{eq:exact_1d_heisenberg_bkp}.
The parameters $u_k$ and $v_k$ are defined in
Eqs.~\eqref{eq:exact_1d_uk} and \eqref{eq:exact_1d_vk}.

As in the case of $t=0$, the Heisenberg representation of each operator satisfies
\begin{align}
 {\hat{A}}_i(t) &= {\hat{a}}^{\dagger}_i(t) + {\hat{a}}_i(t),
\\
 {\hat{B}}_i(t) &= {\hat{b}}^{\dagger}_i(t) - {\hat{b}}_i(t)
\end{align}
with
\begin{align}
 {\hat{a}}_i(t) &=
 \frac{1}{\sqrt{L}} \sum_{k>0}
 \bigl\{
   e^{ikr_j}  [\tilde{u}_k(t) + \tilde{v}_k(t)] {\hat{\gamma}}_k
\nonumber
\\
 &\phantom{=}
 + e^{-ikr_j} [\tilde{u}_k(t) - \tilde{v}_k(t)] {\hat{\gamma}}_{-k}
 \bigr\},
\\
 {\hat{b}}_i(t) &=
 \frac{1}{\sqrt{L}} \sum_{k>0}
 \bigl\{
   e^{ikr_j}  [\tilde{u}_k(t) - \tilde{v}_k(t)] {\hat{\gamma}}_k
\nonumber
\\
 &\phantom{=}
 + e^{-ikr_j} [\tilde{u}_k(t) + \tilde{v}_k(t)] {\hat{\gamma}}_{-k}
 \bigr\}.
\end{align}
Note that
${\hat{a}}_i(t)\not = e^{i{\hat{H}'}t}{\hat{a}}_ie^{-i{\hat{H}'}t}$
and
${\hat{b}}_i(t)\not = e^{i{\hat{H}'}t}{\hat{b}}_ie^{-i{\hat{H}'}t}$
in our notation.
Then, the commutation relations for the operators are
\begin{align}
 &
   \langle \{ {\hat{a}}_i(t), {\hat{a}}_j(t) \} \rangle
 = \langle \{ {\hat{b}}_i(t), {\hat{b}}_j(t) \} \rangle
 = \langle \{ {\hat{a}}_i(t), {\hat{b}}_j(t) \} \rangle
 = 0,
\\
 &
 \langle \{ {\hat{a}}_i(t), {\hat{a}}^{\dagger}_j(t) \} \rangle =: - G^{aa}_{i-j}(t),
\\
 &
 \langle \{ {\hat{b}}_i(t), {\hat{b}}^{\dagger}_j(t) \} \rangle =: - G^{bb}_{i-j}(t),
\\
 &
 \langle \{ {\hat{a}}_i(t), {\hat{b}}^{\dagger}_j(t) \} \rangle =
 \langle \{ {\hat{a}}^{\dagger}_i(t), {\hat{b}}_j(t) \} \rangle
 =: - G^{ab}_{i-j}(t)
\end{align}
with
\begin{align}
 &
 G^{aa}_{i-j}(t) =
 -
 \frac{2}{L} \sum_{k>0}
 \bigl\{
   \cos[k(r_i-r_j)] [|\tilde{u}_k(t)|^2 + |\tilde{v}_k(t)|^2]
\nonumber
\\
 &\phantom{=}
 + i \sin[k(r_i-r_j)] [\tilde{u}_k(t) \tilde{v}^{*}_k(t) + \tilde{v}_k(t) \tilde{u}^{*}_k(t)]
 \bigr\},
\\
 &
 G^{bb}_{i-j}(t) =
 -
 \frac{2}{L} \sum_{k>0}
 \bigl\{
   \cos[k(r_i-r_j)] [|\tilde{u}_k(t)|^2 + |\tilde{v}_k(t)|^2]
\nonumber
\\
 &\phantom{=}
 - i \sin[k(r_i-r_j)] [\tilde{u}_k(t) \tilde{v}^{*}_k(t) + \tilde{v}_k(t) \tilde{u}^{*}_k(t)]
 \bigr\},
\\
 &
 G^{ab}_{i-j}(t) =
 -
 \frac{2}{L} \sum_{k>0}
 \bigl\{
   \cos[k(r_i-r_j)] [|\tilde{u}_k(t)|^2 - |\tilde{v}_k(t)|^2]
\nonumber
\\
 &\phantom{=}
 - i \sin[k(r_i-r_j)] [\tilde{u}_k(t) \tilde{v}^{*}_k(t) - \tilde{v}_k(t) \tilde{u}^{*}_k(t)]
 \bigr\}.
\end{align}
Using these results, we obtain
\begin{align}
 S_{0,r}
 &= \langle {\hat{B}}_0(t) {\hat{B}}_r(t) \rangle
 = - \langle \{{\hat{b}}_0(t), {\hat{b}}^{\dagger}_r(t)\} \rangle
 = + G^{bb}_{-r}(t),
\\
 Q_{0,r}
 &= \langle {\hat{A}}_0(t) {\hat{A}}_r(t) \rangle
 = + \langle \{{\hat{a}}_0(t), {\hat{a}}^{\dagger}_r(t)\} \rangle
 = - G^{aa}_{-r}(t),
\\
 G_{0,r}
 &= \langle {\hat{B}}_0(t) {\hat{A}}_r(t) \rangle
 = - \langle \{{\hat{b}}_0(t), {\hat{a}}^{\dagger}_r(t)\} \rangle
 = + G^{ab}_{+r}(t).
\end{align}

\subsection{Maximum group velocity}
\label{sec:app1_max_vgroup}

In the 1D transverse-field Ising model,
the Lieb-Robinson velocity
is obtained as
the maximum group velocity determined from
the derivative of the band
dispersion~\cite{calabrese2011,cheneau2012,jurcevic2014,gong2022}.
It is given as
\begin{align}
 v^{\mathrm{LR}}
 &=
 \max_{k} \left|\frac{d\omega_k}{dk}\right| 
 =
\begin{cases} 
 2\tilde{g} & \text{if $\tilde{g}\le 1$},\\
 2 & \text{if $\tilde{g}\ge 1$}
\end{cases}
\end{align}
for the Hamiltonian defined in Eq.~\eqref{eq:exact_1d_hamiltonian_g},
and it is obtained as
\begin{align}
 v^{\mathrm{LR}}
 &=
\begin{cases}
 \Gamma & \text{if $\Gamma\le J/2$},\\
 J/2 & \text{if $\Gamma\ge J/2$}
\end{cases}
\end{align}
for the original Hamiltonian given in Eq.~\eqref{eq:tfising_model}.
 \section{Details of the LSWA}
\label{sec:app_sw_corr}

\subsection{Bosonic quadratic Hamiltonian}

We consider a sudden quench within the disordered phase
for the Hamiltonian in Eq.~\eqref{eq:tfising_model}.
We investigate the effect of small quantum fluctuations
around the completely disordered state
at $\Gamma\rightarrow\infty$
using a linear
spin-wave expansion~\cite{henry2012,cevolani2016,buyskikh2016,menu2018,menu2023}.
As long as we consider a quench to a strong transverse field
so that the transverse magnetization is large enough
($\langle S_i^x \rangle \approx 1/2$),
this approach should be a good approximation.
We specifically study the parameter region
$\Gamma \in (\Gamma_{\mathrm{c}}^{\mathrm{classical}},\infty)$,
where
the classical transition point obtained by the mean-field
approximation~\cite{ovchinnikov2003,kato2015,kaneko2021} is
$\Gamma_{\mathrm{c}}^{\mathrm{classical}}=JD$
with $D$ being the spatial dimension.
We review the derivation of the longitudinal correlation
functions~\cite{cevolani2016} and then calculate
the transverse correlation functions,
which have not been investigated in previous studies.

We apply the linearized Holstein-Primakoff transformation,
which is given as
\begin{align}
 {\hat{S}}_i^x &= S - {\hat{b}}^{\dagger}_i {\hat{b}}_i, \quad
 {\hat{S}}_i^z  = \frac{\sqrt{2S}}{2}({\hat{b}}^{\dagger}_i + {\hat{b}}_i)
\end{align}
before the quench and is represented as
\begin{align}
 {{}{\hat{S}}_i^x}' &= S - {\hat{a}}^{\dagger}_i {\hat{a}}_i, \quad
 {{}{\hat{S}}_i^z}'  = \frac{\sqrt{2S}}{2}({\hat{a}}^{\dagger}_i + {\hat{a}}_i)
\end{align}
after the quench.
The prime symbols indicate operators after the quench.
After the Fourier transformation
($
{\hat{b}}_i = \frac{1}{\sqrt{L^D}} \sum_{\bm{k}} e^{-i\bm{k}\cdot\bm{r}_i} {\hat{b}}_{\bm{k}}
$,
$
{\hat{a}}_i = \frac{1}{\sqrt{L^D}} \sum_{\bm{k}} e^{-i\bm{k}\cdot\bm{r}_i} {\hat{a}}_{\bm{k}}
$)
and the Bogoliubov transformation,
we obtain the Hamiltonian for free bosons
before the quench (up to constant terms) as
\begin{align}
 {\hat{H}} &= \sum_{\bm{k}} \Omega_{\bm{k}} {\hat{\beta}}^{\dagger}_{\bm{k}} {\hat{\beta}}_{\bm{k}},
\\
 {\hat{b}}_{\bm{k}} &= s_{\bm{k}} {\hat{\beta}}_{\bm{k}} + t_{\bm{k}} {\hat{\beta}}^{\dagger}_{-\bm{k}},
\end{align}
and that after the quench (up to constant terms) as
\begin{align}
 {\hat{H}'} &= \sum_{\bm{k}} \Omega'_{\bm{k}} {\hat{\alpha}}^{\dagger}_{\bm{k}}
{\hat{\alpha}}_{\bm{k}},
\\
 {\hat{a}}_{\bm{k}} &= s'_{\bm{k}} {\hat{\alpha}}_{\bm{k}} + t'_{\bm{k}} {\hat{\alpha}}^{\dagger}_{-\bm{k}}.
\end{align}
Here the corresponding dispersions and coefficients are defined as
\begin{align}
\label{eq:sw_def_omega}
 \Omega_{\bm{k}} &= {\rm sgn}(A_{\bm{k}}) \sqrt{A_{\bm{k}}^2 - B_{\bm{k}}^2},
\\
\label{eq:sw_def_s}
 s_{\bm{k}} &= {\rm sgn}(A_{\bm{k}})
 \sqrt{\frac{1}{2}\left(\frac{|A_{\bm{k}}|}{|\Omega_{\bm{k}}|}+1\right)},
\\
\label{eq:sw_def_t}
 t_{\bm{k}} &= - {\rm sgn}(B_{\bm{k}})
 \sqrt{\frac{1}{2}\left(\frac{|A_{\bm{k}}|}{|\Omega_{\bm{k}}|}-1\right)},
\\  
\label{eq:sw_def_omega'}
 \Omega'_{\bm{k}} &= {\rm sgn}(A'_{\bm{k}}) \sqrt{{A_{\bm{k}}'}^2 - {B_{\bm{k}}'}^2},
\\
\label{eq:sw_def_s'}
 s'_{\bm{k}} &= {\rm sgn}(A'_{\bm{k}})
 \sqrt{\frac{1}{2}\left(\frac{|A'_{\bm{k}}|}{|\Omega'_{\bm{k}}|}+1\right)},
\\
\label{eq:sw_def_t'}
 t'_{\bm{k}} &= - {\rm sgn}(B'_{\bm{k}})
 \sqrt{\frac{1}{2}\left(\frac{|A'_{\bm{k}}|}{|\Omega'_{\bm{k}}|}-1\right)},
\end{align}
where
\begin{align}
\label{eq:sw_def_AB}
 A_{\bm{k}} &= - \frac{z}{2} J S \gamma_{\bm{k}} + \Gamma,
\quad
 B_{\bm{k}}  = - \frac{z}{2} J S \gamma_{\bm{k}},
\\
\label{eq:sw_def_AB'}
 A'_{\bm{k}} &= - \frac{z}{2} J' S \gamma_{\bm{k}} + \Gamma',
\quad
 B'_{\bm{k}}  = - \frac{z}{2} J' S \gamma_{\bm{k}},
\\
\label{eq:sw_def_gamma}
 \gamma_{\bm{k}} &= \frac{1}{D} \sum_{\nu=1}^{D} \cos k_{\nu}
\end{align}
with
$z=2D$ being the coordination number.
We add the prime symbols to distinguish parameters after the quench.

At $t=0$,
the vacuums of both Hamiltonians are the same,
and bosons before the Holstein-Primakoff transformation satisfy
${\hat{b}}_{\bm{k}} = {\hat{a}}_{\bm{k}}$.
Then, these operators should fulfill
\begin{align}
 \begin{pmatrix}
  {\hat{b}}_{\bm{k}} \\
  {\hat{b}}^{\dagger}_{-\bm{k}}
 \end{pmatrix}
&=
 \begin{pmatrix}
  s_{\bm{k}} & t_{\bm{k}} \\
  t_{\bm{k}} & s_{\bm{k}} 
 \end{pmatrix}
 \begin{pmatrix}
  {\hat{\beta}}_{\bm{k}} \\
  {\hat{\beta}}^{\dagger}_{-\bm{k}}
 \end{pmatrix}
\nonumber
\\
&=
 \begin{pmatrix}
  s'_{\bm{k}} & t'_{\bm{k}} \\
  t'_{\bm{k}} & s'_{\bm{k}} 
 \end{pmatrix}
 \begin{pmatrix}
  {\hat{\alpha}}_{\bm{k}} \\
  {\hat{\alpha}}^{\dagger}_{-\bm{k}}
 \end{pmatrix}
=
 \begin{pmatrix}
  {\hat{a}}_{\bm{k}} \\
  {\hat{a}}^{\dagger}_{-\bm{k}}
 \end{pmatrix}.
\end{align}
This means that Bogoliubov excitations
before and after the quench are connected by
\begin{align}
 \begin{pmatrix}
  {\hat{\alpha}}_{\bm{k}} \\
  {\hat{\alpha}}^{\dagger}_{-\bm{k}}
 \end{pmatrix}
&=
 \begin{pmatrix}
  s'_{\bm{k}} s_{\bm{k}} - t'_{\bm{k}} t_{\bm{k}} & s'_{\bm{k}} t_{\bm{k}} - s_{\bm{k}} t'_{\bm{k}} \\
  s'_{\bm{k}} t_{\bm{k}} - s_{\bm{k}} t'_{\bm{k}} & s'_{\bm{k}} s_{\bm{k}} - t'_{\bm{k}} t_{\bm{k}}
 \end{pmatrix}
 \begin{pmatrix}
  {\hat{\beta}}_{\bm{k}} \\
  {\hat{\beta}}^{\dagger}_{-\bm{k}}
 \end{pmatrix}
\nonumber
\\
\label{eq:sw_uvst}
&=:
 \begin{pmatrix}
  u_{\bm{k}} & v_{\bm{k}} \\
  v_{\bm{k}} & u_{\bm{k}} 
 \end{pmatrix}
 \begin{pmatrix}
  {\hat{\beta}}_{\bm{k}} \\
  {\hat{\beta}}^{\dagger}_{-\bm{k}}
 \end{pmatrix},
\end{align}
where the coefficients satisfy
\begin{align}
   u_{\bm{k}}^2 - v_{\bm{k}}^2
 = {s'_{\bm{k}}}^2 - {t'_{\bm{k}}}^2
 = s_{\bm{k}}^2 - t_{\bm{k}}^2
 = 1.
\end{align}

\subsection{Longitudinal correlation functions}
\label{sec:app_sw_corr_longitudinal}

We evaluate the time-dependent longitudinal correlation functions
defined as
\begin{align}
 C^{zz}(\bm{r},t)
 &=
 \langle\psi_0|
 e^{i{\hat{H}'}t}
 {\hat{S}}^z_{\bm{r}}
 {\hat{S}}^z_{\bm{0}}
 e^{-i{\hat{H}'}t}
 |\psi_0\rangle
\\
 &=
 \frac{S}{2}
 \langle\psi_0|
 e^{i{\hat{H}'}t}
 ({\hat{b}}^{\dagger}_{\bm{r}} + {\hat{b}}_{\bm{r}})
 ({\hat{b}}^{\dagger}_{\bm{0}} + {\hat{b}}_{\bm{0}})
 e^{-i{\hat{H}'}t}
 |\psi_0\rangle.
\end{align}
Writing them in the Fourier space
(${\hat{b}}_i = \frac{1}{\sqrt{L^D}} \sum_{\bm{k}} e^{-i\bm{k}\cdot\bm{r}_i} {\hat{b}}_{\bm{k}}$)
and in the Heisenberg picture,
we obtain
\begin{align}
 C^{zz}(\bm{r},t)
 &=
 \frac{S}{2L^D} \sum_{\bm{k}} e^{i\bm{k}\cdot\bm{r}}
 \langle\psi_0|[
   {\hat{b}}^{\dagger}_{\bm{k}}(t) {\hat{b}}^{\dagger}_{-\bm{k}}(t)
 + {\hat{b}}^{\dagger}_{\bm{k}}(t) {\hat{b}}_{\bm{k}}(t)
\nonumber
\\
 &\phantom{=}
 + {\hat{b}}_{-\bm{k}}(t) {\hat{b}}^{\dagger}_{-\bm{k}}(t)
 + {\hat{b}}_{-\bm{k}}(t) {\hat{b}}_{\bm{k}}(t)
 ]|\psi_0\rangle.
\end{align}
We then replace all operators ${\hat{b}}_{\bm{k}}(t)$ by ${\hat{\beta}}_{\bm{k}}$.
Because
${\hat{b}}_{\bm{k}}(t) = s'_{\bm{k}} {\hat{\alpha}}_{\bm{k}}(t) + t'_{\bm{k}} {\hat{\alpha}}^{\dagger}_{-\bm{k}}(t)$,
${\hat{\alpha}}_{\bm{k}}(t) = e^{-i\Omega'_{\bm{k}}t}{\hat{\alpha}}_{\bm{k}}$,
and 
${\hat{\alpha}}_{\bm{k}} = u_{\bm{k}} {\hat{\beta}}_{\bm{k}} + v_{\bm{k}} {\hat{\beta}}^{\dagger}_{-\bm{k}}$,
the following relation holds:
\begin{align}
 {\hat{b}}_{\bm{k}}(t)
 &= (e^{-i\Omega'_{\bm{k}}t} s'_{\bm{k}} u_{\bm{k}} + e^{+i\Omega'_{\bm{k}}t} t'_{\bm{k}} v_{\bm{k}}) {\hat{\beta}}_{\bm{k}}
\nonumber
\\
 &\phantom{=}
  + (e^{-i\Omega'_{\bm{k}}t} s'_{\bm{k}} v_{\bm{k}} + e^{+i\Omega'_{\bm{k}}t} t'_{\bm{k}} u_{\bm{k}}) {\hat{\beta}}^{\dagger}_{-\bm{k}}.
\end{align}
After straightforward calculations
using ${\hat{\beta}}_{\bm{k}}|\psi_0\rangle = 0$, we obtain
\begin{align}
 C^{zz}(\bm{r},t)
 &=
 \frac{S}{2L^D} \sum_{\bm{k}} e^{i\bm{k}\cdot\bm{r}}
 (s'_{\bm{k}} + t'_{\bm{k}})^2
\nonumber
\\
 &\phantom{=}\cdot
 (u_{\bm{k}}^2 + v_{\bm{k}}^2 + 2 u_{\bm{k}} v_{\bm{k}} \cos 2\Omega'_{\bm{k}} t).
\end{align}
Substituting $u_{\bm{k}}$ and $v_{\bm{k}}$ with
$s_{\bm{k}}$, $s'_{\bm{k}}$, $t_{\bm{k}}$, and $t'_{\bm{k}}$
using Eq.~\eqref{eq:sw_uvst},
we get
\begin{align}
 C^{zz}(\bm{r},0)
 &=
 \frac{S}{2L^D} \sum_{\bm{k}} e^{i\bm{k}\cdot\bm{r}}
 (s'_{\bm{k}} + t'_{\bm{k}})^2
 (u_{\bm{k}} + v_{\bm{k}})^2
\\
 &=
 \frac{S}{2L^D} \sum_{\bm{k}} e^{i\bm{k}\cdot\bm{r}}
 (s_{\bm{k}} + t_{\bm{k}})^2,
\\
 \tilde{C}^{zz}(\bm{r},t)
 &:=
 C^{zz}(\bm{r},t) - C^{zz}(\bm{r},0)
\\
 &=
 \frac{S}{2L^D} \sum_{\bm{k}} e^{i\bm{k}\cdot\bm{r}}
 (s'_{\bm{k}} + t'_{\bm{k}})^2
 2 u_{\bm{k}} v_{\bm{k}} (\cos 2\Omega'_{\bm{k}} t - 1)
\\
 &=
 \frac{S}{2L^D} \sum_{\bm{k}} e^{i\bm{k}\cdot\bm{r}}
 (-2)
 [(s_{\bm{k}}^2 + t_{\bm{k}}^2) s'_{\bm{k}} t'_{\bm{k}}
\nonumber
\\
 &\phantom{=}
 -({s'_{\bm{k}}}^2 + {t'_{\bm{k}}}^2) s_{\bm{k}} t_{\bm{k}}]
 (s'_{\bm{k}} + t'_{\bm{k}})^2
 (\cos 2\Omega'_{\bm{k}} t - 1).
\end{align}
Using the relations defined in
Eqs.~(\ref{eq:sw_def_omega})--(\ref{eq:sw_def_gamma}),
we finally get~\cite{cevolani2016}
\begin{align}
 C^{zz}(\bm{r},0)
 &=
 \frac{S}{2L^D} \sum_{\bm{k}} e^{i\bm{k}\cdot\bm{r}}
 \frac{\Omega_{\bm{k}}}{A_{\bm{k}} + B_{\bm{k}}},
\\
 \tilde{C}^{zz}(\bm{r},t)
 &=
 \frac{S}{2L^D} \sum_{\bm{k}} e^{i\bm{k}\cdot\bm{r}}
 \frac{A_{\bm{k}} B'_{\bm{k}} - A'_{\bm{k}} B_{\bm{k}}}{\Omega_{\bm{k}}(A'_{\bm{k}} + B'_{\bm{k}})}
 (\cos 2\Omega'_{\bm{k}} t - 1).
\end{align}

For $\Gamma\rightarrow\infty$ before the quench,
$\Omega_{\bm{k}}/(A_{\bm{k}}+B_{\bm{k}}) = 1$ is satisfied,
and hence, $C^{zz}(\bm{r},0) = S/2 \times \delta_{\bm{r},L\bm{m}}$
($m_\nu$: integer, $\nu=1,2,\dots,D$) holds.
This means that
$C^{zz}(\bm{r},t) = \tilde{C}^{zz}(\bm{r},t)$
for $1\le r_{\nu}\le L/2$ ($\nu=1,2,\dots,D$).
Besides,
when $J=0$ and $J'\ll \Gamma' < \Gamma$,
the intensity of the correlation would be approximately
$
|C^{zz}(\bm{r},t)|
= \mathcal{O}[
S/L^D \times \sum_{\bm{k}} B'_{\bm{k}} / (A'_{\bm{k}} + B'_{\bm{k}})
]
= \mathcal{O}(
zS^2 J'/\Gamma')
$.

\subsection{Transverse correlation functions}
\label{sec:app_sw_corr_transverse}

We evaluate the time-dependent transverse correlation functions
defined as
\begin{align}
 C^{xx}(\bm{r},t)
 &=
 \langle\psi_0|
 e^{i{\hat{H}'}t}
 {\hat{S}}^x_{\bm{r}}
 {\hat{S}}^x_{\bm{0}} 
 e^{-i{\hat{H}'}t}
 |\psi_0\rangle
\\
 &=
 \langle\psi_0|
 e^{i{\hat{H}'}t}
 (S - {\hat{b}}^{\dagger}_{\bm{r}} {\hat{b}}_{\bm{r}})
 (S - {\hat{b}}^{\dagger}_{\bm{0}} {\hat{b}}_{\bm{0}})
 e^{-i{\hat{H}'}t}
 |\psi_0\rangle
\end{align}
and the connected one defined as
\begin{align}
 C^{xx}_{\rm connected}(\bm{r},t)
 &=
 C^{xx}(\bm{r},t)
 -
 \langle\psi_0|
 e^{i{\hat{H}'}t}
 {\hat{S}}^x_{\bm{r}}
 e^{-i{\hat{H}'}t}
 |\psi_0\rangle
\nonumber
\\
 &\phantom{=}\cdot
 \langle\psi_0|
 e^{i{\hat{H}'}t}
 {\hat{S}}^x_{\bm{0}}
 e^{-i{\hat{H}'}t}
 |\psi_0\rangle.
\end{align}
Writing them in the Fourier space
(${\hat{b}}_i = \frac{1}{\sqrt{L^D}} \sum_{\bm{k}} e^{-i\bm{k}\cdot\bm{r}_i} {\hat{b}}_{\bm{k}}$)
and in the Heisenberg picture,
we obtain
\begin{align}
 &
 C^{xx}(\bm{r},t)
 =
 S^2
 -
 \frac{2S}{L^D} \sum_{\bm{k}}
 \langle\psi_0|
 {\hat{b}}^{\dagger}_{\bm{k}}(t) {\hat{b}}_{\bm{k}}(t)
 |\psi_0\rangle
\nonumber
\\
 &\phantom{=}
 +
 \frac{1}{L^{2D}} \sum_{{\bm{k}},{\bm{l}},{\bm{p}}} e^{i(\bm{k}-\bm{l})\cdot\bm{r}}
 \langle\psi_0|
 {\hat{b}}^{\dagger}_{\bm{k}}(t) {\hat{b}}_{\bm{l}}(t)
 {\hat{b}}^{\dagger}_{\bm{p}}(t) {\hat{b}}_{\bm{k}-\bm{l}+\bm{p}}(t)
 |\psi_0\rangle
\end{align}
and
\begin{align}
 &
 C^{xx}_{\rm connected}(\bm{r},t)
 =
 -
 \left[
 \frac{1}{L^D} \sum_{\bm{k}}
 \langle\psi_0|
 {\hat{b}}^{\dagger}_{\bm{k}}(t) {\hat{b}}_{\bm{k}}(t)
 |\psi_0\rangle
 \right]^2
\nonumber
\\
 &\phantom{=}
 +
 \frac{1}{L^{2D}} \sum_{{\bm{k}},{\bm{l}},{\bm{p}}} e^{i(\bm{k}-\bm{l})\cdot\bm{r}}
 \langle\psi_0|
 {\hat{b}}^{\dagger}_{\bm{k}}(t) {\hat{b}}_{\bm{l}}(t)
 {\hat{b}}^{\dagger}_{\bm{p}}(t) {\hat{b}}_{\bm{k}-\bm{l}+\bm{p}}(t)
 |\psi_0\rangle.
\end{align}
As in the case of longitudinal correlation functions,
we replace ${\hat{b}}_{\bm{k}}(t)$ by ${\hat{\beta}}_{\bm{k}}$
and use ${\hat{\beta}}_{\bm{k}}|\psi_0\rangle = 0$.
Non-vanishing terms contain
$\langle\psi_0| {\hat{\beta}}_{-\bm{k}} {\hat{\beta}}^{\dagger}_{-\bm{l}}
{\hat{\beta}}_{-\bm{p}} {\hat{\beta}}^{\dagger}_{-(\bm{k}-\bm{l}+\bm{p})} |\psi_0\rangle
= \delta_{\bm{k},\bm{l}}$,
$\langle\psi_0| {\hat{\beta}}_{-\bm{k}} {\hat{\beta}}_{\bm{l}}
{\hat{\beta}}^{\dagger}_{\bm{p}} {\hat{\beta}}^{\dagger}_{-(\bm{k}-\bm{l}+\bm{p})} |\psi_0\rangle
= \delta_{\bm{k},-\bm{p}} + \delta_{\bm{l},\bm{p}}$,
and
$\langle\psi_0| {\hat{\beta}}_{-\bm{k}}
{\hat{\beta}}^{\dagger}_{-\bm{k}} |\psi_0\rangle = 1$.
After straightforward calculations, we get
\begin{align}
 &~\phantom{=}~
 \frac{1}{L^D} \sum_{\bm{k}}
 \langle\psi_0|
 {\hat{b}}^{\dagger}_{\bm{k}}(t) {\hat{b}}_{\bm{k}}(t)
 |\psi_0\rangle
\nonumber
\\
 &=
 \frac{1}{L^D} \sum_{\bm{k}}
 ({s'_{\bm{k}}}^2 v_{\bm{k}}^2 + {t'_{\bm{k}}}^2 u_{\bm{k}}^2
 + 2 s'_{\bm{k}} t'_{\bm{k}} u_{\bm{k}} v_{\bm{k}} \cos 2\Omega'_{\bm{k}} t)
\\
 &=
 \frac{1}{L^D} \sum_{\bm{k}}
 \left[
 t_{\bm{k}}^2
 + 2 s'_{\bm{k}} t'_{\bm{k}} u_{\bm{k}} v_{\bm{k}} (\cos 2\Omega'_{\bm{k}} t - 1)
 \right]
\end{align}
and
\begin{widetext}
\begin{align}
 C^{xx}_{\rm connected}(\bm{r},t)
 &=
 \left|
 \frac{1}{L^D} \sum_{\bm{k}} e^{i\bm{k}\cdot\bm{r}}
 \left\{
 [({s'_{\bm{k}}}^2 + {t'_{\bm{k}}}^2) \cos 2\Omega'_{\bm{k}} t + i \sin 2\Omega'_{\bm{k}} t ] u_{\bm{k}} v_{\bm{k}}
 + s'_{\bm{k}} t'_{\bm{k}} ( u_{\bm{k}}^2 + v_{\bm{k}}^2 )
 \right\}
 \right|^2
\nonumber
\\
 &~\phantom{=}~
 +
 \frac{1}{L^{2D}} \sum_{\bm{k},\bm{l}} e^{i(\bm{k}-\bm{l})\cdot\bm{r}}
 \left[
 t_{\bm{k}}^2
 + 2 s'_{\bm{k}} t'_{\bm{k}} u_{\bm{k}} v_{\bm{k}} (\cos 2\Omega'_{\bm{k}} t - 1)
 \right]
 \left[
 s_{\bm{l}}^2
 + 2 s'_{\bm{l}} t'_{\bm{l}} u_{\bm{l}} v_{\bm{l}} (\cos 2\Omega'_{\bm{l}} t - 1)
 \right]
\\
 &=
 \left|
 \frac{1}{L^D} \sum_{\bm{k}} e^{i\bm{k}\cdot\bm{r}}
 \left\{
 [({s'_{\bm{k}}}^2 + {t'_{\bm{k}}}^2) \cos 2\Omega'_{\bm{k}} t + i \sin 2\Omega'_{\bm{k}} t ] u_{\bm{k}} v_{\bm{k}}
 + s'_{\bm{k}} t'_{\bm{k}} ( u_{\bm{k}}^2 + v_{\bm{k}}^2 )
 \right\}
 \right|^2
\nonumber
\\
 &~\phantom{=}~
 +
 \left|
 \frac{1}{L^D} \sum_{\bm{k}} e^{i\bm{k}\cdot\bm{r}}
 \left[
 t_{\bm{k}}^2
 + 2 s'_{\bm{k}} t'_{\bm{k}} u_{\bm{k}} v_{\bm{k}} (\cos 2\Omega'_{\bm{k}} t - 1)
 \right]
 \right|^2
\nonumber
\\
 &~\phantom{==}~
 +
 \frac{1}{L^D} \sum_{\bm{k}}
 \left[
 t_{\bm{k}}^2
 + 2 s'_{\bm{k}} t'_{\bm{k}} u_{\bm{k}} v_{\bm{k}} (\cos 2\Omega'_{\bm{k}} t - 1)
 \right]
 \times
 \delta_{\bm{r},L\bm{m}}
\end{align}
\end{widetext}
with $m_\nu$ ($\nu=1,2,\dots,D$) being integer.
Substituting
the parameters
$u_{\bm{k}}$, $v_{\bm{k}}$,
$s_{\bm{k}}$, $t_{\bm{k}}$,
$s'_{\bm{k}}$, and $t'_{\bm{k}}$
with the parameters
$A_{\bm{k}}$, $B_{\bm{k}}$, $\Omega_{\bm{k}}$,
$A'_{\bm{k}}$, $B'_{\bm{k}}$, and $\Omega'_{\bm{k}}$,
we finally get
\begin{align}
 &~\phantom{=}~
 \frac{1}{L^D} \sum_{\bm{k}}
 \langle\psi_0| 
 {\hat{b}}^{\dagger}_{\bm{k}}(t) {\hat{b}}_{\bm{k}}(t)
 |\psi_0\rangle
\nonumber
\\ 
 &=
 \frac{1}{L^D} \sum_{\bm{k}} e^{i\bm{k}\cdot\bm{r}}
 \Biggl[
 \left(
 \frac{A'_{\bm{k}}}{2}
 \frac{A_{\bm{k}} A'_{\bm{k}} - B_{\bm{k}} B'_{\bm{k}}}{\Omega_{\bm{k}} {\Omega_{\bm{k}}'}^2}
 - \frac{1}{2}
 \right)
\nonumber
\\
 &~\phantom{=}~
 - \frac{B'_{\bm{k}}}{2} \frac{A_{\bm{k}} B'_{\bm{k}} - A'_{\bm{k}} B_{\bm{k}}}{\Omega_{\bm{k}} {\Omega_{\bm{k}}'}^2}
 \cos 2\Omega'_{\bm{k}} t
 \Biggr]
\\
 &=
 \frac{1}{L^D} \sum_{\bm{k}} e^{i\bm{k}\cdot\bm{r}}   
 \Biggl[
 \frac{1}{2}\left(\frac{A_{\bm{k}}}{\Omega_{\bm{k}}} - 1\right)
\nonumber
\\
 &~\phantom{=}~
 - \frac{B'_{\bm{k}}}{2} \frac{A_{\bm{k}} B'_{\bm{k}} - A'_{\bm{k}} B_{\bm{k}}}{\Omega_{\bm{k}} {\Omega_{\bm{k}}'}^2}
 (\cos 2\Omega'_{\bm{k}} t - 1)
 \Biggr]
\end{align}
and
\begin{widetext}
\begin{align}
 C^{xx}_{\rm connected}(\bm{r},t)
 &=
 \left|
 \frac{1}{L^D} \sum_{\bm{k}} e^{i\bm{k}\cdot\bm{r}}
 \left[
 -
 \frac{B'_{\bm{k}}}{2}
 \frac{A_{\bm{k}} A'_{\bm{k}} - B_{\bm{k}} B'_{\bm{k}}}{\Omega_{\bm{k}} {\Omega_{\bm{k}}'}^2}
 +
 \frac{A_{\bm{k}} B'_{\bm{k}} - A'_{\bm{k}} B_{\bm{k}}}{2 \Omega_{\bm{k}} \Omega'_{\bm{k}}}
 \left(\frac{A'_{\bm{k}}}{\Omega'_{\bm{k}}} \cos 2\Omega'_{\bm{k}} t + i \sin 2\Omega'_{\bm{k}} t \right)
 \right]
 \right|^2
\nonumber
\\
 &~\phantom{=}~
 +
 \left|
 \frac{1}{L^D} \sum_{\bm{k}} e^{i\bm{k}\cdot\bm{r}}
 \left[
 \left(
 \frac{A'_{\bm{k}}}{2}
 \frac{A_{\bm{k}} A'_{\bm{k}} - B_{\bm{k}} B'_{\bm{k}}}{\Omega_{\bm{k}} {\Omega_{\bm{k}}'}^2}
 - \frac{1}{2}
 \right)
 - \frac{B'_{\bm{k}}}{2} \frac{A_{\bm{k}} B'_{\bm{k}} - A'_{\bm{k}} B_{\bm{k}}}{\Omega_{\bm{k}} {\Omega_{\bm{k}}'}^2}
 \cos 2\Omega'_{\bm{k}} t
 \right]
 \right|^2
\nonumber
\\
 &~\phantom{==}~
 +
 \frac{1}{L^D} \sum_{\bm{k}}
 \left[
 \left(
 \frac{A'_{\bm{k}}}{2}
 \frac{A_{\bm{k}} A'_{\bm{k}} - B_{\bm{k}} B'_{\bm{k}}}{\Omega_{\bm{k}} {\Omega_{\bm{k}}'}^2}
 - \frac{1}{2}
 \right)
 - \frac{B'_{\bm{k}}}{2} \frac{A_{\bm{k}} B'_{\bm{k}} - A'_{\bm{k}} B_{\bm{k}}}{\Omega_{\bm{k}} {\Omega_{\bm{k}}'}^2}
 \cos 2\Omega'_{\bm{k}} t
 \right]
 \times
 \delta_{\bm{r},L\bm{m}}
\\
 &=
 \left|
 \frac{1}{L^D} \sum_{\bm{k}} e^{i\bm{k}\cdot\bm{r}}
 \left\{
 -
 \frac{B_{\bm{k}}}{2\Omega_{\bm{k}}}
 +
 \frac{A_{\bm{k}} B'_{\bm{k}} - A'_{\bm{k}} B_{\bm{k}}}{2 \Omega_{\bm{k}} \Omega'_{\bm{k}}}
 \left[\frac{A'_{\bm{k}}}{\Omega'_{\bm{k}}} (\cos 2\Omega'_{\bm{k}} t - 1) + i \sin 2\Omega'_{\bm{k}} t \right]
 \right\}
 \right|^2
\nonumber
\\
 &~\phantom{=}~
 +
 \left|
 \frac{1}{L^D} \sum_{\bm{k}} e^{i\bm{k}\cdot\bm{r}}
 \left[
 \frac{1}{2}\left(\frac{A_{\bm{k}}}{\Omega_{\bm{k}}} - 1\right)
 - \frac{B'_{\bm{k}}}{2} \frac{A_{\bm{k}} B'_{\bm{k}} - A'_{\bm{k}} B_{\bm{k}}}{\Omega_{\bm{k}} {\Omega_{\bm{k}}'}^2}
 (\cos 2\Omega'_{\bm{k}} t - 1)
 \right]
 \right|^2
\nonumber
\\
 &~\phantom{==}~
 +
 \frac{1}{L^D} \sum_{\bm{k}}
 \left[
 \frac{1}{2}\left(\frac{A_{\bm{k}}}{\Omega_{\bm{k}}} - 1\right)
 - \frac{B'_{\bm{k}}}{2} \frac{A_{\bm{k}} B'_{\bm{k}} - A'_{\bm{k}} B_{\bm{k}}}{\Omega_{\bm{k}} {\Omega_{\bm{k}}'}^2}
 (\cos 2\Omega'_{\bm{k}} t - 1)
 \right]
 \times
 \delta_{\bm{r},L\bm{m}}
\end{align}
\end{widetext}
with $m_\nu$ ($\nu=1,2,\dots,D$) being integer.

When $J=0$ and $J'\ll \Gamma' < \Gamma$,
the intensity of the correlation would be approximately
$
|C^{xx}_{\rm connected}(\bm{r},t)|
= \mathcal{O}[
|1/L^D \times \sum_{\bm{k}} B'_{\bm{k}} A'_{\bm{k}} /
(\Omega'_{\bm{k}})^2|^2
]
= \mathcal{O}(
z^2S^2 J'^2/\Gamma'^2)
$.

\subsection{Dispersion relation and maximum group velocity}
\label{sec:app_sw_corr_disp_max_group_vel}

\begin{figure}[t]
\centering
\includegraphics[width=1.0\columnwidth]{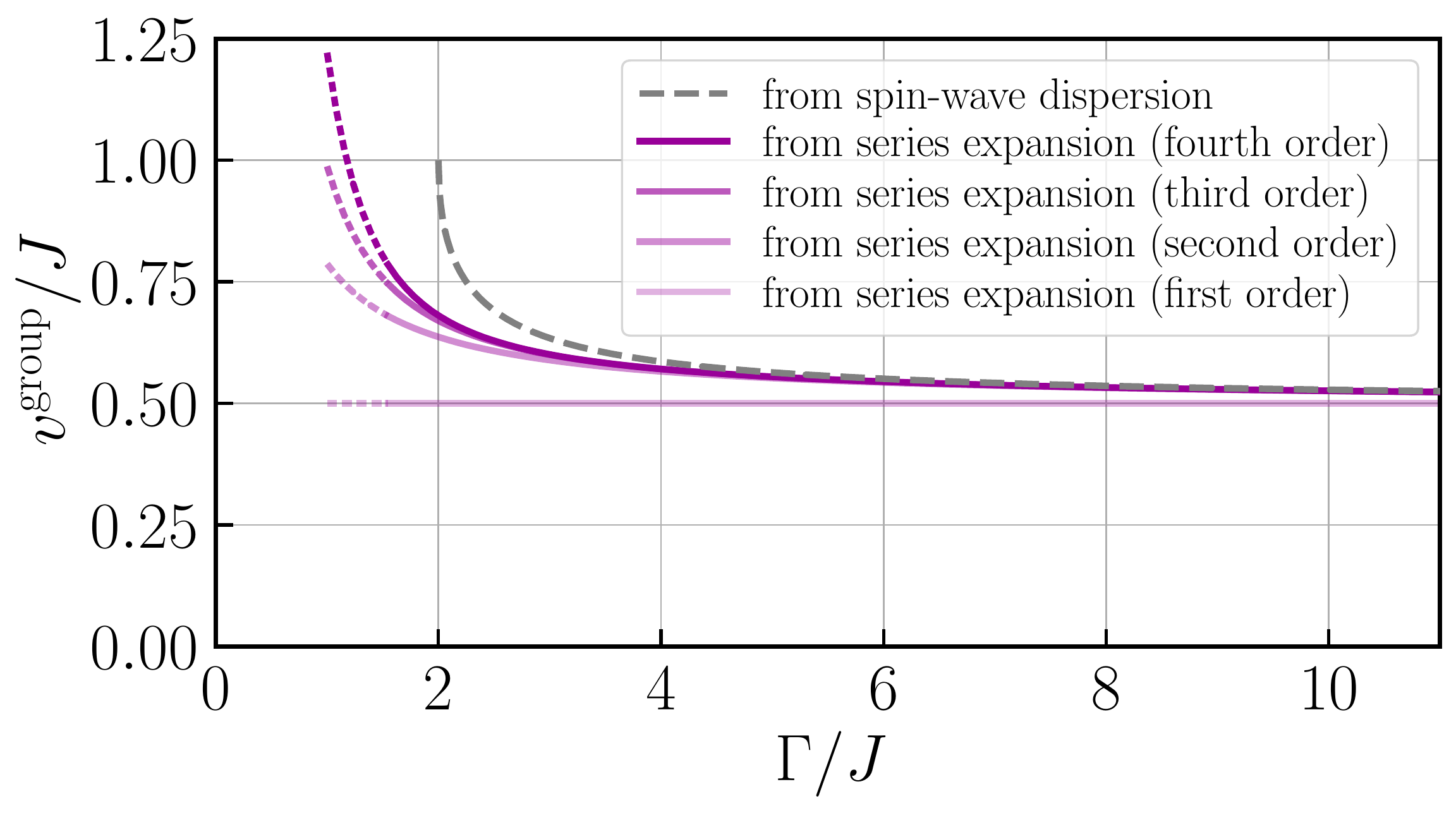}
\caption{Field dependence of the group velocity
along the horizontal axis
estimated from the dispersion relation
obtained by the LSWA
and the series expansion up to
fourth order~\cite{oitmaa2006,hamer2006a,hamer2006b}
in 2D.
The group velocity is represented by the solid (dotted) lines above
(below) the critical transverse field
$\Gamma_{\rm c}/J\approx
1.522$~\cite{rieger1999,bloete2002,kaneko2021}.
Both velocities increase with decreasing the transverse field.
}
\label{fig:2d_perturb}
\end{figure}

Within the LSWA,
the dispersion relation for $S=1/2$ is expressed as
\begin{align}
 \Omega_{\bm{k}}
 &=
 \sqrt{ \Gamma^2 - \frac{z}{2} \Gamma J \gamma_{\bm{k}} },
\\
 &=
 \Gamma
 \left\{
 1
 - \frac{1}{2} \frac{zJ}{2\Gamma} \gamma_{\bm{k}}
 - \frac{1}{8} \left( \frac{zJ}{2\Gamma} \right)^2 \gamma_{\bm{k}}^2
 + \mathcal{O}\left[ \left(\frac{zJ}{2\Gamma}\right)^3 \right]
 \right\}.
\end{align}
This result is consistent with the dispersion relation
\begin{align}
\label{eq:app_disp_perturb}
 &
 \Omega_{\bm{k}}^{\rm perturb}
 =
  \Gamma
 \Biggl\{
 1
 - \frac{1}{2} \frac{zJ}{2\Gamma} \gamma_{\bm{k}}
 - \frac{1}{8} \left( \frac{zJ}{2\Gamma} \right)^2
   \left( \gamma_{\bm{k}}^2 - \frac{2}{9} \right)
\nonumber
\\
 &~\phantom{=}~
 - \frac{1}{16} \left( \frac{zJ}{2\Gamma} \right)^3
   \left( \gamma_{\bm{k}}^3 + \gamma_{\bm{k}} \frac{2-3z}{z^2} \right)
 + \mathcal{O}\left[ \left(\frac{zJ}{2\Gamma}\right)^4 \right]
 \Biggr\}
\end{align}
obtained by the perturbation calculation~\cite{pfeuty1971}
up to $\mathcal{O}\{[(zJ)/(2\Gamma)]^2\}$ terms.

It is widely believed that the Lieb-Robinson velocity
should be the maximum group velocity determined
from the derivative of band
dispersion~\cite{calabrese2011,cheneau2012,jurcevic2014,gong2022}.
Although the dispersion obtained
by the LSWA does not necessarily
offer the exact Lieb-Robinson velocity,
we calculate the reference value using the dispersion.
In 1D, the maximum group velocity of the spin-wave dispersion
is given as
\begin{align}
 v^{\mathrm{SW}}
 = \max_{k} \left|\frac{d\Omega_k}{dk}\right|
 = \frac{J}{\sqrt{2}}
 \left[ 1 + \sqrt{1-\left(\frac{J}{\Gamma}\right)^2} \right]^{-1/2}.
\end{align}
For $\Gamma\rightarrow\infty$,
the group velocity satisfies
$v^{\mathrm{SW}}\rightarrow J/2 = v^{\mathrm{LR}}$,
reproducing the exact maximum group velocity.
On the other hand, for $\Gamma\in (\Gamma_{\mathrm{c}}^{\mathrm{classical}},\infty)$,
the LSWA always gives
$v^{\mathrm{SW}}> J/2 = v^{\mathrm{LR}}$.
Its worst (largest) estimate
$v^{\mathrm{SW}} = J/\sqrt{2} \approx 0.707 J$
at $\Gamma=\Gamma_{\mathrm{c}}^{\mathrm{classical,1D}}(=J)$
is still tighter than the recent bound $1.51J$
obtained by the general formula for the Lieb-Robinson
bound~\cite{wang2020}.

In the same manner,
we can extract the group velocity
as $\bm{v}^{\mathrm{SW}} =
\max_{\bm{k}} \nabla_{\!\bm{k}} \, \Omega_{\bm{k}}$
from the spin-wave dispersion in higher dimensions.
In 2D,
the horizontal and diagonal velocities are given as
\begin{align}
 v^{\mathrm{SW,horizontal}}
 &=
 \frac{J}{\sqrt{2}}
 \left(
 1 - \frac{J}{\Gamma} + \sqrt{1-\frac{2J}{\Gamma}}
 \right)^{-1/2},
\\
 v^{\mathrm{SW,diagonal}}
 &=
 J
 \left[
 1 + \sqrt{1-\left(\frac{2J}{\Gamma}\right)^2}
 \right]^{-1/2},
\end{align}
respectively.
The maximum velocity along the horizontal (diagonal) axis
is estimated to be
$v^{\mathrm{SW,horizontal}} \rightarrow J/2$
($v^{\mathrm{SW,diagonal}} \rightarrow J/\sqrt{2}$)
for $\Gamma\rightarrow\infty$.
On the other hand, for both axes,
it approaches the value $J$
($v^{\mathrm{SW,horizontal}}, v^{\mathrm{SW,diagonal}}
\rightarrow J$) for $\Gamma\rightarrow \Gamma_{\mathrm{c}}
^{\mathrm{classical,2D}}(=2J)$.

The group velocity in 2D obtained by the LSWA
increases with decreasing the transverse field
(see also Sec.~\ref{sec:results_2d_sw}).
As we will see below,
this behavior agrees with that obtained by
a high-order series
expansion~\cite{oitmaa2006,hamer2006a,hamer2006b}.
We extract the group velocity from the dispersion
relation obtained by the series expansion up to fourth order of
$\lambda = J/(2\Gamma)$~\cite{oitmaa2006,hamer2006a,hamer2006b}.
The dispersion relation is
described as
\begin{align}
\label{eq:app_disp_series}
 \Omega^{\mathrm{series}}_{\bm{k}}
 &=
 \Gamma\Biggl[
   \lambda \cdot (-2) \gamma_{\bm{k}}
 + \lambda^2 \cdot (-2) \gamma_{\bm{k}}^2
 + \lambda^3 \left( \frac{5}{2} \gamma_{\bm{k}}
   - 4 \gamma_{\bm{k}}^3 \right)
\nonumber
\\
 &~\phantom{=}~
 + \lambda^4 \left( 7 \gamma_{\bm{k}}^2
   - 10 \gamma_{\bm{k}}^4 \right) 
 + \mathcal{O}(\lambda^5)
 \Biggr]
 +
 \mathrm{const},
\end{align}
where the constant term does not depend on $\bm{k}$
(but depends on $\lambda$ and $\Gamma$).
Note that this relation is consistent with
that in Eq.~\eqref{eq:app_disp_perturb}
for $z=4$ on a square lattice.
We calculate the velocity
$\bm{v}^{\mathrm{series}} =
\max_{\bm{k}} \nabla_{\!\bm{k}} \,
\Omega^{\mathrm{series}}_{\bm{k}}$
numerically and compare it with our result
obtained by the LSWA.
As shown in Fig.~\ref{fig:2d_perturb},
at a fixed transverse field,
the velocity along the horizontal axis
obtained by the series expansion
increases monotonically
as higher-order terms are taken into account.
They are always slower than the velocity obtained by the LSWA.
On the other hand, both velocities
obtained by the LSWA and the series expansion
nearly coincide for strong transverse fields.
They increase with decreasing the transverse field.

\bibliographystyle{apsrev4-2}

\onecolumngrid

\end{document}